\documentclass[10pt,a4paper]{article}
\pdfoutput=1 
\usepackage[usenames,svgnames,xcdraw,table]{xcolor}
\usepackage{todonotes}
\usepackage{amssymb,amsthm,amsmath,mathtools}
\usepackage{thmtools,thm-restate}
\usepackage[square,sort&compress,numbers]{natbib}
\usepackage{pifont}
\newcommand{\cmark}{\ding{51}}
\newcommand{\xmark}{\ding{55}}
\newcommand{\pd}[1]{\frac{\partial}{\partial #1}}
\newcommand{\ppd}[2]{\frac{\partial^2}{\partial#1\partial #2}}
\DeclareMathOperator*{\argmax}{\arg\!\max}

\definecolor{azure}{rgb}{0.0, 0.5, 1.0}
\definecolor{DarkGreen}{rgb}{0.1,0.5,0.1}
\allowdisplaybreaks

\newif\ifremovegamma
\removegammatrue

\newif\ifverbose

\usepackage[ruled,vlined,linesnumbered]{algorithm2e}

\usepackage[backref=page]{hyperref}
\hypersetup{
	colorlinks=true,
	linkcolor=red,
	urlcolor=DarkGreen,
	citecolor=blue
}
\renewcommand*{\backref}[1]{}
\renewcommand*{\backrefalt}[4]{%
    \ifcase #1 (Not cited.)%
    \or        (Cited on page~#2)%
    \else      (Cited on pages~#2)%
    \fi}
\usepackage[capitalise,noabbrev]{cleveref}
\Crefname{theorem}{Theorem}{Theorems}
\Crefname{example}{Example}{Examples}
\Crefname{table}{Table}{Tables}
\Crefname{algorithm}{Algorithm}{Algorithms}
\Crefname{assumption}{Assumption}{Assumptions}
\Crefname{proposition}{Proposition}{Propositions}
\usepackage{tabularx}
\usepackage{nicefrac}
\usepackage[margin=1in]{geometry}
\usepackage[T1]{fontenc}
\usepackage[utf8]{inputenc}
\usepackage{graphicx}
\usepackage{booktabs}

\usepackage[labelfont={normalfont,bf},textfont=it]{caption}
\usepackage{subcaption}
\usepackage[inline]{enumitem}

\theoremstyle{plain}
\newtheorem{theorem}{Theorem}
\newtheorem{lemma}{Lemma}
\newtheorem{corollary}{Corollary}
\newtheorem{proposition}{Proposition}
\newtheorem{definition}{Definition}
\newtheorem{assumption}{Assumption}

\newtheorem{examplex}{Example}
\newenvironment{example}{\pushQED{\qed}\examplex}{\popQED\endexamplex}

\newcommand{\mechfull}{\textrm{\textup{Updated Parameter Best Response Dynamics}}}
\newcommand{\mech}{\texttt{\textup{UPBReD}}}

\newcommand{\mechtwo}{\texttt{\textup{2P-UPBReD}}}
\newcommand{\mechfullthree}{\textrm{\textup{Truthful Updated Parameter Best Response Dynamics}}}
\newcommand{\mechthree}{\texttt{\textup{T-UPBReD}}}
\newcommand{\mechfour}{\texttt{\textup{T-2P-UPBReD}}}
\newcommand{\mechfullfour}{\textrm{\textup{Truthful Two Phase Updated Parameter Best Response Dynamics}}}
\newcommand{\strategic}{\texttt{\textup{FedAvgStrategic}}}
\newcommand{\murhekar}{\texttt{\textup{FedBR-BG}}}
\newcommand{\FedAvg}{\texttt{\textup{FedAvg}}}
\newcommand{\wopt}{w^{\texttt{\textup{OPT}}}}
\newcommand\ddfrac[2]{\frac{\displaystyle #1}{\displaystyle #2}}
\let\displaystyle\textstyle
\title{Incentivize Contribution and Learn Parameters Too: Federated Learning with Strategic Data Owners}
\author{Drashthi Doshi \qquad Avishek Ghosh \qquad Swaprava Nath\\[4pt]
Indian Institute of Technology Bombay\\[2pt]
\texttt{drashthi@iitb.ac.in}, \texttt{avishek\_ghosh@iitb.ac.in}, \texttt{swaprava@iitb.ac.in}}
\date{}

\begin{document}
\maketitle
\begin{abstract}
In federated learning (FL), a center trains a shared model from updates that clients compute on their private data. Collecting and training on data is costly, so a strategic client may contribute little data and free-ride on the model trained by the others. We study mechanisms that incentivize data contribution and learn the model at the same time. Our first mechanism, \mech{}, updates the clients' contributions and the model jointly, and converges at a geometric rate to a Nash equilibrium of the contribution game together with the welfare-maximizing model for those contributions. This equilibrium can involve under-contribution, so our second mechanism, \mechtwo{}, adds budget-balanced transfers that make full contribution a dominant strategy, after which training converges to the optimal model; we also give conditions under which every client prefers to participate. Both mechanisms trust what clients report about the data they use and the gradients they compute. We show that no payment rule can reward unverifiable contributions unless something the center observes depends on the data actually used. Building on this, \mechfour{} checks each client's reported contribution against the other clients' realized valuations of the trained model, and prices gradient reports with a budget-balanced charge. It elicits truthful contribution reports, makes gradient reports approximately truthful, still reaches full contribution and the optimal model, and never runs a deficit. Experiments on CIFAR-10, FEMNIST, and Twitter show that \mechtwo{} matches the welfare of FedAvg trained on all data, while strategic clients without incentives lower the welfare of the learned model.

\end{abstract}
\section{Introduction}
\label{sec:intro}
Accurate machine learning models need large amounts of training data, but in many applications, such as language, vision, and disease modeling, the data are spread across many owners and cannot be pooled. Federated learning (FL)~\cite{konevcny2016federated,mcmahan2017communication,kairouz2021advances,zhang2021survey} trains a shared model without moving the data: a center, the parameter server, maintains the model, and each data owner, which we call an agent, computes updates on its local data and sends only these updates to the center. Since an agent typically holds too little data to learn a good model alone, every agent benefits from the data of the others.

Classical FL assumes that agents contribute all their data voluntarily. In practice, collecting, cleaning, and training on data is costly, while the benefit of a better model is shared by everyone. A rational agent may therefore contribute little data and \emph{free-ride} on the model trained by the others~\citep{karimireddy2022mechanisms}. Consider hospitals in a region that jointly train a disease model on patient records. Sampling and annotating records is expensive, and hospitals differ in how much data they can collect, so a hospital may prefer to contribute little and still receive the shared model. If many hospitals reason this way, the model is trained on little data, and every hospital is worse off. This raises the question:

\smallskip\noindent
{\em ``Can a mechanism incentivize every agent to contribute all of its data and, at the same time, learn the optimal model, even though the center cannot verify how much data an agent actually uses?''}

\smallskip\noindent
Recent work on incentives in FL~\citep{Blum2021OneforOne,murhekar2023incentives,karimireddy2022mechanisms} abstracts training away: the value of a contribution profile is given by a function of the profile, and the model is not learned inside the mechanism. In FL, however, contributions and the model evolve together, round by round, and the center sees only what the agents report, namely how much data they used and the gradients they computed. Neither report can be verified, and an agent may misreport either one to its advantage. This paper treats three questions together: how contributions and the model evolve under joint dynamics, how to remove the inefficiency of the resulting equilibrium with transfers that sum to zero, and how to make the agents' reports truthful.

\begin{table}[t!]
    \centering
    \renewcommand{\arraystretch}{1.15}
    \setlength{\tabcolsep}{4pt}
    \begin{tabularx}{\textwidth}{>{\raggedright\arraybackslash}X|cccccc c}
        \toprule
        & Eq. & Learn. & Transfers & Reports & Truthful & Contrib. & Private\\
        \midrule
        \citet{procaccia2025incentiveshetero} & E & \xmark & BB     & D   & DSIC & P & \xmark \\
        \citet{bornstein2024facttruth}        & E & \xmark & \cmark & C   & DSIC & F & \cmark \\
        \citet{Blum2021OneforOne}             & E & \xmark & \xmark & S   & \xmark & P & \cmark \\
        \citet{murhekar2023incentives}        & C & \xmark & BB     & S,C & \xmark & P & \cmark \\
        \citet{karimireddy2022mechanisms}     & E & \xmark & \cmark & S,C & \xmark & F & \cmark \\
        \midrule
        \mech{} (Thm.~\ref{thm:convergence-genl})
            & C & J   & \xmark   & S,G & \xmark & P & \cmark \\
        \mechthree{} (Thm.~\ref{thm:EPIC})
            & C & J   & BB$^*$   & S,G & EPIC (S)$^\dagger$, $\varepsilon$ (G) & P & \cmark \\
        \mechtwo{} (Thm.~\ref{thm:two_phase}, Prop.~\ref{prop:IR})
            & C & OPT & BB, IR   & S,G & \xmark & F & \cmark \\
        \mechfour{} (Thm.~\ref{thm:truthful_T2P})
            & C & OPT & BB$^*$   & S,G & EPIC (S)$^\dagger$, $\varepsilon$ (G) & F & \cmark \\
        \bottomrule
    \end{tabularx}
    \caption{\textbf{Comparison with incentive mechanisms for FL; our mechanisms are in the
    last four rows.} \textit{Eq.}: existence (E) of, or convergence (C) to, a Nash
    equilibrium. \textit{Learn.}: joint convergence of the model and the contributions (J),
    or convergence to the optimal parameter $\wopt$ (OPT). \textit{Transfers}: budget
    balanced (BB), budget balanced on the truthful path and never in deficit (BB$^*$),
    individually rational (IR). \textit{Reports}: data distribution (D), cost (C),
    contribution (S), gradient (G). \textit{Truthful}: guarantee for the reported
    quantity; $\varepsilon$ denotes $\varepsilon$-truthfulness with a vanishing gain from
    misreporting, and $^\dagger$ marks guarantees that rely on agents observing realized valuations (\Cref{assump:realized}); \Cref{prop:no-evidence} shows that evidence of this kind is necessary. \textit{Contrib.}: partial (P) or full (F) contribution
    at equilibrium. \textit{Private}: data are not shared with the center.}
    \label{tab:comparison_literature}
\end{table}

\subsection{Our contributions}
\begin{itemize}[leftmargin=10pt, itemsep=2pt, parsep=0pt, topsep=2pt]
    \item \textbf{Joint dynamics of contributions and learning.} We model FL as a game in which each agent's valuation depends on the learned model and on all agents' contributions. In \mech{} (\Cref{algo:main_algo}), agents adjust their contributions by projected gradient steps while the center updates the model from their gradients. Under smoothness and concavity conditions, \mech{} converges at a geometric rate to a point at which the contributions form a Nash equilibrium and the model maximizes welfare for those contributions (\Cref{thm:convergence-genl}). The best-response dynamics of \citet{murhekar2023incentives} converge in the contributions alone, with no model being trained.

    \item \textbf{Free-riding at equilibrium, and a budget-balanced remedy.} The equilibrium of \mech{} can involve under-contribution: in \Cref{exam:NEnotsocial}, both agents contribute $7$ of their $9$ data points and the social welfare falls from $3$ to $2.75$. \mechtwo{} (\Cref{algo:2-phase}) adds a linear transfer that pays each agent in proportion to how much its contribution exceeds the average contribution of the other agents, and charges it when the difference is negative. The transfers sum to zero and make full contribution a dominant strategy, and training then converges to $\wopt$ at a linear rate (\Cref{thm:two_phase}). We give an explicit range of the transfer rate under which every agent prefers participating to training alone, and zero-sum rebates that restore participation otherwise (\Cref{prop:IR}).

    \item \textbf{Truthful contribution requires evidence.} \mechtwo{} pays for reported contributions, which the center cannot verify. We show that no transfer rule can reward contributions unless something the center observes depends on the data actually used (\Cref{prop:no-evidence}), a reporting analogue of the informativeness principle of \citet{holmstrom1979moral}. In particular, checking an agent's reports only against each other cannot make the subsidy of \mechtwo{} truthful.

    \item \textbf{A truthful mechanism that never runs a deficit.} \mechfour{} (\Cref{algo:truthful_welfare}) obtains this evidence from the other agents: after training, it checks each agent's reported contribution against the other agents' realized valuations of the trained model, and it prices gradient reports with a budget-balanced charge derived from VCG on a surrogate welfare. Truthful contribution and valuation reports are ex-post incentive compatible, the gain from misreporting a gradient vanishes as training converges, full contribution remains a dominant strategy, the model converges to $\wopt$ whether agents report gradients truthfully or play the equilibrium of each round, and the transfers sum to zero on the truthful path and never to a deficit (\Cref{thm:truthful_T2P}). Applied in every round, the same transfers make the reports of \mech{} truthful as well (\mechthree{}, \Cref{thm:EPIC}).

    \item \textbf{Experiments.} On CIFAR-10, FEMNIST, and Twitter, with deep networks that lie beyond the scope of our theory, \mechtwo{} matches the welfare of \FedAvg{} trained on all data, already for small transfer rates, while \mech{}, a strategic variant of FedAvg (\strategic{}), and \murhekar{}~\citep{murhekar2023incentives} reach lower welfare (\Cref{sec:experiments}).
\end{itemize}
\Cref{tab:comparison_literature} compares these guarantees with those of prior incentive mechanisms for FL. Proofs, additional remarks, and further experiments are in the appendix.

\subsection{Related work}
Federated learning~\citep{konevcny2016federated} has received wide attention over the last decade, largely owing to the FedAvg algorithm~\citep{mcmahan2017communication}, which reduces communication by taking several local steps between aggregations. Subsequent work addressed data heterogeneity~\citep{karimireddy2020scaffold,ghosh2020efficient}, Byzantine robustness~\citep{yin2018byzantine,karimireddy2020byzantine,ghosh2021communication}, communication overhead~\citep{stich2018sparsified,karimireddy2019error,ghosh2020communication}, and privacy~\citep{wei2020federated,truex2020ldp,Kumar_2025_CVPR}.

\paragraph{Incentives through the model.} Several works reward participation through the model itself: by weighting client updates so that the global model outperforms local ones~\citep{cho2022to}, by valuing contributions with the Shapley value~\citep{Tastan24contriSV}, by contract-theoretic accuracy shaping that elicits full contribution~\citep{karimireddy2022mechanisms}, by excluding malicious clients and rewarding helpful ones~\citep{gao2021}, and by differentiated model rewards during training~\citep{wu2024incentive}. \citet{procaccia2025incentiveshetero} maximize utilitarian welfare in a heterogeneous setting under PAC learning constraints.

\paragraph{Monetary incentives.} \citet{Yu2020FairnessAware} allocate a dynamic budget to maximize utility and reduce inequality. Blockchain-based methods reward high-quality contributions under budget constraints~\citep{pandey2022fedtoken} or distribute profits by the Shapley value~\citep{liu2020fedcoin}, and client rewards have been shown to improve the final model~\citep{yang2023}. \citet{Georgoulaki2023} show a price of anarchy of two and a price of stability of one for utility sharing with budget-balanced payments. In auction-based designs, agents adjust their bids by reinforcement learning~\citep{TangAuction2023auctionbasedRL}, and the center optimizes its budget to improve utility and reduce delays~\citep{tang2024AuctionbasedmultisessionRL}.

\paragraph{Coalitions and free-riding.} \citet{Donahue_Kleinberg_2021} model FL as a coalitional game, \citet{chaudhury2022Fairnesscoreset} ensure core stability with CoreFed, later extended to ordinal utilities~\citep{chaudhury2024fair}, and \citet{murhekar2025you} use the Shapley value for reciprocal fairness. \citet{meng2024federated} study how free-riding among competing firms affects the incentive to form FL coalitions, \citet{chen2024free} and \citet{tan2024fedcompetitors} propose conflict-aware collaboration among competitors, and \citet{Mao2024gameanalysis} show that subgame perfect equilibria of repeated FL games can be inefficient and design budget-balanced mechanisms that restore efficiency and individual rationality.

\paragraph{Truthful mechanisms and contribution games.} \citet{bornstein2024facttruth} elicit truthful cost reports, and \citet{chakarov2025incentivizingtruthfulcollaborationheterogeneous} give a budget-balanced, Bayesian incentive compatible payment rule for heterogeneous agents. Closest to our work, \citet{Blum2021OneforOne} and \citet{murhekar2023incentives} study equilibria of contribution games, the latter with best-response dynamics and budget-balanced payments. These works do not train the model inside the mechanism, and they treat contributions as observable. We analyze contributions and training jointly, and we make both the unverifiable contribution reports and the gradient reports incentive compatible, building on two-stage mechanisms for interdependent valuations~\citep{mezzetti2004mechanism,NATH2013321}.

\section{Preliminaries}
Consider an FL setup where a set of data contributors, given by $N = \{1,2,\ldots,n\}$, is interacting with a center. Each data contributor (agent) $i$ has access to a private labeled dataset $D_i$ sampled from distribution $\mathcal{D}_i$, with the size of the dataset given by $s_i^{\max} = |D_i|$. The agents are interested in learning a common parameter $w \in \mathbb{R}^m$ from their data for a common learning task (e.g., to perform a classification task). As the agents have limited individual data, the model is trained on the consolidated data of all the $n$ users. However, sampling data is costly, and agent $i$ incurs a cost, given by a function $c_i(s_i)$, when it trains the model locally on its chosen dataset of size $s_i \in S_i := [s_i^{\min}, s_i^{\max}]$, where $s_i^{\min}$ is the smallest amount of data that the agent can contribute. Along with these costs, there may be some other arbitrary costs that an agent incurs based on the mechanism and other agents' contributions. Under this setup, each agent $i$ gets a utility based on how much data $s$ has been chosen by the agents to sample and train on. Hence, the utility is given by 

\begin{equation}
    \label{eq:utility-model}
    u_i(w, s_i, s_{-i}) = v_i(w, s_i, s_{-i}) - z_i(s_i, s_{-i}),
\end{equation}
where $s_{-i}$ is the data chosen by the agents other than $i$ in this FL process. The function $v_i(w, s_i, s_{-i})$ is the {\em valuation function}, which denotes the {\em benefit} to agent $i$ if the parameter learned by the center is $w$ and the agents contribute by running the FL algorithm on their dataset sizes given by the vector $s = (s_i, s_{-i})$. The function $z_i(s_i, s_{-i})$ is the {\em effective cost} to agent $i$, which may depend on the data contributions of all agents. 

FL is an iterative process: the center updates a parameter $w^t$ at round $t$, and agents use this to choose their dataset size $s_i^{t+1}$ and to compute and share gradients $d_i^{t+1} \in \mathbb{R}^m$ with the center. The center aggregates gradients to update the parameter to $w^{t+1}$. We use the superscript $t$ when referring to these terms in a specific iteration of the training round. Agents are myopic: in each round, agent $i$ updates its contribution by a gradient step on its current-round utility given the broadcast $(w^t,s^t)$, and does not anticipate the effect on future models. Myopic adjustment is a standard model of learning in games \citep{fudenberg1998theory}.

\noindent\textsc{Remark (Arguments of the $u_i$ function).}
    Note that we have explicitly assumed that the utility function $u_i$ depends on the parameter learned and the data contributions $(w, s_i, s_{-i})$. We now motivate this dependence. In FL problems, we typically run iterative algorithms to optimize the accuracy (or loss) function. Note that, with this, the weight $w^t$ at time $t$ depends on all the $s_{i}^\ell$, $\ell<t, \forall i \in N$. However, the utility $u_i$ at time $t$ depends on $(s_i^t,s_{-i}^t)$ as well, which is not captured through $w^t$. Hence, we require the said explicit dependencies. We represent this dependency via two functions: the valuation function $v_i$ gives the benefit the agent gets from the parameter learned and the data contributions, while the effective cost is only dependent on all players' data contributions. 

\noindent\textsc{Remark (Realizing the $v_i$ function).}
$v_i(w,s)$ denotes the benefit agent $i$ derives from the learned parameters at a given contribution vector. It can be realized by an agent as the negative of the loss computed on its data for the given parameter $w$. For experiments in \Cref{sec:experiments}, we use the cross-entropy loss. 

Since the agents are strategic, every agent $i$'s aim is to maximize its utility by appropriately choosing its strategy $s_i$ given the strategies $s_{-i}$ of the other players and the parameter $w$ chosen by the center. The center, on the other hand, is interested in learning the optimal parameter $w$ that maximizes the sum of the valuations as follows, when all agents contribute their maximum data sizes, i.e., $s_i^{\max}, i \in N$.

\begin{equation}
    \label{eq:social-accuracy}
    \wopt \in \argmax_{w} \ \sum_{i \in N}v_i(w,s^{\max})
\end{equation}
As in prior work on data sharing in federated learning~\citep{karimireddy2022mechanisms}, the center's goal is the quality of the learned model, that is, the aggregate valuation $\sum_{i\in N}v_i$ obtained when all available data is used; the agents' data-collection costs are private and are not part of this objective. We therefore use \emph{social welfare} to mean this learning welfare, not the net utility.
We are interested in the question of whether we can design an FL algorithm that can make $(s_i^{\max}, s_{-i}^{\max})$ a Nash equilibrium of the underlying game, and the center can learn $\wopt$. In this context, the Nash equilibrium is defined as follows.
\begin{definition}[Nash equilibrium]
\label{def:nash}
    A {\em pure strategy Nash equilibrium} (PSNE) for a given parameter $w$ is a strategy profile $(s_i^{*}, s_{-i}^{*})$ of the agents such that $u_i(w, s_i^*, s_{-i}^*) \geqslant u_i(w, s_i, s_{-i}^*), \ \forall s_i \in S_i, \forall i \in N$.
\end{definition}
This definition is a modification of the standard definition of Nash equilibrium since the equilibrium profile $(s_i^{*}, s_{-i}^{*})$ depends on the parameter $w$ (we do not write it explicitly for notational cleanliness). Consequently, we have the limit pair $(w^*, s^*)$ as a fixed point where $s^*$ is a PSNE of the game parameterized by $w^*$, and $w^* \in \arg\max_w \sum_{i \in N} v_i(w, s^*)$. We will refer to the term PSNE as Nash equilibrium (NE) in the rest of the paper.

In order to ensure convergence towards the optimal parameter, we introduce monetary transfers. We also use monetary transfers to satisfy two design objectives, defined below: incentive compatibility, so that truthful reporting is optimal for the agents, and budget balance, so that the transfers only redistribute money among the agents.

In every round, agent $i$ takes an \emph{action}, the contribution $s_i^t\in S_i$ that it
actually uses, and sends \emph{reports} to the center: a contribution report
$\hat s_i^t$, a gradient report $\hat d_i^t\in\mathbb{R}^m$, and, in the truthful
mechanisms, a report $\hat v_i$ of its valuation of a model it has observed. The center
computes the model from the reported gradients, $w^{t+1} = w^t+\frac{\eta}{n}\sum_{j\in
N}\hat d_j^t$, as in \Cref{eq:update_u_w}, and computes any transfers $p_i$ from the
reports. Reporting is \emph{truthful} if $\hat s_i^t = s_i^t$, if
$\hat d_i^t = \nabla_w v_i(w^t,s^t)$ is the gradient computed on the data actually used,
and if $\hat v_i = v_i(w,s)$ is the realized valuation of the evaluated model $w$ at the
contributions $s$ actually used. Until \Cref{sec:truthful-welfare} we assume truthful
reporting. When reports may differ from the truth, agent $i$'s utility is
\begin{equation}
\label{eq:truthful_utility}
    u_i = v_i(w,s) - c_i(s_i) + p_i(\hat s,\hat d,\hat v),
\end{equation}
where $w$ is the model computed from the reports and $s$ is the profile of contributions
actually used. The valuation and the cost are realized by the agent and are evaluated at
the actual contributions, whereas the transfers are computed by the center and are
evaluated at the reports. Under truthful reporting, \Cref{eq:truthful_utility} reduces to
\Cref{eq:utility-model} with $z_i = c_i - p_i$.

\paragraph{Truthfulness notions.}
Valuations here are \emph{interdependent}: agent $i$'s realized valuation depends on the
model, which is computed from all agents' reports, and on the contributions actually used
by all agents. With interdependent valuations, dominant-strategy implementation of
efficient outcomes is generally impossible \citep{jehiel2001efficient}, so we target
ex-post incentive compatibility, the standard goal in this setting. Following the
two-stage construction of \citet{mezzetti2004mechanism}, our truthful mechanisms first
collect contribution and gradient reports, compute the model, and then elicit the agents'
realized valuations, which enter the transfers. A \emph{reporting stage} is a step of a
mechanism at which agents send reports; as elsewhere in the paper, agents are myopic, so
the utility relevant to a report is the one realized at the stage in which it is sent.

\begin{definition}[Ex-post incentive compatibility]
\label{def:epic}
A reporting stage of a mechanism is \emph{ex-post incentive compatible} (EPIC) if, for
every agent $i\in N$, every history of the mechanism up to that stage, and every report of
agent $i$ at that stage, agent $i$'s utility \Cref{eq:truthful_utility} when all agents
report truthfully is at least its utility when it sends that report and all other agents
report truthfully.
\end{definition}

\begin{definition}[Approximate truthfulness]
\label{def:eps-truthful}
A reporting stage is \emph{$\varepsilon$-truthful} for agent $i$ if, for every history up
to that stage, whenever all other agents report truthfully, no report of agent $i$ raises
its utility \Cref{eq:truthful_utility} by more than $\varepsilon$ over truthful reporting.
\end{definition}

\noindent In \mechfour{}, the elicitation stage is EPIC for contribution and valuation reports, and every round of training is $\varepsilon_i^t$-truthful for gradient reports (\Cref{thm:truthful_T2P}). We use \Cref{eq:truthful_utility} only when establishing truthfulness, in \Cref{sec:truthful-welfare} and \Cref{app:T-UPBReD}; elsewhere reports are truthful and we use the simpler form \Cref{eq:utility-model}. 

Besides incentive compatibility, we want the transfers to only redistribute money among the agents. Transfers are computed from the reports, not from the actions the agents actually take, so we require this for every profile of admissible reports (for example, $\hat s_i\in S_i$), not only for truthful ones.

\begin{definition}[Budget balance]
\label{def:bb}
Let $r$ denote the profile of reports sent during a run of the mechanism (contribution
reports $\hat s$, gradient reports $\hat d$, and valuation reports $\hat v$, as
applicable), and let $p_i(r)$ be the total transfer paid \emph{to} agent $i$ over the
run. The mechanism is
\begin{itemize}[leftmargin=15pt, itemsep=0pt, parsep=0pt, topsep=0pt]
    \item {\em budget balanced} (BB) if $\sum_{i\in N}p_i(r)=0$ for every report profile $r$;
    \item {\em weakly budget balanced} (W-BB) if $\sum_{i\in N}p_i(r)\leqslant 0$ for every
    report profile $r$, that is, the center never runs a deficit.
\end{itemize}
\end{definition}
Budget balanced ensures that the net monetary inflow or outflow is zero and the mechanism only allows monetary redistribution among the agents while weak budget balanced ensures that the center is not running a deficit.
In the following section we discuss some realistic forms of the utility function along with some properties that satisfy the assumptions we make in future sections.

\section{Examples of valuation functions}
\label{sec:examples_utility}
 
We now give a concrete form of the utilities that fits our framework and satisfies our
assumptions. All agents learn the same model, parameterized by $w\in\mathbb{R}^m$, and
each agent $i$ contributes $s_i\in S_i=[1,s_i^{\max}]$, that is, $s_i^{\min}=1$. A natural
form of agent $i$'s utility is
\begin{equation}
    \label{eq:utility-example}
    u_i(w,s) = \underbrace{r_i - L_i(w,s)}_{\text{valuation}}
              - \underbrace{c_i \cdot s_i}_{\text{cost}}
              + \underbrace{p_i(s)}_{\text{payments}},
\end{equation}
where $L_i(w,s)$ is agent $i$'s loss when the model is $w$ and the agents contribute $s$,
$c_i>0$ is a per-sample cost, and $r_i$ is a constant large enough to keep utilities
non-negative over a bounded set containing the iterates of our algorithms. A lower loss
thus means a higher valuation $v_i(w,s) = r_i - L_i(w,s)$. Motivated by the additive
scaling-law form of \citet{hoffmann2022trainingcomputeoptimallargelanguage} and
\citet{Kaplan2020scalinglaw}, we model
\begin{equation}
    \label{eq:loss-decomposition}
    L_i(w,s) = L_i(w) + \frac{1}{n}\sum_{j\in N} s_j^{-\alpha_i}, \qquad \alpha_i>0,
\end{equation}
where the first term is the loss achievable at the parameter $w$ and the second is a
finite-sample correction that decays as each agent contributes more data. Since the
center aggregates the agents' updates with uniform weights $\frac1n$
(\Cref{eq:update_u_w}), the finite-sample error decomposes across data sources:
specializing the result of \citet{Donahue_Kleinberg_2021} to uniform weights yields a
term proportional to $\sum_{j\in N}s_j^{-1}$, which is the case $\alpha_i=1$ of
\Cref{eq:loss-decomposition}.

The first term $L_i(w)$ can be any standard loss, for example, the regularized cross-entropy and mean squared error losses of linear models, or the quadratic distance to the parameter that best fits agent $i$'s own data. These functions satisfy the properties that the convergence results of \Cref{sec:without-payment,sec:with-payment} require of the learning problem; the losses and the resulting constants are given {below}. The quadratic distance also makes concrete the knowledge of the valuations that the truthful mechanism of \Cref{sec:truthful-welfare} requires of the center: agent $i$'s valuation is determined by $A_i$, $w_i^*$ and $\alpha_i$, and truthful gradients identify $A_i$ and $w_i^*$ ({see below}). In \mechtwo{} the payments are $p_i = p_i^1$ of \Cref{eq:payment}; the transfers of the truthful mechanisms are defined in \Cref{sec:truthful-welfare} and \Cref{app:T-UPBReD}. 
 
This decomposition, with linear models, serves only to show that our assumptions are satisfiable under a realistic functional form. Our experiments (\Cref{sec:experiments}) use deep networks and compute $L_i$ directly as the empirical loss on each agent's data, with the cross-entropy loss, without imposing this form.

We now describe the losses $L_i(w)$ referred to above. For each, we record the two properties that our assumptions use, namely a Hessian that
is bounded for every $w\in\mathbb{R}^m$ and strong convexity; the resulting constants
are derived in the verification remarks of \Cref{assump:conv-NE} and
\Cref{assump:conv-opt-NE}.
 
\paragraph{Cross-entropy loss.} For classification tasks with $K$ classes, the
regularized cross-entropy loss for agent $i$ is
\begin{equation}
    \label{eq:cross-entropy}
    L_i(w) = \mathbb{E}_{(x,y)\sim\mathcal{D}_i}\big[-\log p(y \mid x;\, w)\big] + \frac{\mu}{2}\|w\|_2^2,
\end{equation}
where $p(y \mid x;\, w)$ is the softmax output of a linear model and $\mu>0$ is the
regularization weight. Under the standard assumption that inputs are bounded,
$\|x\|_2 \leqslant B_x$, the Hessian of $L_i$ is bounded for every $w\in\mathbb{R}^m$,
and the regularization term makes $L_i$ $\mu$-strongly convex. Both properties are
standard in the federated learning literature \citep{karimireddy2020scaffold}.
 
\paragraph{Mean squared error loss.} For regression tasks with a linear predictor
$\hat y(w,x) = w^\top x$, the regularized MSE loss for agent $i$ is
\begin{equation}
    \label{eq:mse}
    L_i(w) = \mathbb{E}_{(x,y)\sim\mathcal{D}_i}\big[\big(y - \hat{y}(w, x)\big)^2\big] + \frac{\mu}{2}\|w\|_2^2,
\end{equation}
whose Hessian $2\,\mathbb{E}[xx^\top]+\mu\mathbb{I}$ is bounded under
$\|x\|_2\leqslant B_x$, and whose regularization term again makes it $\mu$-strongly
convex.
 
\paragraph{Distance to the local optimum.} Let $w_i^*$ denote the parameter that
minimizes agent $i$'s loss on its own data, and let $A_i\succ 0$ be a positive definite
matrix. The loss
\begin{equation}
    \label{eq:quadratic-loss}
    L_i(w) = (w-w_i^*)^\top A_i\,(w-w_i^*)
\end{equation}
measures how far the learned parameter is from the one that best fits agent $i$'s data.
The MSE loss of \Cref{eq:mse} is of this form up to an additive constant, with
$A_i = \mathbb{E}_{x\sim\mathcal{D}_i}[xx^\top] + \frac{\mu}{2}\mathbb{I}$ and $w_i^*$ the
regularized least-squares solution on $\mathcal{D}_i$; for other losses, such as
\Cref{eq:cross-entropy}, it is the local quadratic approximation of $L_i$ around $w_i^*$.
Since this loss carries no regularization term, \Cref{assump:conv-opt-NE} then holds with
$\nu = 2\lambda_{\min}(\bar A)$ and $M = 2\lambda_{\max}(\bar A)$, where $\bar A := \frac1n\sum_{i\in N}A_i$, rather than with the regularization weight $\mu$ as for the losses \Cref{eq:cross-entropy,eq:mse}.
 
\paragraph{Evaluating the valuations.} The truthful mechanisms require the center to evaluate each agent's valuation at the models it computes (\Cref{assump:realized,assump:realized-round}); under \Cref{eq:loss-decomposition} this amounts to knowing $L_i$ at those models and the exponent $\alpha_i$. The loss \Cref{eq:quadratic-loss} shows one way in which such knowledge can arise without access to the agents' data. Its gradient $\nabla L_i(w) = 2A_i(w-w_i^*)$ is affine in $w$, so gradients at $m+1$ affinely independent parameters determine $A_i$ and $w_i^*$; when $A_i=\mathbb{I}$ a single gradient suffices, giving $w_i^* = w-\tfrac12\nabla L_i(w)$. This reconstruction requires the gradients to be exact and reported truthfully, as in \mech{} and \mechtwo{}. In \mechfour{} gradient reports are strategic, so \Cref{thm:truthful_T2P} uses \Cref{assump:realized} as stated, under which the center's evaluation does not depend on the gradient reports. A gradient-based estimate would be biased: with $A_i=\mathbb{I}$, an agent playing the round equilibrium of part~4 of \Cref{thm:truthful_T2P} reports $\nabla_w v_i(w^{t+1},s^{\max}) = -2(w^{t+1}-w_i^*)$, so the estimate $w^t+\tfrac12\hat d_i^t = w_i^*+(w^t-w^{t+1})$ is off by one step. Since the steps vanish as training converges, such an estimate, like an estimate from mini-batch gradients, is compatible with the tolerance version of the tests (remark after \Cref{thm:truthful_T2P}), but not with the exact tests of \Cref{eq:payment_truthful_2P}, which would penalize truthful agents. Finally, knowing $(A_i,w_i^*)$ for every agent is more than the mechanisms need, since it determines $\wopt = \bar A^{-1}\frac1n\sum_{i\in N}A_iw_i^*$ in closed form; the mechanisms require only the losses at the models they compute.
\color{black}

\section{Learning with arbitrary effective costs} 
\label{sec:without-payment}

In this section, we consider the effective cost of agent $i$, i.e., $z_i(s_i, s_{-i})$, to be an arbitrary function of $s = (s_i, s_{-i})$. We know that an alternative interpretation of NE (\Cref{def:nash}) is a strategy profile $(s_i^{*}, s_{-i}^{*})$ where every agent's {\em best response} to the strategies of the other players is its own strategy in that profile \citep[see, e.g.,][]{maschler2020game}, i.e., $s_i^* \in \argmax_{s_i \in S_i} \ u_i(w, s_i, s_{-i}^*), \ \forall i \in N$.
Hence, an algorithm that simultaneously updates all agents' strategies with the best responses to the current strategies of the other players is called a {\em best response dynamics} of a strategic form game \citep[see, e.g.,][for a detailed description]{fudenberg1991game}. In our problem, this approach cannot be directly employed since the center also needs to learn and update the model parameters $w$ as the agents choose their data contributions. We, therefore, propose a mechanism for FL that simultaneously updates both the agents' strategies and the center's choice of $w$. 
FL protocols iteratively use local gradients to update a global model; the mechanism we describe below captures strategic behavior at every iteration by allowing agents to update their strategy (data set sizes used for gradient computation) at every realization of the global model based on \Cref{eq:utility-model}.

In this mechanism, each agent $i$ starts with some initial choices of $s_i^{0}$ and shares that with the center.\footnote{Note that only the number of data points, $s_i$, is shared with the center and not the data, which is consistent with the principle of FL.} The center initializes a $w^{0}$ and broadcasts that and the entire initial data-contribution vector $s^{0}$ to all the agents. In every subsequent iteration $t$, each agent $i$ locally computes two quantities: (i) its updated contribution $s_i^{t+1}$ by taking one gradient ascent step w.r.t.\ its own contribution $s_i$, and (ii) the local gradient of agent $i$'s valuation component of the social welfare w.r.t.\ $w$. Both are evaluated at the current values of $w^{t}$ and $s^t$ and sent back to the center. The center calculates an updated $w^{t+1}$ that averages (in the spirit of the FedAvg algorithm~\citep{mcmahan2017communication}) all the local gradients sent by the agents. The center then shares $w^{t+1}$ and $s^{t+1}$ with all the agents. Formally, the updates are given as follows.
\begin{align}
    s^{t+1} = s^t + \gamma g(w^t,s^t,\mu^t), &\qquad w^{t+1} = w^t + \eta \tilde{g}(w^t,s^t), \label{eq:update_u_w}
\end{align}

where the function $\tilde{g}(w^t,s^t) = \frac{1}{n} \sum_{i \in N} \nabla_w v_i(w^t,s^t)$
and $[g(w^t,s^t,\mu^t)]_i = \pd{s_i}u_i(w^t,s^t) + \mu_i^t$, with $\mu_i^t$ defined below.
Let $s_i' := s_i^t + \gamma\,\pd{s_i}u_i(w^t,s^t)$ denote the unconstrained
gradient step. Then
\begin{equation}
    \label{eq:mu-def}
    \mu_i^t =
    \begin{cases}
        \big(s_i^{\max} - s_i'\big)/\gamma & \text{if } s_i' > s_i^{\max},\\[2pt]
        \big(s_i^{\min} - s_i'\big)/\gamma & \text{if } s_i' < s_i^{\min},\\[2pt]
        0 & \text{otherwise},
    \end{cases}
\end{equation}
so that $s_i^{t+1} = s_i' + \gamma\mu_i^t = \Pi_{S_i}(s_i')$, where
$S_i=[s_i^{\min},s_i^{\max}]$ and
$\Pi_{S_i}(x) := \min\{s_i^{\max},\max\{s_i^{\min},x\}\}$.
Equivalently, $[g(w^t,s^t,\mu^t)]_i=\big(\Pi_{S_i}(s_i')-s_i^t\big)/\gamma$.

The mechanism is detailed in \Cref{algo:main_algo}.
 To guarantee the existence of and convergence to a Nash equilibrium, we require standard curvature and smoothness properties on the agents' utility functions. Intuitively, each agent's utility is strongly concave in its own contribution, the average valuation is strongly concave in $w$, and the cross-derivatives between $w$ and $s$ are bounded relative to these curvatures formalized in \Cref{assump:conv-NE}. Under these conditions, our main result shows that \Cref{algo:main_algo} always converges to a Nash equilibrium.

We need a few matrices, defined as follows for every $w$ and $s$. For $i,j \in N$ and $k,\ell \in \{1,\ldots,m\}, G(w,s)_{ij} = \ppd{s_j}{s_i}u_i(w,s); \tilde{G}(w,s)_{k \ell} = \frac{1}{n} \sum_{i\in N} \ppd{w_\ell}{w_k} v_i(w,s); H(w,s)_{ik} = \ppd{w_k}{s_i}u_i(w,s)$; $\tilde{H}(w,s)_{kj} = \frac{1}{n}\sum_{i \in N}\ppd{s_j}{w_k} v_i(w,s)$.
\begin{assumption}
\label{assump:conv-NE}
    Consider the utility functions given by \Cref{eq:utility-model} where functions $v_i, z_i, i \in N$ are such that the following properties hold for every $w \in \mathbb{R}^m$ and $s \in \prod_{i \in N} S_i$, $\forall i,j \in N,\forall k,\ell \in \{1,\ldots,m\}$ (the matrices below are as defined above).
    \begin{enumerate}[leftmargin=7pt, itemsep=0pt, parsep=0pt, topsep=0pt]
       \item\label{cond:concavity} The matrices $G(w,s)+\lambda \mathbb{I}$ and $\tilde{G}(w,s)+\tilde{\lambda} \mathbb{I}$ are negative semi-definite, i.e., $v^\top\big(G(w,s)+\lambda\mathbb{I}\big)v \leqslant 0$ for all $v\in\mathbb{R}^n$ and $v'^\top\big(\tilde G(w,s)+\tilde\lambda\mathbb{I}\big)v' \leqslant 0$ for all $v'\in\mathbb{R}^m$.
        \item\label{cond:bound-deriv} We assume the following bounds: 
        (i)~$|G(w,s)_{ij}| \leqslant L$, (ii)~$|\tilde{G}(w,s)_{k \ell}| \leqslant \tilde{L}$,\\ (iii)~$\|H(w,s)\|_{op} \leqslant P$, where $\|A\|_{op} := \inf \{c \geqslant 0: \|A v\| \leqslant c\|v\|, \forall v\}$,  (iv)~$\|\tilde{H}(w,s)\|_{op}\leqslant \tilde{P}$.
    
    \end{enumerate}
\end{assumption}

\noindent\textsc{Remark.} While the conditions on $G$ align with assumptions on contribution dynamics of \citet{murhekar2023incentives}, the joint conditions on $(w, s)$ via $\tilde{G}$ are novel to our work. These additional joint assumptions are used to establish a stronger guarantee: the simultaneous convergence of both the strategic contributions and the model parameter $w$.

\noindent \textsc{Remark (Verification for \Cref{eq:utility-example}).}
We verify \Cref{assump:conv-NE} for the utility form of \Cref{eq:utility-example} with the decomposition \Cref{eq:loss-decomposition}, in the setting of \mech{}, where no payments are made. The valuation is additively separable in $w$ and $s$: $L_i(w)$ does not depend on $s$, and the finite-sample correction does not depend on $w$. Every mixed derivative $\ppd{w_k}{s_j}$ therefore vanishes, so $H\equiv 0$ and $\tilde H\equiv 0$, conditions~\ref{cond:bound-deriv}(iii) and~(iv) hold with $P=\tilde P=0$, and the requirements $\lambda>\tilde P$, $\tilde\lambda>P$ of \Cref{thm:convergence-genl} reduce to $\lambda,\tilde\lambda>0$.
 
For the matrix $G$, differentiating \Cref{eq:utility-example} with respect to $s_i$, the cost contributes the constant $-c_i$, so that
\begin{equation*}
    \pd{s_i}u_i(w,s) = \frac{\alpha_i}{n}\,s_i^{-\alpha_i-1} - c_i .
\end{equation*}
This depends on $s_i$ alone, because the correction in \Cref{eq:loss-decomposition} decomposes across agents. Differentiating once more, $G(w,s)_{ij} = 0$ for $j\neq i$ and
\begin{equation*}
    G(w,s)_{ii} = -\frac{\alpha_i(\alpha_i+1)}{n}\,s_i^{-\alpha_i-2},
\end{equation*}
so $G$ is diagonal, and in particular symmetric. On $S_i=[1,s_i^{\max}]$ the diagonal entries lie between $-\frac{\alpha_i(\alpha_i+1)}{n}$ and $-\frac{\alpha_i(\alpha_i+1)}{n}(s_i^{\max})^{-\alpha_i-2}$. Hence condition~\ref{cond:concavity} holds with $\lambda = \min_{i\in N}\frac{\alpha_i(\alpha_i+1)}{n}(s_i^{\max})^{-\alpha_i-2} > 0$, and condition~\ref{cond:bound-deriv}(i) holds with $L = \max_{i\in N}\frac{\alpha_i(\alpha_i+1)}{n}$. The lower endpoint $1$ of $S_i$ is what keeps $L$ finite.
 
For the matrix $\tilde G = -\frac1n\sum_{i\in N}\nabla^2L_i(w)$, the losses \Cref{eq:cross-entropy,eq:mse} satisfy $\nabla^2L_i\succeq\mu\mathbb{I}$, so condition~\ref{cond:concavity} holds with $\tilde\lambda=\mu$. For the cross-entropy loss, every entry of the softmax Hessian $\operatorname{diag}(p)-pp^\top$ has absolute value at most $\tfrac14$, so the entries of $\nabla^2L_i$ are bounded by $\tilde L = \tfrac14B_x^2+\mu$; for the MSE loss, by $\tilde L = 2B_x^2+\mu$. These bounds hold for every $w\in\mathbb{R}^m$, so condition~\ref{cond:bound-deriv}(ii) requires no bound on the weights.
 
\begin{equation}
    \label{eq:expressions-1}
    \begin{split}
    W_1 &= \sqrt{1+\gamma^2n^2L^2-2\gamma\lambda}+\sqrt{\tilde{P}^2\gamma^2}, \\
    W_2 &= \sqrt{1+\eta^2m^2\tilde{L}^2-2\eta\tilde{\lambda}}+\sqrt{P^2\eta^2}, \\
 E &=\|g(w^0,s^0,\mu^0)\|_2+\|\tilde{g}(w^0,s^0)\|_2, \\
    T_0&(w^0,s^0) = \nicefrac{\left(\ln{\ddfrac{E}{\epsilon}}\right)}{\left(\ln{\ddfrac{1}{W}}\right)},
    \end{split}
\end{equation}
where $w^0 \in \mathbb{R}^m, s^0 \in \prod_{i \in N} S_i$ and $W = \max\{W_1,W_2\}$.

    \begin{theorem}[\mech{} convergence to NE]
    \label{thm:convergence-genl}
    Suppose \Cref{assump:conv-NE} holds with $\lambda>\tilde P$ and $\tilde\lambda>P$, and the
    step sizes satisfy
    $\gamma<\frac{2(\lambda-\tilde P)}{n^2L^2-\tilde P^2}$ and
    $\eta<\frac{2(\tilde\lambda-P)}{m^2\tilde L^2-P^2}$.
    Then for every $(w^0,s^0)$ the iterates of \Cref{algo:main_algo} converge to a limit
    $(w^*,s^*)$ such that $s^*$ is a PSNE at $w^*$ and $w^*$ is the unique maximizer of
    $\frac1n\sum_{i\in N}v_i(\cdot,s^*)$. Moreover, for every $\epsilon>0$ and every
    $T>T_0(w^0,s^0)$, we have $\|g(w^T,s^T,\mu^T)\|_2<\epsilon$,
    $\|\tilde g(w^T,s^T)\|_2<\epsilon$, and
    $\|s^T-s^\star(w^T,s^T)\|_2<\frac{1+\gamma L}{\lambda}\,\epsilon$,
    where $s^\star(w,s)$ is the profile of best responses.
    \end{theorem}

\begin{proof}[Proof sketch]
The proof tracks the gradient mappings $g^t$ and $\tilde g^t$. Since the projection onto $S=\prod_{i\in N}S_i$ is non-expansive, the integral form of the mean value theorem bounds $\|g^{t+1}\|_2$ and $\|\tilde g^{t+1}\|_2$ through averages of $G$, $H$, $\tilde G$, $\tilde H$, which inherit \Cref{assump:conv-NE}; strong concavity and the bounds on the cross terms give $\|g^{t+1}\|_2+\|\tilde g^{t+1}\|_2\leqslant W(\|g^t\|_2+\|\tilde g^t\|_2)$, and the step-size conditions are exactly those that make $W<1$. The increments are then summable and the iterates converge, since $\mathbb{R}^m\times\mathbb{R}^n$ is complete. A zero of the projected step is a Nash equilibrium because each $u_i$ is concave in $s_i$, a stationary point of the strongly concave average valuation is its unique maximizer, and an error bound for the projected step gives the distance to the best responses. The complete proof is in \Cref{sec:convergence-genl-app}.
\end{proof}
\color{black}



%

\begin{algorithm}[t]
\DontPrintSemicolon
  \caption{\mechfull{} (\mech{})}
  \label{algo:main_algo}

  \KwIn{Step size $\gamma, \eta$, initialization $w^0$, $s_i^0$ for $i \in N$, number of iterations $T$}
  \KwOut{$w^T$}

  \For{$t=0$ \textbf{to} $T-1$}{
    \textbf{Center} broadcasts $w^t, s^t$\;
    
    \ForEach{\textbf{agent} $i \in N$ in parallel}{
      $s^{t+1}_i = s^t_i + \gamma [g(w^t,s^t,\mu^t)]_i$\;
      Compute local gradient: $d_i^{t+1} = \nabla_w \:v_i(w^{t},s_i^{t},s_{-i}^{t})$\;
      Send $s_i^{t+1}, d_i^{t+1}$ to the center\;
    }
    
    \textbf{Center} updates: $w^{t+1} = w^{t} + \frac{\eta}{n} \sum_{i\in N} d_i^{t+1} = w^t + \eta \tilde{g}(w^t,s^t)$\;
  }

  \Return $w^T$\;
  
\end{algorithm}

\begin{figure}[ht]
    \centering
    \begin{subfigure}[t]{0.5\linewidth}
        \centering
        \includegraphics[width=\linewidth]{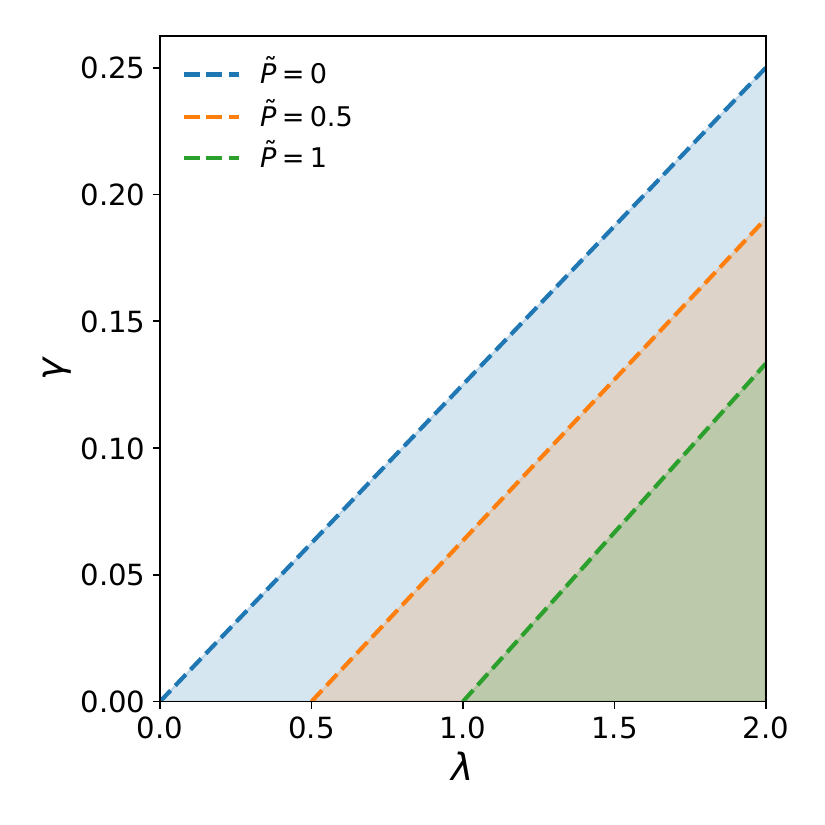}
        \caption{$\tilde{P}=1$}
    \end{subfigure}%
    ~ 
    \begin{subfigure}[t]{0.5\linewidth}
        \includegraphics[width=\linewidth]{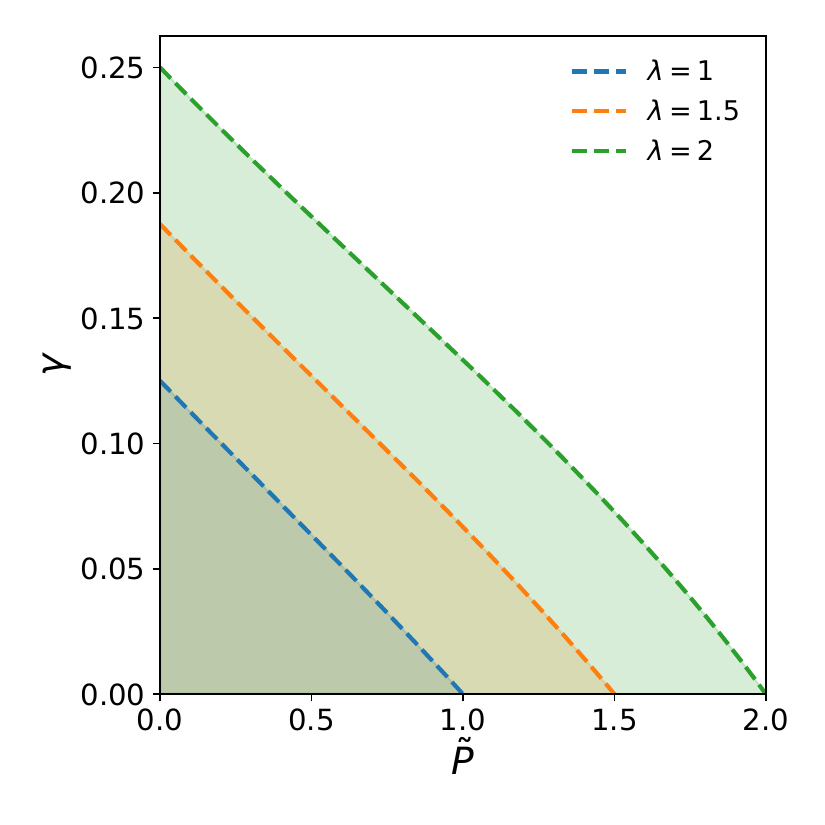}
        \caption{$\lambda=1.1$}
    \end{subfigure}
\caption{Shaded regions show the feasible choices of $\gamma$ in \Cref{thm:convergence-genl} for $n=2$ and $L=2$: (a)~as a function of $\lambda$ for the values of $\tilde P$ shown, and (b)~as a function of $\tilde P$ for the values of $\lambda$ shown. Dashed lines mark the boundary $\gamma = 2(\lambda-\tilde P)/(n^2L^2-\tilde P^2)$. The region is empty when $\lambda\leqslant\tilde P$. The feasible choices of $\eta$ are obtained by replacing $(n,L,\lambda,\tilde P)$ with $(m,\tilde L,\tilde\lambda,P)$.}
    \label{fig:region}
\end{figure}

The theorem provides a guarantee of convergence of \Cref{algo:main_algo} for any arbitrary initial condition $(w^0,s^0)$. It needs to run for a minimum number of iterations $T_0(w^0,s^0)$, with appropriate parameters of $\gamma$ and $\eta$ that govern the gradient ascent rates of the agents' and center's objectives, respectively. Note that for a better convergence, i.e., a smaller $\epsilon$, the algorithm needs to run for more iterations. The parameter $W$ determines the contraction factor of the recurrence of $\|g(w^t,s^t)\|_2 + \|\tilde{g}(w^t,s^t)\|_2$ and is chosen to be smaller than unity by choosing $\gamma$ and $\eta$ appropriately. 
\Cref{fig:region} shows an example for such a feasible regions of $\gamma$ for certain $n,L,\lambda,\tilde{P}$.

\noindent\textsc{Remark.}
While the step sizes $\gamma$ and $\eta$ required by \Cref{thm:convergence-genl} scale inversely with $m$ and $n$, these restrictions arise as sufficient conditions within our convergence analysis. In our experimental evaluations, we observe robust convergence even with small constant values of $\gamma$ and $\eta$ (see \Cref{sec:experiments}). Furthermore, when monetary transfers are permitted (\Cref{sec:with-payment}), \mechtwo{} leverages its two-phase structure to establish convergence guarantees with step sizes that are naturally independent of $m$ and $n$.

While agents strategically choose $s_i$ to maximize their utility, we assume truthful reporting of these choices. In \mech{}, agents report $s_i^t$ and $d_i^t$ in each round. As \Cref{exam:misreport} shows, misreporting can improve an agent's utility. In \Cref{app:T-UPBReD} we introduce payments that make the reports of $s_i^t$ and $d_i^t$ truthful.


%
\color{black}

\subsection*{Application of \Cref{thm:convergence-genl}: effective costs are only personal}
\label{subsec:application}
One special but practical case of the above setup is when the effective cost is only borne by the agent. Mathematically, it is represented as $z_i(s_i,s_{-i}) = c_i(s_i), \forall s_i \in S_i, s_{-i} \in S_{-i}, \forall i \in N$. 

We assume convex cost functions $c_i, \forall i \in N$, which is a commonly used assumption in the literature~\citep{Li2014Pricing_convex}. This setup is consistent with the assumptions of \Cref{assump:conv-NE}, and according to \Cref{thm:convergence-genl}, \Cref{algo:main_algo} converges to a Nash equilibrium, which we denote as $(w^*, s^*)$. However, it is possible that either of the following happens: (i)~$s^*\neq s^{\max}$, i.e., the agents do not contribute their entire data for the federated learning process, leading to a suboptimal learning, or (ii)~$w^*\neq\wopt$ (see \Cref{eq:social-accuracy} for the definition), which is neither the objective of the center nor of the agents. The following example shows such an instance.
\begin{example}[NE different from socially optimal]
\label{exam:NEnotsocial}
Consider two agents with linear costs who learn a model with two parameters, where agent
$1$'s data informs the first parameter and agent $2$'s data informs the second. Both
agents have the valuation
\begin{equation*}
    v_i(w,s) = 2-(1-w_1)^2-(2-w_2)^2-\frac{w_1^2}{s_1}-\frac{w_2^2}{s_2},\qquad i=1,2,
\end{equation*}
in which the term $w_k^2/s_k$ acts as a regularizer on the parameter informed by agent
$k$, whose weight shrinks as agent $k$ contributes more data. The costs are
$c_1(s_1)=\frac{1}{64}s_1$ and $c_2(s_2)=\frac{1}{16}s_2$, and the strategy sets are
$S_i=[1,9]$.
 
For a fixed contribution profile $s$, maximizing $v_i$ over $w$ gives
$w_1 = s_1/(s_1+1)$ and $w_2 = 2s_2/(s_2+1)$, so the learned parameter approaches the
target $(1,2)$ only as the contributions grow. At $s=s^{\max}=(9,9)$ this yields
$\wopt=(0.9,1.8)$ and the optimal social welfare $\sum_i v_i(\wopt,s^{\max})=3$. This
profile is not an NE: agent $i$'s marginal utility at $(\wopt,s^{\max})$ is
$w_i^2/(s_i^{\max})^2-c_i'$, which equals $0.01-\frac{1}{64}<0$ for agent $1$ and
$0.04-\frac{1}{16}<0$ for agent $2$, so each agent prefers to reduce its contribution.
 
Starting from $(\wopt,s^{\max})$ with $\gamma=\eta=0.25$, \Cref{algo:main_algo}
converges to $w^*=(0.875,1.75)$ and $s^*=(7,7)$, where the social welfare is $2.75<3$.
The agents contribute less than their full data, and as a consequence the center learns
a parameter different from $\wopt$. Full contribution is efficient here because each
agent's data benefits both agents, whereas each agent weighs only its own benefit
against its cost.
\end{example}
\color{black}

 The example shows that the Nash equilibrium reached by \mech{} need not be socially optimal. The same effect appears on
real data: in \Cref{fig:combined_plots}, the social welfare of \mech{} lies below that of
\FedAvg{}, in which every agent trains on its entire dataset. This motivates the payments of
the next section, which make full contribution optimal for every agent.

\section{Welfare maximization at Nash equilibrium} 
\label{sec:with-payment}

In this section, we consider the scenario where the center can make a payment $p_i(s_i, s_{-i})$ {\em to} agent $i$ to alter its utility function. Therefore, the effective cost of agent $i$ becomes 
    $z_i(s_i,s_{-i})=c_i(s_i)-p_i(s_i,s_{-i})$.
Note that the payment can be either positive or negative, which determines whether the agent is {\em subsidized} or {\em taxed} respectively. It is reasonable to expect that the {\em data contributors} of the FL process are subsidized for their data contribution while the {\em data consumers} are charged a payment for obtaining the learned parameter. We assume that the derivatives of the valuation $v_i$ and cost function $c_i$ are bounded.
We choose the following payment function:
\begin{equation}
\label{eq:payment}
    \textstyle p_i^1(s) = \beta (s_i-\frac{1}{n-1} \sum_{j \neq i} s_j ),
\end{equation}
where $\beta$ is a parameter of choice. Note that this payment mechanism is budget balanced by design, since $\sum_{i \in N} p_i^1(s_i, s_{-i}) = 0, \forall s_i \in S_i, s_{-i} \in S_{-i}$. With payment, the utility of agent $i$ becomes $u_i(w,s)=v_i(w,s)-c_i(s_i) + p_i^1(s).$
We now discuss how the payment can be used to maximize the social welfare at a NE.

\subsection{Convergence to the NE $(w^{\textnormal{OPT}},s^{\max})$} 
In this setup, we first prove that the utility of every agent $i$ can be made strictly increasing in its own contribution $s_i$ in the following manner.
\begin{assumption}
\label{assump:derivatives}
The partial derivatives of $v_i(w,s)$ are bounded below, i.e., $\pd{s_j}v_i(w,s)\geqslant-\tau$ for some $0\leqslant\tau<\infty$, for all $i,j\in N$, all $w$ and all $s$. Moreover, the cost functions $c_i(\cdot)$ have bounded derivatives, $c_i'(s_i)\leqslant\zeta$ for all $s_i$ and $i$.
\end{assumption}
Note that the above assumptions are mild. Previous works make stronger assumptions (like convexity) on $c_i$ and $v_i$ with respect to $s_i$ \citep[see][]{murhekar2023incentives}. Considering the utility form as described in \Cref{eq:utility-example}, \Cref{assump:derivatives} holds with $\tau = 0$ as 
$\partial v_i/\partial s_j = \frac{\alpha_i}{n}\,s_{j}^{-\alpha_i-1} > 0$. We show that the above bounds on the derivatives of  $v_i$ and $c_i$ are sufficient to claim our result.
\begin{lemma}[Increasing utility]
\label{lem:incr_utility}
Suppose \Cref{assump:derivatives} holds and the payment function is given by
\Cref{eq:payment} with $\beta>\zeta+\tau$. Then for every $i\in N$, every $w$, and every
$s\in\prod_{j}S_j$
\[
  \pd{s_i}u_i(w,s_i,s_{-i}) \;>\;0 .
\]
\end{lemma}

\begin{proof}[Proof sketch]
By \Cref{eq:payment}, $\pd{s_i}u_i = \pd{s_i}v_i - c_i' + \beta\geqslant\beta-\tau-\zeta>0$ by \Cref{assump:derivatives}. The complete proof is in \Cref{app:proof-incr}.
\end{proof}
\color{black}


\noindent\textsc{Remark (Knowledge of $\beta$).}
The knowledge of $\beta$ is required for theoretical tractability only. In the experiments, we do not assume any knowledge of $\beta$; rather, we tune this hyperparameter. We observe that uniformly across all datasets, if $\beta$ is chosen larger than a threshold, we obtain the best performance of our proposed algorithms, which also validates the above theoretical requirement. 

\noindent\textsc{Remark ($s^{\max}$ is a dominant strategy).}
     The result of \Cref{lem:incr_utility} is stronger than the Nash property invoked in \Cref{thm:two_phase}, since \Cref{assump:derivatives} is already quantified over all $w$ and all $s$. It is worth recording because it removes any equilibrium selection question from Phase 1: the agents need not coordinate, need not hold correct beliefs about one another, and need not observe the parameter, since increasing one's own contribution is optimal for any choice of the above.

 This lemma implies that even in \Cref{algo:main_algo} if we apply the above payment, $s$ will converge to the maximum value $s^{\max}$. However, unlike \Cref{algo:main_algo}, in this section we provide an algorithm which gives the step sizes of the gradient ascents in a more concrete manner and is, therefore, a superior one. In order to keep the convergence rate the same as \Cref{thm:convergence-genl}, we assume that the negative\footnote{We negate the welfare so that we can consider functions as convex to apply the results of convex analysis easily.} social welfare function at $s^{\max}$, given by $f(w,s^{\max})=-\frac{1}{n}\sum_{i \in N}v_i(w,s^{\max})$, is $M$-smooth and $\nu$-strongly convex in $w$. {These properties are defined as follows.}

\begin{definition}[$M$-smoothness]
    \label{def:M-smooth}
    A function $f:\mathbb{R}^m \to \mathbb{R}$ is $M$-smooth if for all $x, x' \in \mathbb{R}^m$, we have
    $f(x') \leqslant f(x)+\langle \nabla_x f(x),x'-x\rangle + \frac{M}{2}\|x-x'\|_2^2$.
\end{definition}
\begin{definition}[$\nu$-strongly convex]
    \label{def:m-strict-convex}
    A function $f:\mathbb{R}^m \to \mathbb{R}$ is $\nu$-strongly convex if for all $x, x' \in \mathbb{R}^m$, we have
    $f(x') \geqslant f(x)+\langle \nabla_x f(x),x'-x\rangle + \frac{\nu}{2}\|x-x'\|_2^2$.
\end{definition}
\color{black}

%
%

\begin{algorithm}[t]
\DontPrintSemicolon
\LinesNumbered
\SetStartEndCondition{ }{}{}
\caption{\mechtwo{}}
\label{algo:2-phase}
\KwIn{Step sizes $\gamma,\eta$; initialization $w^0$, $s_i^0\in S_i$ for $i\in N$;
      iteration counts $\kappa,T_0$ as given by \Cref{thm:two_phase}}
\KwOut{$w^{T_0}$}
\tcp{Phase 1: contributions, with the model fixed at $w^0$}
\textbf{Center} broadcasts $w^0,s^0$; set $k=0$\;
\While{$k<\kappa$}{
  \textbf{Center} broadcasts $s^k$\;
  \ForEach{\textbf{agent} $i\in N$ in parallel}{
    $s_i^{k+1}=s_i^k+\gamma\,[g(w^0,s^k,\mu^k)]_i$; send $s_i^{k+1}$ to center\;
  }
  $k\gets k+1$\;
}
\tcp{Under \Cref{eq:payment}, $s^\kappa=s^{\max}$}
\tcp{Phase 2: model learning at $s^{\max}$}
\For{$t=0$ \KwTo $T_0-1$}{
  \textbf{Center} broadcasts $w^t$\;
  \ForEach{\textbf{agent} $i\in N$ in parallel}{
    Compute $d_i^{t+1}=\nabla_w v_i(w^t,s^{\max})$; send $d_i^{t+1}$ to center\;
  }
  \textbf{Center} updates $w^{t+1}=w^t+\frac{\eta}{n}\sum_{i\in N}d_i^{t+1}=w^t+\eta\,\tilde g(w^t,s^{\max})$\;
}
\Return $w^{T_0}$\;
\end{algorithm}

We propose a two-phase algorithm given by \Cref{algo:2-phase}. The algorithm, in the first phase, incentivizes the agents to contribute $s^{\max}$, and in the second phase, converges to $\wopt$. Note that in the first phase of \Cref{algo:2-phase}, every agent runs a gradient ascent step. Thanks to the increasing utility property (\Cref{lem:incr_utility}), we have an increasing sequence of $s_i^t$ for all $i \in N$. Moreover, from the definition of $[g(w^0,s^t,\mu^t)]_i$ it is ensured that $s_i^{\max}$ is the fixed point of this gradient ascent update. Hence after a finite number of iterations, every agent reaches $s_i^{\max}$. Note that in this phase the model parameter $w^0$ remains unchanged.

In the second phase of the algorithm, we update the model parameter $w^t$. Note that since all the agents contribute the maximum amount of data they own, this phase is simply \emph{pure} FL (without incentive design). As such, each agent now computes the local gradient $d_i^{t+1}$ and sends it to the center in each round.

We now provide the convergence guarantees of \Cref{algo:2-phase}. 
 Before that, let us discuss the necessary assumptions.
\begin{assumption}
\label{assump:conv-opt-NE}
The negative social welfare function at $s^{\max}$, $f(w,s^{\max})$, is $M$-smooth and $\nu$-strongly convex in $w$, with $M > \nu$.
\end{assumption} 
\noindent\textsc{Remark (Strong convexity and smoothness). } The smoothness and strong convexity assumptions have featured in several previous papers on FL~\citep{karimireddy2020scaffold,yin2018byzantine}. Considering the utility function described in \Cref{eq:utility-example}, \Cref{assump:conv-opt-NE} holds under L2 regularization, consistent with 
standard convergence analyses of FedAvg and SCAFFOLD 
\citep{karimireddy2020scaffold}, and more recent analyses of more complex FL frameworks such as split FL \citep{han2024convergence}.
We show in the following theorem that the convergence rate under \Cref{assump:conv-opt-NE} is $\mathcal{O}(\ln{\nicefrac{1}{\epsilon}})$ which is the same as that of \Cref{algo:main_algo}. However, \Cref{algo:2-phase} is cleaner in terms of the phases of convergence and the choices of the step sizes $\gamma,\eta$. These structural assumptions of strong convexity and smoothness may be relaxed at the expense of weaker rates. For example, $O(1/\epsilon)$ iterations for convex smooth $f$, and $O(1/\epsilon^2)$ iterations to reach an $\epsilon$-stationary point for non-convex smooth $f$. Since the second phase of \mechtwo{} is pure FL, any convergence result in the FL literature with the corresponding assumptions on $f(w,s^{\max})$ can be used in place of the second phase of this algorithm. 
The main result of this section is as follows.
\begin{theorem}[\mechtwo{} convergence to the optimal NE]
\label{thm:two_phase}
  Suppose \Cref{assump:derivatives,assump:conv-opt-NE} hold and we consider the utility function with payment scheme given in \Cref{eq:payment} with $\beta > \zeta+\tau$. Then under truthful reports, for every $w^0 \in \mathbb{R}^m$ and $s^0 \in \prod_{i \in N} S_i$, \mechtwo{} (\Cref{algo:2-phase}) converges to the Nash equilibrium $(\wopt,s^{\max})$ and is budget balanced. In particular, for any given $\epsilon > 0$, we have 
  $\frac{1}{n}\sum_{i \in N}v_i(\wopt,s^{\max})-\frac{1}{n}\sum_{i \in N}v_i(w^T,s^{\max})<\epsilon$,
      and $s^T = s^{\max}$ in $T= \kappa + T_0$ iterations provided we choose $\gamma = c>0$ (a universal constant) and $\eta = 1/M$ with
        \begin{equation}
            \label{eq:2-phase_strong-T_1}
            \begin{aligned}
            \kappa &\geqslant \max_i\left\{\frac{(s^{\max}_i-s^0_i)}{c \, \Delta}\right\},\\
            T_0 &> \left(\ln{\frac{f(w^0,s^{\max})-f(\wopt,s^{\max})}{\epsilon}}\right)\Big/\left(\ln \frac{1}{{1-\frac{\nu}{M}}}\right)
            \end{aligned}
         \end{equation}
        where $f(w,s) := -\frac{1}{n} \sum_{i \in N} v_i(w,s)$ and $\Delta=\beta-\tau-\zeta$.
\end{theorem}

\begin{proof}[Proof sketch]
By \Cref{lem:incr_utility}, $\pd{s_i}u_i(w^0,s^k)\geqslant\Delta$, so each projected step raises $s_i$ by at least $\gamma\Delta$ until it reaches $s_i^{\max}$, which is a fixed point of the step; hence $s^\kappa=s^{\max}$, which is each agent's unique best response at every $w$ because its utility is strictly increasing in its own contribution. Phase~1 uses no concavity, only the lower bound $\Delta$ on the marginal utility. Budget balance holds because $\sum_{i\in N}p_i^1(s)=0$ for every $s$. Phase~2 is gradient descent with $\eta=1/M$ on the $M$-smooth, $\nu$-strongly convex $f(\cdot,s^{\max})$, which reduces $f(w^t)-f(\wopt)$ by the factor $1-\nu/M$ in every round~\citep{Wright_Recht_2022}, and strong convexity gives $\|w^{T_0}-\wopt\|_2^2\leqslant\frac{2}{\nu}\big(f(w^{T_0})-f(\wopt)\big)$, so the iterates converge to $\wopt$. The complete proof is in \Cref{app:proof-two-phase}.
\end{proof}
\color{black}

\noindent\textsc{Remark ($\gamma$ and rationality). }
    Phase 1 of \Cref{algo:2-phase} achieves the optimal contribution $s^{\max}$ without updating the learning parameter $w$. The speed of convergence of this phase depends on the parameter $\gamma = c$, which can be any positive constant. A larger $c$ ensures faster convergence, however the algorithm converges for any positive $c>0$. In Phase 2 of the algorithm, the agents stop updating their contributions and only update the gradients $d_i$ and the center accumulates them to update $w^t$.

Thanks to the smoothness and strong convexity properties of $f(\cdot,s^{\max})$, we can show that running \Cref{algo:2-phase} guarantees that the model iterate $w^T$ converges to $\wopt$ at an exponential speed as well.

\begin{corollary}[Iterate convergence]
\label{cor:iterate}
With the same setup as above and $\gamma = c$ (a universal constant), $\eta = \frac{2}{M+\nu}$, we obtain $\|w^T - \wopt\|_2 < \epsilon, \text{ and } s^T = s^{\max}$,
where $T = \kappa + \tilde{T}_0$, with $\kappa$ as in \Cref{eq:2-phase_strong-T_1} and $\tilde{T}_0 > \left(\ln{\frac{\|w^0-\wopt\|_2}{\epsilon}}\right)/\left(\ln \frac{ {1+\frac{\nu}{M}}}{{1-\frac{\nu}{M}}}\right)$. 
\end{corollary}

\noindent The proof follows that of Phase~2 of \Cref{thm:two_phase} with the step size $\eta=\frac{2}{M+\nu}$, for which gradient descent contracts the distance to $\wopt$ by the factor $\frac{1-\nu/M}{1+\nu/M}$ in every round; it is in \Cref{app:proof-iterate}.

\subsection{Individual rationality}
\label{sec:IR}

Participation in federated learning is voluntary. \Cref{thm:two_phase} shows that, once the agents have joined, the payment $p^1$ of \Cref{eq:payment} makes full contribution optimal for every agent. This guarantee is only useful if the agents join in the first place: an agent that expects to be worse off inside the mechanism than outside it will not participate, and free-riding then reappears as non-participation rather than as under-contribution. We therefore ask whether every agent prefers taking part in \mechtwo{} to training a model on its own data alone. This property is called \emph{individual rationality}.

Two features of \mechtwo{} make the question non-trivial. First, the transfers are budget balanced, so they redistribute money among the agents rather than inject it from the center. Since $p_i^1$ rewards contributing more than the average of the other agents, the redistribution runs from agents with small datasets to agents with large ones: an agent with below-average capacity pays, and it participates only if the shared model is worth more to it than both its payment and its data cost, compared with training alone. Second, the shared model need not benefit every agent equally. An agent whose data distribution differs from the others' may value the federated model less than a model trained on its own data; if this agent has a large dataset, it must be compensated through its subsidy.

The parameter $\beta$ governs both effects. \Cref{lem:incr_utility} requires $\beta>\zeta+\tau$ so that contributing more is always profitable, and compensating large contributors whose gain from the shared model is small may require $\beta$ to be larger still. On the other hand, the payment of each small contributor grows linearly in $\beta$, so a large $\beta$ can drive small contributors out. Individual rationality therefore holds only for $\beta$ in a window, whose endpoints depend on how much each agent gains from collaboration and on how far its capacity is from the average.

The rest of this subsection makes this precise. We first fix the cost accounting, the individual learning outcome and the notion of individual rationality. \Cref{prop:IR} then gives each agent's exact gain from participating at the outcome of \mechtwo{} and characterizes the window of $\beta$ for which every agent participates, and \Cref{cor:IR-finite} shows that the guarantee holds after finitely many rounds. Finally, when the window is empty, \Cref{cor:IR-rebate} shows that zero-sum lump-sum rebates restore individual rationality, without affecting incentives or budget balance, exactly when the agents gain from collaboration in aggregate.


\paragraph{Individual learning outcome.} An agent that does not participate pays and receives no
transfers and has no access to the federated model; it can only train on its own data.
For $x\in S_i$, let $v_i^{\mathrm{loc}}(x)$ denote the valuation that agent $i$ derives
from the best model it can train alone on $x$ of its own data points. Agent $i$'s
\emph{individual learning outcome} is the best stand-alone utility,
\begin{equation}
\label{eq:outside-option}
    \bar u_i := \sup_{x\in S_i}\big[v_i^{\mathrm{loc}}(x) - c_i(x)\big].
\end{equation}
We use the stand-alone benchmark rather than the value $0$ because the valuation
$v_i = r_i - L_i$ contains the arbitrary normalizing constant $r_i$, so comparing
utilities with $0$ would carry no meaning.

\paragraph{Outcome and individual rationality.} By \Cref{thm:two_phase}, \mechtwo{}
reaches the full-contribution profile $s^{\max}$ and learns $\wopt$. The utility of
agent $i$ at this \emph{limit outcome} is
\begin{equation}
\label{eq:limit-utility}
    U_i := v_i(\wopt,s^{\max}) - c_i(s_i^{\max}) + p_i^1(s^{\max}),
\end{equation}
and its utility at the outcome after $T$ rounds is
$U_i^T := v_i(w^T,s^T) - c_i(s_i^T) + p_i^1(s^T)$.

\begin{definition}[Individual rationality]
\label{def:IR}
The limit outcome of \mechtwo{} is \emph{individually rational} (IR) if
$U_i\geqslant\bar u_i$ for every $i\in N$. The outcome after $T$ rounds is
individually rational if $U_i^T\geqslant\bar u_i$ for every $i\in N$.
\end{definition}

\paragraph{Gain from collaboration and average capacity.} We write
\begin{equation}
\label{eq:collab-gain}
    \pi_i := v_i(\wopt,s^{\max}) - c_i(s_i^{\max}) - \bar u_i
\end{equation}
for agent $i$'s \emph{gain from collaboration}: the value of the federated model at full
contribution, net of the cost of contributing all of its data, in excess of what the
agent achieves alone. The gain $\pi_i$ may be negative, for instance for an agent with
high per-sample cost. We also write
\begin{equation}
\label{eq:avg-capacity}
    \bar s^{\max} := \frac1n\sum_{j\in N}s_j^{\max}
\end{equation}
for the average capacity, and we split the agents into those with below-average,
average and above-average capacity,
\begin{equation*}
    N_- := \{i : s_i^{\max}<\bar s^{\max}\},\qquad
    N_0 := \{i : s_i^{\max}=\bar s^{\max}\},\qquad
    N_+ := \{i : s_i^{\max}>\bar s^{\max}\}.
\end{equation*}
Finally, we define the two thresholds
\begin{equation}
\label{eq:beta-window}
    \beta_{\mathrm{low}} := \max\Big\{0,\ \max_{i\in N_+}
        \frac{-(n-1)\,\pi_i}{n\,(s_i^{\max}-\bar s^{\max})}\Big\},
    \qquad
    \beta_{\mathrm{high}} := \min_{i\in N_-}
        \frac{(n-1)\,\pi_i}{n\,(\bar s^{\max}-s_i^{\max})},
\end{equation}
with the conventions that a maximum over the empty set is $0$ and a minimum over the
empty set is $+\infty$.

\begin{proposition}[Individual rationality of \mechtwo{}]
\label{prop:IR}
Suppose \Cref{assump:derivatives,assump:conv-opt-NE} hold and the payment is
$p^1$ of \Cref{eq:payment} with $\beta>\zeta+\tau$. Then the following hold.
\begin{enumerate}[leftmargin=15pt, itemsep=0pt, parsep=0pt, topsep=0pt]
    \item At the limit outcome, agent $i$'s gain from participating is
    \begin{equation}
    \label{eq:IR-gain}
        U_i - \bar u_i \;=\; \pi_i + \frac{n\beta}{n-1}\big(s_i^{\max} - \bar s^{\max}\big).
    \end{equation}
    \item The limit outcome is individually rational if and only if
    $\pi_i\geqslant 0$ for every $i\in N_0$ and
    $\beta_{\mathrm{low}}\leqslant\beta\leqslant\beta_{\mathrm{high}}$.
    \item Consequently, there exists a $\beta$ for which \Cref{thm:two_phase} applies
    and the limit outcome is individually rational if and only if $\pi_i\geqslant0$ for
    every $i\in N_0$, $\beta_{\mathrm{low}}\leqslant\beta_{\mathrm{high}}$, and
    $\zeta+\tau<\beta_{\mathrm{high}}$. In particular, if $\pi_i\geqslant 0$ for every
    $i\in N$, such a $\beta$ exists if and only if $\zeta+\tau<\beta_{\mathrm{high}}$.
\end{enumerate}
\end{proposition}

\begin{proof}[Proof sketch]
Evaluating \Cref{eq:payment} at $s^{\max}$ gives $p_i^1(s^{\max})=\frac{n\beta}{n-1}(s_i^{\max}-\bar s^{\max})$, which yields Part~1. The condition of Part~1 is then linear in $\beta$: it is an upper bound on $\beta$ for agents in $N_-$, a lower bound for agents in $N_+$, and independent of $\beta$ for agents in $N_0$, which gives Parts~2 and~3. The complete proof is in \Cref{app:proofs-IR}.
\end{proof}

\noindent\textsc{Remark (Interpretation and choice of $\beta$).}
\Cref{prop:IR} shows that individual rationality can fail only for agents with
below-average capacity whose gain from collaboration is small: the tax such an agent
pays at the limit outcome, $\frac{n\beta}{n-1}(\bar s^{\max}-s_i^{\max})$, grows linearly
in $\beta$. Since \Cref{thm:two_phase} only needs $\beta>\zeta+\tau$, the center should
choose $\beta$ just above this threshold, which makes the window of
\Cref{prop:IR} as easy to satisfy as possible.

The proposition concerns the limit outcome. The next corollary shows that the conclusion
carries over to every sufficiently late round whenever the gains are strict.

\begin{corollary}[Individual rationality after finitely many rounds]
\label{cor:IR-finite}
In addition to the hypotheses of \Cref{prop:IR}, suppose that each valuation
$v_i(\cdot,s^{\max})$ is continuous at $\wopt$, that \mechtwo{} is run with
$\eta=1/M$, and that $U_i>\bar u_i$ for every $i\in N$. Then there exists $\bar T$ such
that the outcome after $T$ rounds is individually rational for every $T\geqslant\bar T$.
\end{corollary}

\begin{proof}[Proof sketch]
After Phase~1, the contributions and the transfers no longer change, and $w^T\to\wopt$ by strong convexity, so $U_i^T\to U_i$ by continuity; the strict inequality $U_i>\bar u_i$ then holds after finitely many rounds. The complete proof is in \Cref{app:proofs-IR}.
\end{proof}

\paragraph{Lump-sum rebates.} When the window of \Cref{prop:IR} is empty, the center
can restore individual rationality without affecting incentives or budget balance. The
reason is that \Cref{lem:incr_utility} depends only on the slope
$\pd{s_i}p_i^1 = \beta$ of the payment, not on its level. For constants
$\rho_1,\ldots,\rho_n\in\mathbb{R}$, consider the payment
\begin{equation}
\label{eq:payment-rebate}
    p_i^{\rho}(s) := p_i^1(s) - \rho_i,
    \qquad\text{where}\qquad \sum_{i\in N}\rho_i = 0
    \quad\text{and each $\rho_i$ is fixed before Phase~1.}
\end{equation}
Requiring $\rho_i$ to be fixed in advance, and in particular independent of agent $i$'s
own reports, is essential: if $\rho_i$ were computed from agent $i$'s reported final
contribution, for instance as $\rho_i = p_i^1(\hat s)$, it would cancel $p_i^1$ exactly
and remove the incentive to contribute.

\begin{corollary}[Individual rationality with lump-sum rebates]
\label{cor:IR-rebate}
Suppose the hypotheses of \Cref{thm:two_phase} hold, and the payment $p^1$ is replaced
by $p^{\rho}$ of \Cref{eq:payment-rebate}. Then the following hold.
\begin{enumerate}[leftmargin=15pt, itemsep=0pt, parsep=0pt, topsep=0pt]
    \item \mechtwo{} remains budget balanced, \Cref{lem:incr_utility} still holds, and
    all conclusions of \Cref{thm:two_phase} remain valid.
    \item There exists a choice of rebates for which the limit outcome is individually
    rational if and only if $\sum_{i\in N}\pi_i\geqslant 0$.
    \item If the capacities $s_1^{\max},\ldots,s_n^{\max}$ are known to the center
    before Phase~1, the choice $\rho_i := p_i^1(s^{\max})$ satisfies
    \Cref{eq:payment-rebate}, makes every transfer at the limit outcome zero, and makes
    the limit outcome individually rational if and only if $\pi_i\geqslant 0$ for every
    $i\in N$.
\end{enumerate}
\end{corollary}

\begin{proof}[Proof sketch]
The rebates are constants that sum to zero, so they change neither the slope of the payment, on which \Cref{lem:incr_utility} and \Cref{thm:two_phase} rely, nor its budget balance. Summing the gains $\pi_i+p_i^1(s^{\max})-\rho_i$ over the agents gives $\sum_i\pi_i$, which proves the necessity in Part~2, and the rebates that split $\sum_i\pi_i$ equally prove its sufficiency. The complete proof is in \Cref{app:proofs-IR}.
\end{proof}

\noindent\textsc{Remark (Information required by the rebates).}
Part~2 of \Cref{cor:IR-rebate} is a full-information statement: to set the rebates, the
center must know each agent's gain $\pi_i$, which depends on private costs and
stand-alone values. 
Part~3 needs only the capacities, which agents disclose when registering; since an agent
could under-declare its capacity, this part assumes the declared capacities are correct.
\color{black}

\subsection{Truthfulness and welfare maximization at NE}
\label{sec:truthful-welfare}

\paragraph{Why a truthful version of \mechtwo{}.} Both \mech{} and \mechtwo{} rely on what the agents report about the data they use and the gradients they compute. We make \mechtwo{} truthful because it is the mechanism that reaches full contribution and learns $\wopt$; truthful reports in \mech{} would only certify an equilibrium that may involve under-contribution, as in \Cref{exam:NEnotsocial}. The idea developed below, a cross-check of each agent's contribution report against the other agents' realized valuations together with a budget-balanced charge on gradient reports, applies to \mech{} as well: used in every round of \mech{}, it gives \mechthree{}, whose exposition and formal guarantees (\Cref{thm:EPIC}) are in \Cref{app:T-UPBReD}.

\mechtwo{} reaches full contribution and learns $\wopt$, but it trusts two kinds of reports: the contributions $\hat s_i$, on which the subsidy $p^1$ is computed, and the gradients $\hat d_i^t$, from which the model is computed. In this subsection we ask whether both kinds of reports can be made incentive compatible (truthful) without giving up budget balance. The answer has three parts. First, \Cref{prop:no-evidence} shows that no payment rule can reward contributions unless something the center observes depends on the data an agent actually used. Second, we introduce two assumptions under which the other agents' realized valuations provide such evidence, and a payment rule that exploits it through a cross-check, together with another payment that ensures the gradient reports are approximately truthful. Third, \Cref{thm:truthful_T2P} shows that the resulting mechanism, \mechfour{}, elicits contributions truthfully, reaches full contribution, learns $\wopt$ under truthful as well as equilibrium gradient reporting, and never runs a deficit.

\paragraph{Why truthful contributions need evidence.} \Cref{thm:two_phase} assumes that the contributions $\hat s_i$ on which $p^1$ is computed are the contributions actually used in training. The next result shows that this assumption cannot be removed by a cleverer payment rule alone.

\paragraph{Mechanisms with the structure of \mechtwo{}.}
We consider mechanisms that run in two phases, as \mechtwo{} does. Each agent $i$ takes
one \emph{action}, its contribution $s_i\in S_i=[s_i^{\min},s_i^{\max}]$, which is the
amount of data it actually uses in training, and it incurs the one-time cost
$c_i(s_i)$, in line with the cost accounting of \Cref{sec:IR}. Throughout the run,
agent $i$ also sends \emph{reports}: contribution reports in Phase~1, gradient reports
in Phase~2, and possibly further reports, such as valuations, at the end. A \emph{reporting strategy} $m_i$ specifies all of these reports as functions of the information broadcast by the center so far, and $M_i$ denotes the set of reporting strategies available to agent $i$. Reporting strategies do not take agent $i$'s own contribution, or observations determined by it, as inputs; since agent $i$ chooses the pair $(s_i,m_i)$ jointly, it can still coordinate its reports with its contribution, so this is without loss of generality. The \emph{record} $R$ is everything the center observes during the run: all
agents' reports and any additional observations the center may make. The center
computes the final parameter $\bar w = W(R)$ from the record, for instance by
aggregating the reported gradients as in \Cref{algo:2-phase}, and pays each agent a
transfer $p_i(R)$. The transfer rule $p$ is arbitrary: it need not be budget balanced,
and it may depend on any part of the record. The record may be random, for instance
through mini-batch sampling, in which case utilities are understood in expectation.
 
We fix the strategies of all agents other than $i$, both their actions and their
reporting strategies, and we study agent $i$'s choice of a pair $(s_i,m_i)$. Agent $i$'s
utility is
\begin{equation}
\label{eq:app-ne-utility}
    u_i = v_i(\bar w, s) - c_i(s_i) + p_i(R),
\end{equation}
where $s=(s_i,s_{-i})$ is the profile of contributions actually used.

If agents also value intermediate models, as in the myopic Phase~2 utility of
\mechfour{}, the term $\varphi_i(W(R))$ below is replaced by a sum of terms
$\varphi_i(w^t)$, each of which is computed from the record; the proof is unchanged. 

\begin{definition}[Separable valuations]
\label{def:separable}
The valuations are \emph{separable} if, for every agent $i$, there are functions
$\varphi_i$ and $\psi_{ij}$, $j\in N$, such that
\begin{equation*}
    v_i(w,s) = \varphi_i(w) + \sum_{j\in N}\psi_{ij}(s_j)
    \qquad\text{for all } w\in\mathbb{R}^m \text{ and } s\in\textstyle\prod_{j}S_j .
\end{equation*}
\end{definition}
 
The valuations of \Cref{sec:examples_utility} are separable: with the loss
\Cref{eq:loss-decomposition}, $\varphi_i(w) = r_i - L_i(w)$ and
$\psi_{ij}(s_j) = -\frac1n s_j^{-\alpha_i}$, and this holds for each of the losses
\Cref{eq:cross-entropy,eq:mse,eq:quadratic-loss}. Separability has a consequence that we
use repeatedly: the gradient $\nabla_w v_i(w,s) = \nabla\varphi_i(w)$ does not depend on
the contributions, so a gradient report carries no information about how much data was
used, whether it is truthful or not.
 
\begin{definition}[No evidence]
\label{def:no-evidence-app}
Agent $i$'s contribution leaves \emph{no evidence} if
\begin{enumerate}[leftmargin=20pt, itemsep=0pt, parsep=0pt, topsep=0pt]
    \item[(E1)] the set $M_i$ of reporting strategies does not depend on $s_i$; and
    \item[(E2)] for every reporting strategy $m_i\in M_i$, the joint distribution of the
    record $R$ and the other agents' contributions $s_{-i}$ under $(s_i,m_i)$ does not
    depend on $s_i$.
\end{enumerate}
\end{definition}
 
Condition (E1) says that agent $i$ can send the same reports whatever amount of data it
actually used, and condition (E2) says that neither the center nor the other agents
receive any information about agent $i$'s contribution except through agent $i$'s own
reports.
 
\noindent\textsc{Remark (When the conditions hold and when they fail).}
In \mechtwo{} with separable valuations, both conditions hold when no agent observes
another agent's contribution. Condition (E1) holds because the reports that matter are
numbers that agent $i$ can choose freely, and because, by separability, even a truthful
gradient $\nabla\varphi_i(w^t)$ is the same for every contribution. Condition (E2)
holds because the center's record consists of reports only, and because the other agents
neither observe $s_i$ nor, by separability, have their own best contributions depend on
it. Each of the following violates a condition and therefore escapes the proposition:
\begin{itemize}[leftmargin=15pt, itemsep=0pt, parsep=0pt, topsep=0pt]
    \item \Cref{assump:realized}, under which the other agents report valuations that
    depend on agent $i$'s actual contribution, violates (E2);
    \item another agent that observes $s_i$ and reports it violates (E2);
    \item if gradients are truthful by assumption and the valuations are not separable,
    so that $\nabla_w v_i(w,s)$ depends on $s_i$, then the truthful gradient differs
    across contributions and (E1) fails; such gradients act as evidence in the sense of
    \citet{green1986partially}.
\end{itemize}
By contrast, a public upper bound on each agent's capacity does \emph{not} violate (E1):
it restricts the reports in the same way whatever the agent actually contributes, so it
rules out claims beyond $s_i^{\max}$ but not using less data than claimed.
 
We state the proposition below.
 
\begin{proposition}[No evidence, no incentive]
\label{prop:no-evidence}
Suppose the valuations are separable, and agent $i$'s contribution leaves no evidence.
For a reporting strategy $m_i$, let
\begin{equation}
\label{eq:app-ne-Pi}
    \Pi_i(m_i) := \mathbb{E}\Big[\varphi_i\big(W(R)\big)
        + \sum_{j\neq i}\psi_{ij}(s_j) + p_i(R)\Big],
    \qquad
    A_i(s_i) := \psi_{ii}(s_i) - c_i(s_i),
\end{equation}
where the expectation is taken under the pair $(s_i,m_i)$, and assume that these
expectations are finite. Then, for every transfer rule $p$, the following hold.
\begin{enumerate}[leftmargin=15pt, itemsep=0pt, parsep=0pt, topsep=0pt]
    \item \textnormal{(Transfers cannot depend on the contribution.)} $\Pi_i(m_i)$ does
    not depend on $s_i$; in particular, neither does the expected transfer
    $\mathbb{E}[p_i(R)]$.
    \item \textnormal{(Contributions are as without transfers.)} Agent $i$'s expected
    utility under $(s_i,m_i)$ equals $A_i(s_i)+\Pi_i(m_i)$. A pair $(s_i,m_i)$ is a best
    response if and only if $s_i$ maximizes $A_i$ over $S_i$ and $m_i$ maximizes $\Pi_i$
    over $M_i$. Since $A_i$ does not involve $p$, agent $i$'s best contributions are the
    same under every transfer rule, including the rule $p\equiv 0$.
    \item \textnormal{(Full contribution.)} Some transfer rule makes $s_i^{\max}$ part of
    a best response if and only if every transfer rule does, if and only if
    $s_i^{\max}$ maximizes $A_i$ over $S_i$.
    \item \textnormal{(Approximate truthfulness.)} For each $x\in S_i$, let
    $m_i^{T}(x)\in M_i$ be a reporting strategy that reports the contribution $x$. If
    reporting truthfully is $\varepsilon$-optimal, that is,
    $\Pi_i\big(m_i^T(s_i)\big)\geqslant\sup_{m\in M_i}\Pi_i(m)-\varepsilon$ for every
    $s_i\in S_i$, then
    $\big|\Pi_i\big(m_i^T(x)\big)-\Pi_i\big(m_i^T(y)\big)\big|\leqslant\varepsilon$ for
    all $x,y\in S_i$.
\end{enumerate}
\end{proposition}

\begin{proof}[Proof sketch]
By separability, agent $i$'s realized utility splits into $A_i(s_i)$, which involves neither the transfers nor the reports, and a part that is a function of the record $R$ and the other agents' contributions $s_{-i}$ only. Condition (E2) makes the joint distribution of $(R,s_{-i})$, and hence the expectation of the second part, independent of $s_i$, which gives Part~1, and condition (E1) lets agent $i$ choose $s_i$ and $m_i$ independently, so its expected utility $A_i(s_i)+\Pi_i(m_i)$ is maximized separately in each, which gives Parts~2 and~3. Part~4 follows because all values $\Pi_i(m_i^T(x))$ lie in an interval of length $\varepsilon$. The complete proof is in \Cref{app:proofs-no-evidence}.
\end{proof}

We now derive two consequences for \mechtwo{}.

\begin{corollary}[Consequences for \mechtwo{}]
\label{cor:no-evidence-2P}
In the setting of \Cref{prop:no-evidence}, consider \mechtwo{} with truthful gradient
reports and the payment $p^1$ of \Cref{eq:payment} computed on the reported
contributions.
\begin{enumerate}[leftmargin=15pt, itemsep=0pt, parsep=0pt, topsep=0pt]
    \item If, in addition, the valuations are those of \Cref{sec:examples_utility} and the
    costs are linear, $c_i(s_i)=c_is_i$, then agent $i$ contributes
    $s_i^{\mathrm{nt}} := \min\big\{s_i^{\max},\max\{s_i^{\min},(\alpha_i/(nc_i))^{1/(\alpha_i+1)}\}\big\}$,
    whatever the value of $\beta$; so full contribution fails whenever
    $(\alpha_i/(nc_i))^{1/(\alpha_i+1)}<s_i^{\max}$.
    \item Truthful contribution reporting under $p^1$ is $\varepsilon$-optimal only if
    $\varepsilon\geqslant\beta(s_i^{\max}-s_i^{\min})$; with $\beta>\zeta+\tau$, as
    \Cref{thm:two_phase} requires, this exceeds $(\zeta+\tau)(s_i^{\max}-s_i^{\min})$.
\end{enumerate}
\end{corollary}

\begin{proof}[Proof sketch]
Part~1 maximizes $A_i(x)=-\frac1n x^{-\alpha_i}-c_ix$, which is strictly concave, over $S_i$. For Part~2, under separability the reported contribution changes neither the model nor the other agents' contributions, so two reporting strategies that differ only in the reported contribution differ in $\Pi_i$ by exactly $\beta(x-y)$, and the comparison of $s_i^{\min}$ with $s_i^{\max}$ gives the bound. The complete proof is in \Cref{app:proofs-no-evidence}.
\end{proof}

\noindent\textsc{Remark (Relation to the literature).} \Cref{prop:no-evidence} is closely related to the informativeness principle of \citet{holmstrom1979moral}, which states that an optimal incentive contract for an unobserved action should depend on a signal if and only if the signal is informative about that action. Our statement differs in three respects: it applies to every transfer rule rather than to optimal contracts; the center's information consists largely of reports that the agents choose strategically, rather than of exogenous signals; and it includes an approximate version (Part~4).  Any truthful version of \mechtwo{} must therefore rely on some observation that depends on the actual contributions.
\color{black}

A natural design follows the two-stage mechanisms for interdependent valuations of \citet{mezzetti2004mechanism} and \citet{NATH2013321}: pay each agent the other agents' reported valuations, and check only its own valuation report against the reported contributions. Here this fails in two ways. An agent that over-reports its contribution and reports the matching valuation passes its own check and keeps the extra subsidy, as \Cref{prop:no-evidence} predicts; and paying the others' valuations aligns each agent's utility with the welfare, so an agent gains by bending its reported gradient to move the model towards a better outcome. \mechfour{} therefore checks each agent's contribution against the \emph{other} agents' valuations, and controls the gradients with a separate payment.

\paragraph{The payments.}
\mechfour{} uses three transfers. The first is the budget-balanced subsidy of \mechtwo{},
\begin{equation}
\label{eq:2PT_part1}
    p^1_i(\hat s)=\beta\Big(\hat s_i-\frac{1}{n-1}\sum_{j \neq i}\hat s_j\Big).
\end{equation}
The second is computed once, after training, at the final model $\bar w := w^{T_0}$.
Each agent reports the contribution $\hat s_i$ it used and its realized valuation
$\hat v_i$ of $\bar w$, and pays
\begin{equation}
\label{eq:payment_truthful_2P}
    p_i^{2}(\hat s,\hat v)
    = -\big(\hat v_i - v_i(\bar w,\hat s)\big)^2
      - \chi_i(\hat s,\hat v)\,\big(\beta\hat s_i + K\big),
    \qquad
    \chi_i(\hat s,\hat v) =
    \begin{cases}
        1 & \text{if } \hat v_j \neq v_j(\bar w,\hat s) \text{ for some } j\neq i,\\
        0 & \text{otherwise,}
    \end{cases}
\end{equation}
where $K>0$ is a constant. The first term is the own-report test of \citet{NATH2013321}, which makes truthful valuation reports strictly optimal. The second term is a \emph{cross-check}: agent $i$ is penalized whenever the \emph{other} agents' valuation reports are inconsistent with the reported contributions. The penalty $\beta\hat s_i$ cancels the subsidy that a false contribution report would earn, and $K$ makes the deviation strictly unprofitable. The third transfer is charged in every round $t$ of Phase~2 on the reported gradients,
\begin{equation}
\label{eq:grad-charge}
    p_i^3(\hat d^t)
    = -\frac{\eta}{2n}\,\|\hat d_i^t\|_2^2
      + \frac{\eta}{2n(n-1)}\sum_{j\neq i}\|\hat d_j^t\|_2^2 .
\end{equation}

The gradient step is the exact maximizer of a surrogate welfare built from the reports,

\begin{equation}
\label{eq:surrogate}
    w^{t+1}
    = w^t + \frac{\eta}{n}\sum_{j\in N}\hat d_j^t
    = \argmax_{w\in\mathbb{R}^m}\Big\{\sum_{j\in N}(\hat d_j^t)^\top (w-w^t)
      - \frac{n}{2\eta}\,\|w-w^t\|_2^2\Big\},
\end{equation}

and the first term of \Cref{eq:grad-charge} is the Clarke payment of VCG for this
surrogate. The second term depends only on the other agents' reports, so it does not
affect incentives, and it makes $\sum_i p_i^3 = 0$ for every report profile. Note that
$p^3$ uses only the reported gradients. On the truthful path both penalties in
\Cref{eq:payment_truthful_2P} vanish, so agent $i$'s total transfer is
$p_i^1(s)+\sum_t p_i^3(d^t)$, and agent $i$'s utility is
\begin{equation}
\label{eq:2PT_total_payment}
    v_i(\bar w,s)-c_i(s_i)+ p^1_i(\hat s)+p_i^2(\hat s,\hat v)+\sum_{t<T_0}p_i^3(\hat d^t),
\end{equation}
where the valuation and the cost are evaluated at the true contributions and the transfers at the reports.

The cross-check supplies the evidence that \Cref{prop:no-evidence} requires, provided two properties hold: another agent's valuation must react to agent $i$'s contribution, and
agents must experience that reaction.

\begin{assumption}[Strict interdependence]
\label{assump:strict-interdep}
For every agent $i\in N$, every parameter $w$, every $s_{-i}$ and every pair $\hat s_i \neq s_i$ in $S_i$, there exists an agent $j\neq i$ such that $v_j(w,\hat s_i,s_{-i}) \neq v_j(w,s_i,s_{-i})$.
\end{assumption}

\begin{assumption}[Realized payoffs]
\label{assump:realized}
After training, each agent $j$ observes its realized valuation $v_j(\bar w,s)$ at the contribution profile $s$ that was actually used in training, whatever the agents reported, and the center can evaluate $v_j(\bar w,\hat s)$ at any reported profile $\hat s$.
\end{assumption}

\noindent\textsc{Remark (Black-box valuations).} We view each valuation as a black box: the world produces agent $j$'s realized valuation from the model and the data actually used to train it, whatever the agents report. Agents observe only these realized outputs, while the center knows how the box operates and can evaluate it at any input, in particular at the reported profile, reflecting that it observes outcomes across many agents and deployments. This is the standard primitive of interdependent-value mechanisms \citep{mezzetti2004mechanism,NATH2013321}, and it is why valuations are elicited after training, when the effect of the data actually used is experienced. In our setup, the deterministic $\bar w$ of the analysis stands for a model trained on stochastic gradients computed from the samples actually used, and the finite-sample term represents the expected excess loss of that realized model \citep{Donahue_Kleinberg_2021}. In the terms of \Cref{prop:no-evidence}, the black box supplies the required evidence: by \Cref{assump:strict-interdep}, the other agents' realized valuations depend on agent $i$'s actual contribution, so the cross-check can tie the subsidy to the data actually used. Since realized valuations are noisy in practice, the tests in \Cref{eq:payment_truthful_2P} should use a tolerance (see the remark after \Cref{thm:truthful_T2P}).

\noindent\textsc{Remark (On strict interdependence).}
\Cref{assump:strict-interdep} requires $n\geqslant 2$. It holds for the utility form of \Cref{eq:utility-example} with the loss \Cref{eq:loss-decomposition}: for every
$j\neq i$, $\pd{s_i}v_j(w,s) = \frac{\alpha_j}{n}s_i^{-\alpha_j-1}>0$, so $v_j$ is strictly
increasing in $s_i$.


\mechfour{} runs in three stages, given in \Cref{algo:truthful_welfare}.
\begin{enumerate}[leftmargin=15pt, itemsep=0pt, parsep=0pt, topsep=0pt]
    \item \emph{Phase~1 (contributions), rounds $k=0,\ldots,\kappa-1$.} The model stays at $w^0$. The center broadcasts the current contribution reports $\hat s^k$; each agent $i$ updates its actual contribution by a projected gradient step on $v_i-c_i+p_i^1$, evaluated at $(w^0,\hat s^k)$, and reports $\hat s_i^{k+1}$. No
    transfers are made in this phase.
    
    \item \emph{Phase~2 (training), rounds $t=0,\ldots,T_0-1$.} The center broadcasts $w^t$; each agent sends a gradient report $\hat d_i^t$, whose truthful value is the gradient computed on the data it actually uses ($s^{\max}$ on the equilibrium path); the center updates $w^{t+1}=w^t+\frac{\eta}{n}\sum_{j\in N}\hat d_j^t$ and charges $p^3(\hat d^t)$ of \Cref{eq:grad-charge}.
   
    \item \emph{Elicitation, once, at $\bar w = w^{T_0}$.} The center broadcasts $\bar w$; each agent observes its realized valuation $v_i(\bar w,s)$, and all agents send their reports $(\hat s_i,\hat v_i)$ simultaneously, without seeing each other's reports. The center then pays $p^1(\hat s)+p^2(\hat s,\hat v)$ of \Cref{eq:2PT_part1,eq:payment_truthful_2P}.
\end{enumerate}
This order has three consequences, which the proof of \Cref{thm:truthful_T2P} uses repeatedly: (T1)~contribution reports sent in Phase~1 enter no transfer and do not affect the model; (T2)~a gradient report affects only the next model and the charge of its own round; and (T3)~the elicitation reports are sent after $\bar w$ is computed and all
Phase~2 charges are paid, so they affect only $p^1$ and $p^2$. Because the elicitation reports are simultaneous, no agent can tailor its valuation report to another agent's contribution report. Under truthful reporting, Phases~1 and~2 coincide with those of \mechtwo{}.
\begin{algorithm}
\DontPrintSemicolon
\caption{\mechfullfour{} (\mechfour{})}
\label{algo:truthful_welfare}
\KwIn{Step sizes $\gamma,\eta$; constant $K>0$; initialization $w^0$, $s_i^0\in S_i$ for $i\in N$;
      iteration counts $\kappa,T_0$ as per \Cref{thm:truthful_T2P}}
\KwOut{$w^{T_0}$}
\tcp{Phase 1: contributions, with the model fixed at $w^0$}
\textbf{Center} broadcasts $w^0,\hat s^0$; set $k=0$\;
\While{$k<\kappa$}{
  \textbf{Center} broadcasts $\hat s^k$\;
  \ForEach{\textbf{agent} $i\in N$ in parallel}{
    $s_i^{k+1}=s_i^k+\gamma\,[g(w^0,\hat s^k,\mu^k)]_i$; send $\hat s_i^{k+1}$ to center\;
  }
  $k\gets k+1$\;
}
\tcp{Phase 2: training, with gradient charges}
\For{$t=0$ \KwTo $T_0-1$}{
  \textbf{Center} broadcasts $w^{t}$\;
  \ForEach{\textbf{agent} $i\in N$ in parallel}{
    Send gradient report $\hat d_i^{t}$ (truthful: $\nabla_w v_i(w^{t},s^{\max})$)\;
  }
  \textbf{Center} updates $w^{t+1}=w^{t}+\frac{\eta}{n}\sum_{i\in N}\hat d_i^{t}$ and pays $p_i^3(\hat d^t)$ of \Cref{eq:grad-charge}\;
}
\tcp{Elicitation, at the final model $\bar w=w^{T_0}$}
\ForEach{\textbf{agent} $i\in N$ in parallel}{
  Observe $v_i(\bar w,s)$; send $\hat s_i$ and $\hat v_i$ to center\;
}
\textbf{Center} pays $p_i^1(\hat s)+p_i^2(\hat s,\hat v)$ of \Cref{eq:2PT_part1,eq:payment_truthful_2P}\;
\Return $w^{T_0}$\;
\end{algorithm}

The gradient guarantees need one more assumption on the individual valuations.

\begin{assumption}
\label{assump:agent-smooth}
Each $v_i(\cdot,s^{\max})$ is differentiable with $M_i$-Lipschitz gradient, and $\eta M_i < n$ for all $i\in N$.
\end{assumption}

In Phase~2 agents are myopic, as elsewhere in the paper: in round $t$, agent $i$'s
utility is $v_i(w^{t+1},s^{\max})+p_i^3(\hat d^t)$. Write
$f(w) := -\frac1n\sum_{i\in N}v_i(w,s^{\max})$,
$\tilde g^t := \frac1n\sum_{j\in N}\nabla_w v_j(w^t,s^{\max})$, and
$\Delta := \beta-\tau-\zeta$.
\begin{theorem}[\mechfour{}: truthfulness, full contribution, and learning]
    \label{thm:truthful_T2P}
    Suppose \Cref{assump:derivatives,assump:conv-opt-NE,assump:strict-interdep,assump:realized,assump:agent-smooth} hold, $\beta>\zeta+\tau$ and $\eta = 1/M$. Then \mechfour{} (\Cref{algo:truthful_welfare}) satisfies the following.
    \begin{enumerate}[leftmargin=15pt, itemsep=0pt, parsep=0pt, topsep=0pt]
        \item  \textnormal{(Truthful contribution and valuation reports.)} The elicitation stage is EPIC: if the other agents report truthfully, then for \emph{every} final model $\bar w$, and hence for every history of gradient reports, truthful reporting of $(\hat s_i,\hat v_i)$ is optimal. In particular, an agent that trains on fewer samples than it reports is penalized.
        
        \item  \textnormal{(Full contribution.)} With truthful elicitation, for every $w$, contributing $s_i^{\max}$ maximizes $v_i(w,s)-c_i(s_i)+p_i^1(s)$ whatever the other agents do, and Phase~1 reaches $s^{\kappa}=s^{\max}$ for every $w^0\in\mathbb{R}^m$ and $s^0\in\prod_{i\in N}S_i$, where
        \begin{equation}
            \label{eq:kappa-prime}
            \textstyle \kappa = \max_{i\in N}\left\lceil \frac{s_i^{\max}-s_i^0}{\gamma\,\Delta}\right\rceil .
        \end{equation}
        \item  \textnormal{(Approximately truthful gradients.)} In every round $t$ of Phase~2, if the other agents report truthfully, no gradient report gains more than
        $\varepsilon_i^t := \frac{\eta^3 M_i^2}{2(n-\eta M_i)}\|\tilde g^t\|_2^2$ over truthful reporting. 
         Along the truthful path, $\varepsilon_i^t\leqslant C_i\big(1-\nicefrac{\nu}{M}\big)^t$ with $C_i := \frac{M_i^2(f(w^0)-f(\wopt))}{M(nM-M_i)}$, so the sum of these one-shot bounds over the rounds of Phase~2 is at most $\frac{M_i^2(f(w^0)-f(\wopt))}{\nu(nM-M_i)}$.
        
        \item \textnormal{(Optimal parameter.)} If agents report gradients truthfully, $f(w^{T_0})-f(\wopt)<\epsilon$ after $T_0$ rounds, with $T_0$ as in \Cref{eq:2-phase_strong-T_1}. If instead, in every round, each agent reports the gradient that maximizes its round utility given the other agents' reports, such a profile of reports exists and is unique, and $\|w^{t}-\wopt\|_2\leqslant(1+\eta\nu)^{-t}\|w^0-\wopt\|_2$.
        
        \item \textnormal{(Budget.)} On the truthful path, the transfers sum to zero. For every report profile, $\sum_i p_i^3=0$ in every round and $\sum_i(p_i^1+p_i^2)\leqslant 0$, so the center never runs a deficit.
    \end{enumerate}
\end{theorem}

\noindent\emph{Proof sketch.}
Part~1: gradient reports do not depend on $\hat s_i$, so agent $i$'s contribution report changes only $p^1+p^2$. If $\hat s_i\neq s_i$, the other agents' realized valuations
differ from $v_j(\bar w,\hat s)$ for some $j$ (\Cref{assump:strict-interdep,assump:realized}), so $\chi_i=1$ and the extra subsidy $\beta(\hat s_i - s_i)$ is outweighed by the penalty $\beta\hat s_i + K$. 

Part~2 is \Cref{lem:incr_utility}, which applies with $\beta>\zeta+\tau$ because no term $\sum_{j\neq i}v_j$ is paid. 

Part~3: agent $i$'s round utility is $\mu$-strongly concave in its report with $\mu=\frac{\eta}{n}(1-\frac{\eta M_i}{n})$, and its gradient at the truthful report is at most $\frac{\eta^2}{n}M_i\|\tilde g^t\|_2$. 

Part~4: at the round equilibrium every agent reports $\nabla_w v_i(w^{t+1},s^{\max})$, which makes the update a proximal point step on $f$. 

Part~5: $p^1$ and $p^3$ are budget balanced, and both penalties in $p^2$ are non-negative and vanish on the truthful path. The full proof is in \Cref{app:T2P-proof}.
\hfill$\square$ 

\Cref{example:T-2P-necessary} illustrates two guarantees of \mechfour{}: checking an agent's valuation report only against its own contribution report does not prevent misreporting, and the charge \Cref{eq:grad-charge} reduces the gain from misreporting a gradient.

\begin{example}
\label{example:T-2P-necessary}
\emph{Contributions.} Take $n=3$, $\beta=2$, $K=1$, $S_i=[1,15]$ and the utility form of
\Cref{eq:utility-example}, and let agent $i$ train on $s_i=5$ samples. Under a rule that
checks only agent $i$'s own valuation report, agent $i$ could report $\hat s_i = 12$ with
the matching valuation, pass its own test, and gain $\beta(12-5)=14$. Under
\Cref{eq:payment_truthful_2P}, the other two agents' realized valuations are inconsistent
with the reported profile, so $\chi_i=1$ and the same deviation loses at least
$\beta s_i+K=11$. Likewise, an agent that reports $5$ but trains on only $3$ samples is
caught, since the other agents' realized valuations reflect the data actually used, and it
loses at least $\beta\cdot 3+K=7$ compared with reporting truthfully.

\emph{Gradients.} True contributions are fixed at $s^{\max}$ in Phase~2, so the costs and the finite-sample term
of the valuations are constant and do not affect gradient reports; we therefore specify
only the parameter-dependent part of the valuations. Take $n=4$, $m=2$ and the quadratic valuations of
\Cref{eq:quadratic-loss}, $v_i(w) = -\frac12(w-w_i^*)^\top A_i(w-w_i^*)$, with
$A_1=A_2=\mathrm{diag}(4,\frac14)$, $A_3=\mathrm{diag}(\frac12,1)$,
$A_4=\mathrm{diag}(1,\frac12)$, own optima $w_1^*=(0,0)$, $w_2^*=(2,0)$, $w_3^*=(0,2)$,
$w_4^*=(2,2)$, $w^0=(-1,-1)$ and $\eta=1/M$. Here $\nu=0.5$, $M=2.375$ and $\eta M_i\leqslant 1.69<n$.
Without $p^3$, an agent's best gradient misreport steers the model towards its own
optimum $w_i^*$; it gains between $1.82$ and $3.34$ in round~$0$, and the gain does not
vanish: it is still between $0.41$ and $2.49$ in round~$20$, since at $\wopt$ every agent
still prefers its own optimum. With $p^3$, the best gain is at most $2.50$ in round~$0$,
when the update is large, but falls to at most $8.2\times10^{-3}$ in round~$1$,
$1.2\times10^{-4}$ in round~$10$ and $1.0\times10^{-6}$ in round~$20$, always within the
bound $\varepsilon_i^t$ of part~3 of \Cref{thm:truthful_T2P}. The charges sum to zero in
every round, and both truthful and equilibrium reporting converge to
$\wopt=(1.053,\,1.5)$.
\end{example}
\color{black}

\noindent\textsc{Remark (Scope and limitations).} (i)~The guarantees are per round and assume myopic agents; deviations planned across rounds, and rounds in which some agents report gradients truthfully while others best-respond, are not covered. (ii)~Realized valuations are noisy in practice, so the tests in \Cref{eq:payment_truthful_2P} need a tolerance; if every misreport larger than $\varepsilon$ moves some other agent's valuation by more than the tolerance, such misreports remain unprofitable, while smaller ones gain at most $\beta\varepsilon$. (iii)~A truthful agent can be penalized when another agent misreports; this does not affect EPIC, which concerns unilateral deviations. (iv)~The Phase~2 charges sum to zero across agents but need not vanish for an individual agent: in each round, agent $i$'s net Phase~2 transfer lies in $\big[-\frac{\eta}{2n}\max_j\|\hat d_j^t\|_2^2,\ \frac{\eta}{2n}\max_j\|\hat d_j^t\|_2^2\big]$, so its total over Phase~2 is at most $B := \frac{\eta T_0}{2n}\max_{t,j}\|\hat d_j^t\|_2^2$ in absolute value. Hence the limit outcome of \mechfour{} is individually rational whenever the gain of \Cref{prop:IR} exceeds this amount, that is, whenever $\pi_i+\frac{n\beta}{n-1}\big(s_i^{\max}-\bar s^{\max}\big)\geqslant B$ for every $i\in N$.

\noindent\textsc{Remark (Robustness to adversarial gradients).} \Cref{thm:truthful_T2P} handles \emph{strategic} gradient reports, which aim to improve the reporting agent's own utility. \emph{Byzantine} agents, which may send arbitrary gradients, are a different threat model; for them, our framework integrates with adversary-robust aggregation rules~\citep{yin2018byzantine,karimireddy2020byzantine,ghosh2021communication}. We describe one such method, trimmed mean~\citep{yin2018byzantine}, in \Cref{sec:adversarial} and implement \mechtwo{} with trimmed mean in \Cref{app:experiments_adversarial}.

\subsection{\mechthree{}: truthful reports in the joint dynamics}
\label{app:T-UPBReD}

We now show that an agent can profit from misreporting its contribution in \mech{}, even when gradients are reported truthfully and the center aggregates them with uniform weights. Since the allocation is then unaffected by the reports, the manipulation works through the other agents' strategy updates, which use the broadcast contribution profile.
 
\begin{example}[Profitable misreporting in \mech{}]
\label{exam:misreport}
Consider two agents who learn a model with two parameters and target $w^\circ=(1,2)$.
Both agents have the valuation
\begin{equation*}
    v_i(w,s) = -\|w-w^\circ\|_2^2 + 10\,(s_1+s_2) - \tfrac12(s_1+s_2)^2,\qquad i=1,2,
\end{equation*}
the costs $c_i(s_i)=s_i+\frac12 s_i^2$, and the strategy sets $S_i=[0,5]$. The valuation
increases with the total contribution with diminishing returns on the whole domain,
because the total never exceeds $10$; the agents' data are therefore substitutes. The given valuation form satisfies \Cref{assump:conv-NE}.
 
 
\emph{Truthful reporting.} Agent $i$'s marginal utility is
$\pd{s_i}u_i = 10-(s_1+s_2)-1-s_i$. Setting it to zero for both agents gives the
symmetric NE $s^*=(3,3)$, and the parameter converges to $w^*=w^\circ$. Each agent's
utility is $10\cdot6-\frac12\cdot 6^2-3-\frac12\cdot3^2=34.5$.
 
\emph{Misreporting.} Suppose agent $1$ reports $\hat s_1=0$ in every round, while it
updates its true contribution using its true value. Agent $2$'s gradient step is
evaluated at the broadcast profile, so its marginal utility becomes
$10-(0+s_2)-1-s_2 = 9-2s_2$, and agent $2$ converges to $s_2=4.5$. Agent $1$'s marginal
utility, evaluated at its true contribution and agent $2$'s truthful report, is
$9-s_2-2s_1$, so agent $1$ converges to $s_1=(9-4.5)/2=2.25$. The total contribution
is $6.75$, and agent $1$'s utility is
\begin{equation*}
    10\cdot 6.75-\tfrac12\cdot 6.75^2-2.25-\tfrac12\cdot 2.25^2 \;=\; 39.94 \;>\; 34.5,
\end{equation*}
while agent $2$'s utility falls to $30.09$. By under-reporting, agent $1$ makes agent $2$
believe that the total contribution is lower than it is; this raises agent $2$'s
perceived marginal benefit, so agent $2$ contributes more and agent $1$ free-rides on
the additional data while contributing less itself.
\end{example}




    



      


 
\mech{} relies on two kinds of reports in every round: the contributions $\hat s_i^t$,
which are broadcast and enter the other agents' contribution updates, and the gradients
$\hat d_i^t$, from which the model is computed. \Cref{exam:misreport} shows that an agent
can gain by misreporting its contribution, even when gradients are truthful, because
the broadcast report changes how the other agents contribute. In this subsection we
construct \mechthree{}, which adds to \mech{} per-round transfers that are analogous to
those of \mechfour{} (\Cref{sec:truthful-welfare}): a cross-check on the reported
contributions and a budget-balanced charge on the reported gradients. Unlike \mechfour{},
\mechthree{} pays no subsidy for contributions, so it does not change the equilibrium
contributions of \mech{}; its purpose is to make the reports on which \mech{} relies
reliable. By \Cref{prop:no-evidence}, this requires some observation that depends on the
contributions actually used, and we use the per-round analogue of \Cref{assump:realized}.
 
\paragraph{Timing of a round.}
\mechthree{} keeps the update rules of \mech{} (\Cref{eq:update_u_w}) and inserts
reports and transfers into every round. Round $t\in\{0,\ldots,T-1\}$ proceeds as follows.
\begin{enumerate}[leftmargin=15pt, itemsep=0pt, parsep=0pt, topsep=0pt]
    \item The center broadcasts the current model $w^t$ and the contribution reports
    $\hat s^t$ collected in the previous round (for $t=0$, the initial reports
    $\hat s^0$).
    \item Each agent $i$ updates its actual contribution by a projected gradient step,
    $s_i^{t+1} = \Pi_{S_i}\big(s_i^t+\gamma\,\pd{s_i}u_i(w^t,s_i^t,\hat s_{-i}^t)\big)$,
    using its own actual contribution and the others' reports. It then sends a gradient
    report $\hat d_i^{t+1}$, whose truthful value is $\nabla_w v_i(w^t,s^t)$, together with
    a report $\hat s_i^{t+1}$ of its new contribution.
    \item The center updates the model, $w^{t+1} = w^t+\frac{\eta}{n}\sum_{j\in N}\hat
    d_j^{t+1}$, charges the gradient transfer $p^{3,t}$ of \Cref{eq:tupbred-grad} below,
    and broadcasts $w^{t+1}$. It does \emph{not} yet broadcast $\hat s^{t+1}$.
    \item Each agent $j$ observes its realized valuation $v_j(w^{t+1},s^{t+1})$ and reports
    $\hat v_j^{t+1}$. The center then charges the transfer $p^{2,t}$ of
    \Cref{eq:tupbred-cc} below.
\end{enumerate}
Under truthful reporting, steps 2 and 3 are exactly the updates of \Cref{algo:main_algo},
so \mechthree{} induces the same dynamics as \mech{}. The contribution reports
$\hat s^{t+1}$ are revealed only in step~1 of the next round, after all valuation
reports of round $t$ have been collected, so no agent can tailor its valuation report to
another agent's contribution report.
 
\paragraph{The payments.}
\mechthree{} uses two transfers in every round. The first is computed on the
contribution and valuation reports of the round,
\begin{equation}
\label{eq:tupbred-cc}
    p_i^{2,t}\big(\hat s^{t+1},\hat v^{t+1}\big)
    = -\big(\hat v_i^{t+1} - v_i(w^{t+1},\hat s^{t+1})\big)^2
      - K\,\chi_i^t,
    \qquad
    \chi_i^t =
    \begin{cases}
        1 & \text{if } \hat v_j^{t+1} \neq v_j(w^{t+1},\hat s^{t+1}) \text{ for some } j\neq i,\\
        0 & \text{otherwise,}
    \end{cases}
\end{equation}
where $K>0$ is a constant. As in \Cref{eq:payment_truthful_2P}, the first term is the
own-report test of \citet{NATH2013321}, which makes truthful valuation reports strictly
optimal, and the second term is the cross-check, which penalizes agent $i$ whenever the
other agents' valuation reports are inconsistent with the reported contributions. Since
\mechthree{} pays no subsidy, the penalty consists of the constant $K$ alone, instead of
$\beta\hat s_i+K$ in \mechfour{}. The second transfer is the gradient charge of
\Cref{eq:grad-charge}, applied in every round,
\begin{equation}
\label{eq:tupbred-grad}
    p_i^{3,t}\big(\hat d^{t+1}\big)
    = -\frac{\eta}{2n}\,\|\hat d_i^{t+1}\|_2^2
      + \frac{\eta}{2n(n-1)}\sum_{j\neq i}\|\hat d_j^{t+1}\|_2^2 .
\end{equation}
As explained after \Cref{eq:surrogate}, the model update is the exact maximizer of a
surrogate welfare built from the reported gradients, the first term of
\Cref{eq:tupbred-grad} is the Clarke payment of VCG for this surrogate, and the second
term depends only on the other agents' reports and makes the charges sum to zero.
 
\paragraph{Utilities.}
As in \mech{}, agents are myopic. In round $t$, agent $i$'s utility is
\begin{equation}
\label{eq:tupbred-utility}
    u_i^t = v_i(w^{t+1},s^{t+1}) - c_i(s_i^{t+1})
          + p_i^{2,t}\big(\hat s^{t+1},\hat v^{t+1}\big) + p_i^{3,t}\big(\hat d^{t+1}\big),
\end{equation}
where the valuation and the cost are evaluated at the actual contributions and the
transfers at the reports. On the truthful path both penalties in \Cref{eq:tupbred-cc}
vanish, so the only transfer is the gradient charge.
 
\begin{algorithm}[t]
\DontPrintSemicolon
\caption{\mechfullthree{} (\mechthree{})}
\label{algo:truthful}
\KwIn{Step sizes $\gamma,\eta$; constant $K>0$; initialization $w^0$, $s_i^0\in S_i$ for
      $i\in N$; number of rounds $T$}
\KwOut{$w^T$}
Each agent $i$ sends an initial report $\hat s_i^0$\;
\For{$t=0$ \KwTo $T-1$}{
  \textbf{Center} broadcasts $w^t$ and $\hat s^t$\;
  \ForEach{\textbf{agent} $i\in N$ in parallel}{
    $s_i^{t+1}=\Pi_{S_i}\big(s_i^t+\gamma\,\pd{s_i}u_i(w^t,s_i^t,\hat s_{-i}^t)\big)$\;
    Send gradient report $\hat d_i^{t+1}$ (truthful: $\nabla_w v_i(w^t,s^t)$) and
    contribution report $\hat s_i^{t+1}$\;
  }
  \textbf{Center} updates $w^{t+1}=w^t+\frac{\eta}{n}\sum_{i\in N}\hat d_i^{t+1}$,
  pays $p_i^{3,t}(\hat d^{t+1})$ of \Cref{eq:tupbred-grad}, and broadcasts $w^{t+1}$\;
  \ForEach{\textbf{agent} $i\in N$ in parallel}{
    Observe $v_i(w^{t+1},s^{t+1})$; send $\hat v_i^{t+1}$\;
  }
  \textbf{Center} pays $p_i^{2,t}(\hat s^{t+1},\hat v^{t+1})$ of \Cref{eq:tupbred-cc}\;
}
\Return $w^T$\;
\end{algorithm}
 
\paragraph{Assumptions.}
The contribution guarantee uses \Cref{assump:strict-interdep} and the following
per-round version of \Cref{assump:realized}; the black-box interpretation of
\Cref{assump:realized} applies unchanged.
 
\begin{assumption}[Realized payoffs in every round]
\label{assump:realized-round}
In every round $t$, each agent $j$ observes its realized valuation
$v_j(w^{t+1},s^{t+1})$ at the contribution profile $s^{t+1}$ actually chosen, whatever
the agents reported, and the center can evaluate $v_j(w^{t+1},\hat s)$ at any reported
profile $\hat s$.
\end{assumption}
 
The gradient guarantee uses the following analogue of \Cref{assump:agent-smooth}. Since
contributions change from round to round in \mechthree{}, it also bounds how agent $i$'s
gradient responds to the contributions.
 
\begin{assumption}
\label{assump:agent-smooth-joint}
For every agent $i\in N$ there are constants $M_i,Q_i\geqslant 0$ with $\eta M_i<n$ such
that, for all $w,w'\in\mathbb{R}^m$ and $s,s'\in\prod_{j\in N}S_j$,
\begin{equation*}
    \big\|\nabla_w v_i(w',s)-\nabla_w v_i(w,s)\big\|_2\leqslant M_i\|w'-w\|_2,
    \qquad
    \big\|\nabla_w v_i(w,s')-\nabla_w v_i(w,s)\big\|_2\leqslant Q_i\|s'-s\|_2 .
\end{equation*}
\end{assumption}
 
\noindent For separable valuations (\Cref{def:separable}), and in particular for the
valuations of \Cref{sec:examples_utility}, the gradient $\nabla_w v_i$ does not depend on
the contributions, so $Q_i = 0$ and only the first condition remains.
 
We write $g^t := g(w^t,s^t,\mu^t)$ and $\tilde g^t := \tilde g(w^t,s^t)$ for the
quantities of \Cref{eq:update_u_w} along the truthful path, and $E$, $W$ for the
constants of \Cref{eq:expressions-1}.
 
\begin{theorem}[\mechthree{}: truthfulness and learning]
\label{thm:EPIC}
Suppose \Cref{assump:conv-NE,assump:strict-interdep,assump:realized-round,assump:agent-smooth-joint}
hold, and the step sizes satisfy the conditions of \Cref{thm:convergence-genl}. Then
\mechthree{} (\Cref{algo:truthful}) satisfies the following.
\begin{enumerate}[leftmargin=15pt, itemsep=0pt, parsep=0pt, topsep=0pt]
    \item \textnormal{(Truthful contribution and valuation reports.)} In every round $t$,
    if the other agents report their contributions and valuations truthfully, then for
    every history and every profile of gradient reports, truthful reporting of
    $(\hat s_i^{t+1},\hat v_i^{t+1})$ maximizes agent $i$'s round utility, and every report
    $\hat s_i^{t+1}\neq s_i^{t+1}$ loses at least $K$.
    \item \textnormal{(Approximately truthful gradients.)} In every round $t$, if the
    other agents report their gradients truthfully, no gradient report gains more than
    \begin{equation*}
        \varepsilon_i^t := \frac{\eta\,\big(\eta M_i\|\tilde g^t\|_2 + Q_i\|s^{t+1}-s^t\|_2\big)^2}{2(n-\eta M_i)}
    \end{equation*}
    over truthful reporting. Along the truthful path,
    $\varepsilon_i^t\leqslant\frac{\eta\,\max\{\eta M_i,\gamma Q_i\}^2E^2}{2(n-\eta M_i)}\,W^{2t}$,
    so the sum of these one-shot bounds over all rounds is at most
    $\frac{\eta\,\max\{\eta M_i,\gamma Q_i\}^2E^2}{2(n-\eta M_i)(1-W^2)}$.
    \item \textnormal{(Learning.)} If all reports are truthful, the iterates of
    \mechthree{} coincide with those of \mech{}, and all conclusions of
    \Cref{thm:convergence-genl} hold. If instead the agents report contributions
    truthfully and gradients according to the Nash equilibrium of every round, this
    equilibrium exists and is unique; if, moreover, the valuations are separable, then
    $s^t$ converges to the same PSNE as under truthful reporting, and
    $\|w^t-w^*\|_2\leqslant(1+\eta\tilde\lambda)^{-t}\|w^0-w^*\|_2$, where $w^*$ is the
    unique maximizer of $\frac1n\sum_{i\in N}v_i(\cdot,s)$, which does not depend on $s$.
    \item \textnormal{(Budget.)} On the truthful path, the transfers sum to zero in every
    round. For every report profile, $\sum_i p_i^{3,t}=0$ and $\sum_i p_i^{2,t}\leqslant 0$
    in every round, so the center never runs a deficit.
\end{enumerate}
\end{theorem}

\begin{proof}[Proof sketch]
Part~1 repeats Part~1 of \Cref{thm:truthful_T2P} in every round, with the penalty $K$ alone. In Part~2, both the parameter and the contributions change between the two gradients that determine the gain, which adds the term with $Q_i$, and the contraction in the proof of \Cref{thm:convergence-genl} gives the decay $W^{2t}$. In Part~3, under separability the contribution updates depend neither on the model nor on the other agents' contributions, so they are those of \mech{}, and the round equilibrium of gradient reports is again a proximal point step. Part~4 follows as in \Cref{thm:truthful_T2P}. The complete proof is in \Cref{app:EPIC-proof}.
\end{proof}

\noindent\textsc{Remark (Scope of the guarantees).}
(i) The guarantees are per round and assume myopic agents. A contribution report also
enters the other agents' contribution updates in the next round, which a forward-looking
agent could exploit, as in \Cref{exam:misreport}. Under separable valuations this channel
does not exist, since each agent's contribution update does not depend on the others'
contributions, so Part~1 rules out every manipulation of contribution reports. Otherwise
the penalty $K$ must exceed the per-round gain from such a manipulation; in
\Cref{exam:misreport}, a sustained misreport gains about $5.44$ per round once the
dynamics have settled, so any $K>5.44$ makes it unprofitable under
\Cref{assump:realized-round}.
(ii) As in \mechfour{}, realized valuations are noisy in practice, so the tests in
\Cref{eq:tupbred-cc} should use a tolerance, which turns Part~1 into an approximate
guarantee.
(iii) \mechthree{} pays no subsidy, so it leaves the equilibrium contributions of \mech{}
unchanged; when these are below $s^{\max}$, as in \Cref{exam:NEnotsocial}, the
subsidy of \mechfour{} is needed.
(iv) Under separable valuations the gradient charges $p^{3,t}$ do not depend on the
contributions, so they do not affect the contribution incentives either; for
non-separable valuations they can, and the contribution dynamics of \mechthree{} then
follow \Cref{algo:truthful} as prescribed rather than a best response to the full
round utility \Cref{eq:tupbred-utility}.
\color{black}


\section{Experiments}
\label{sec:experiments}
\begin{figure}[t]
  \centering
  \begin{subfigure}{\linewidth}
  \hspace*{0pt}
    \centering
    \includegraphics[width=0.32\linewidth]{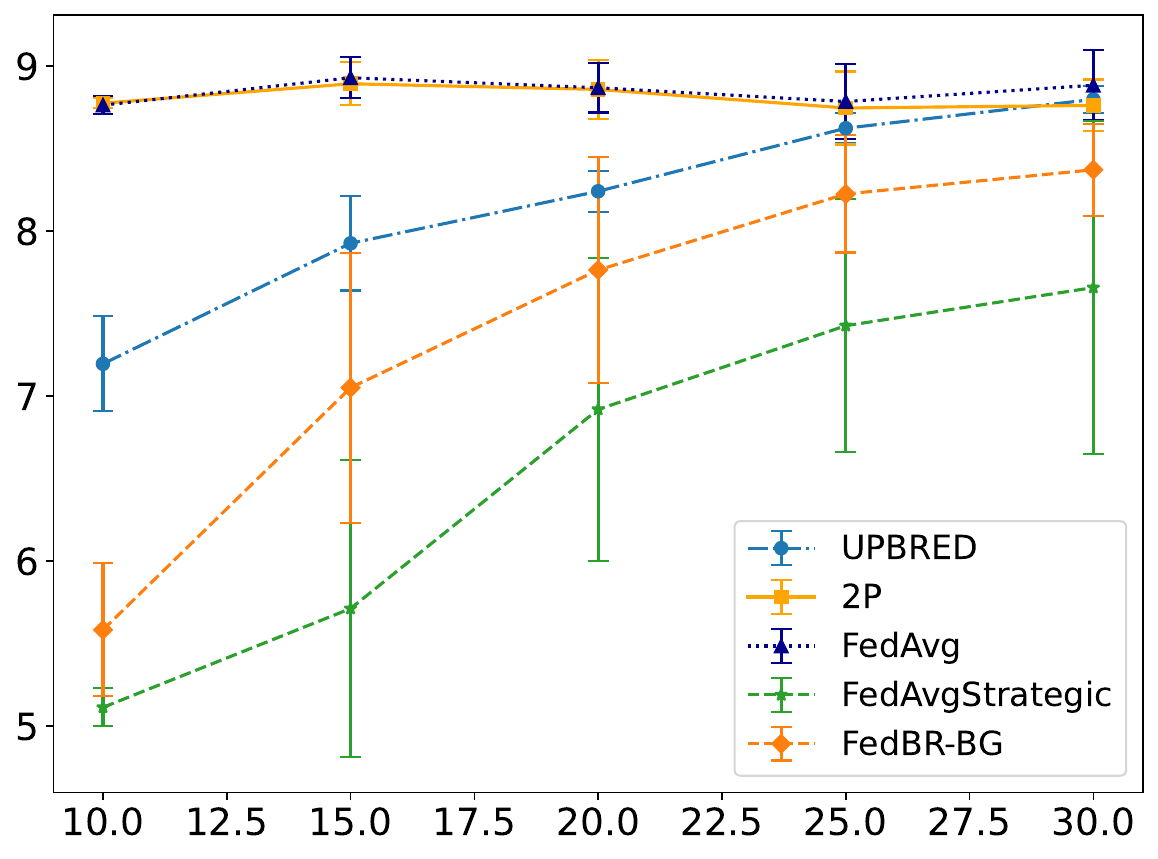}
    \includegraphics[width=0.32\linewidth]{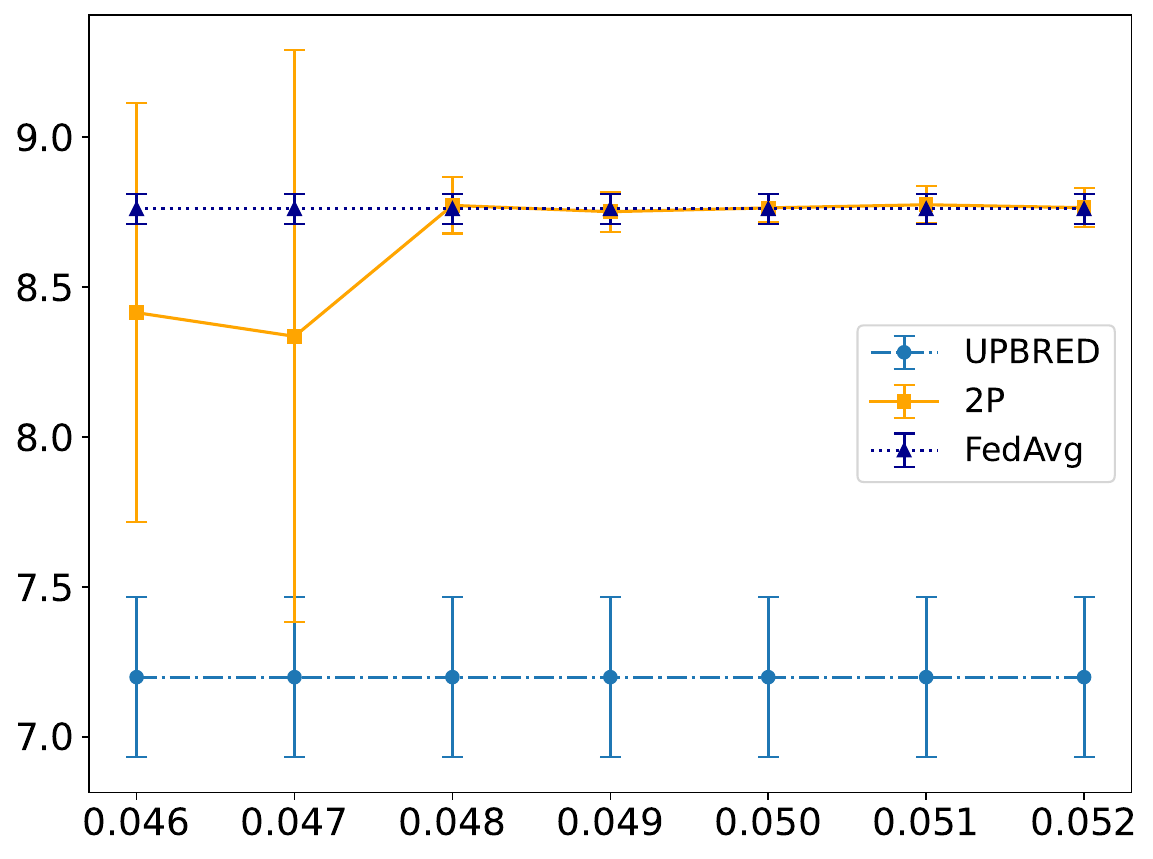}
    \includegraphics[width=0.32\linewidth]{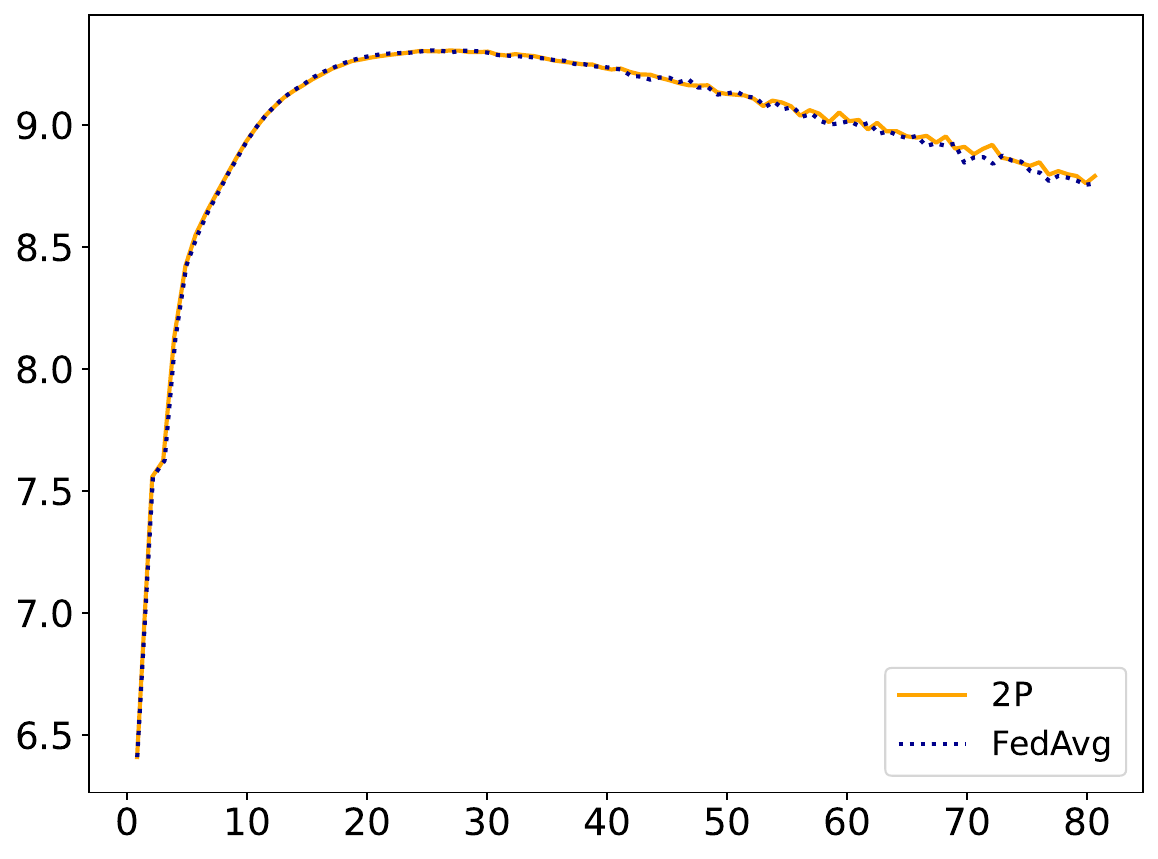}
    \caption{CIFAR-10}
  \end{subfigure}


  \begin{subfigure}{\linewidth}
  \hspace*{0pt}
    \centering
    \includegraphics[width=0.32\linewidth]{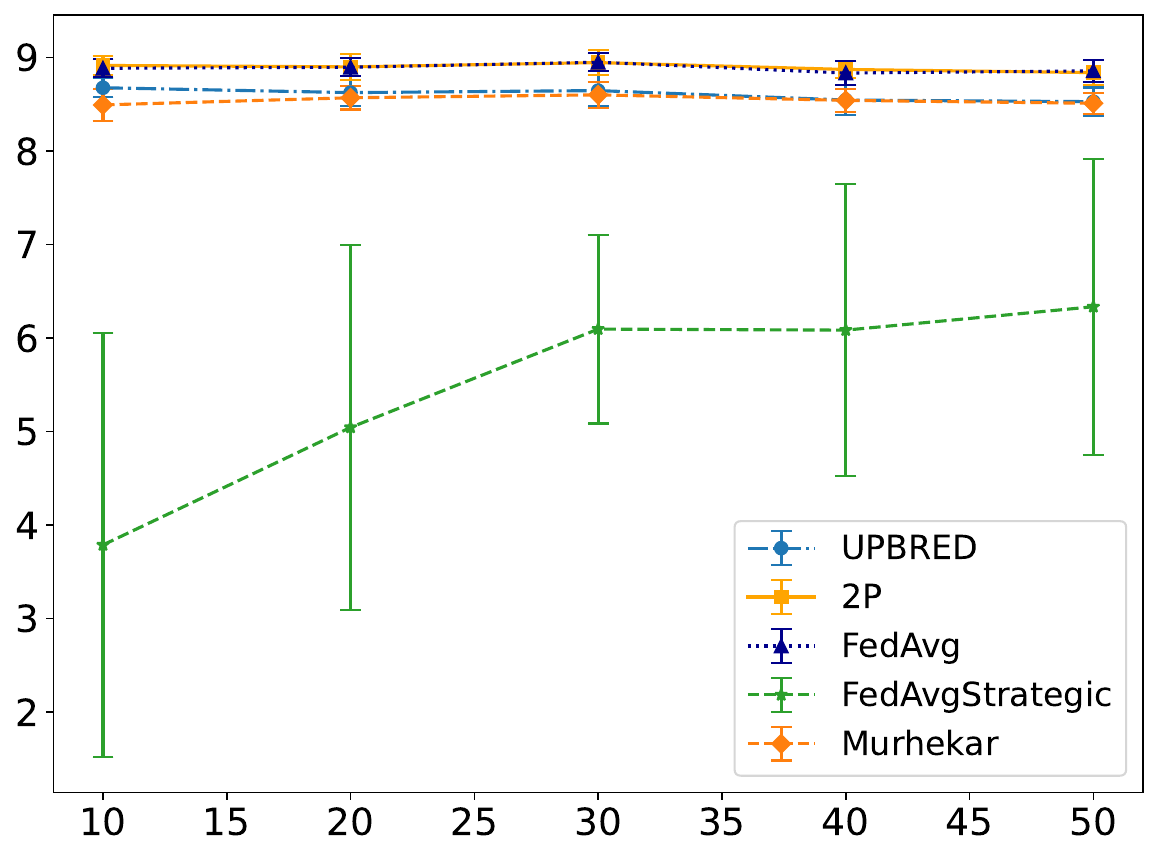}
    \includegraphics[width=0.32\linewidth]{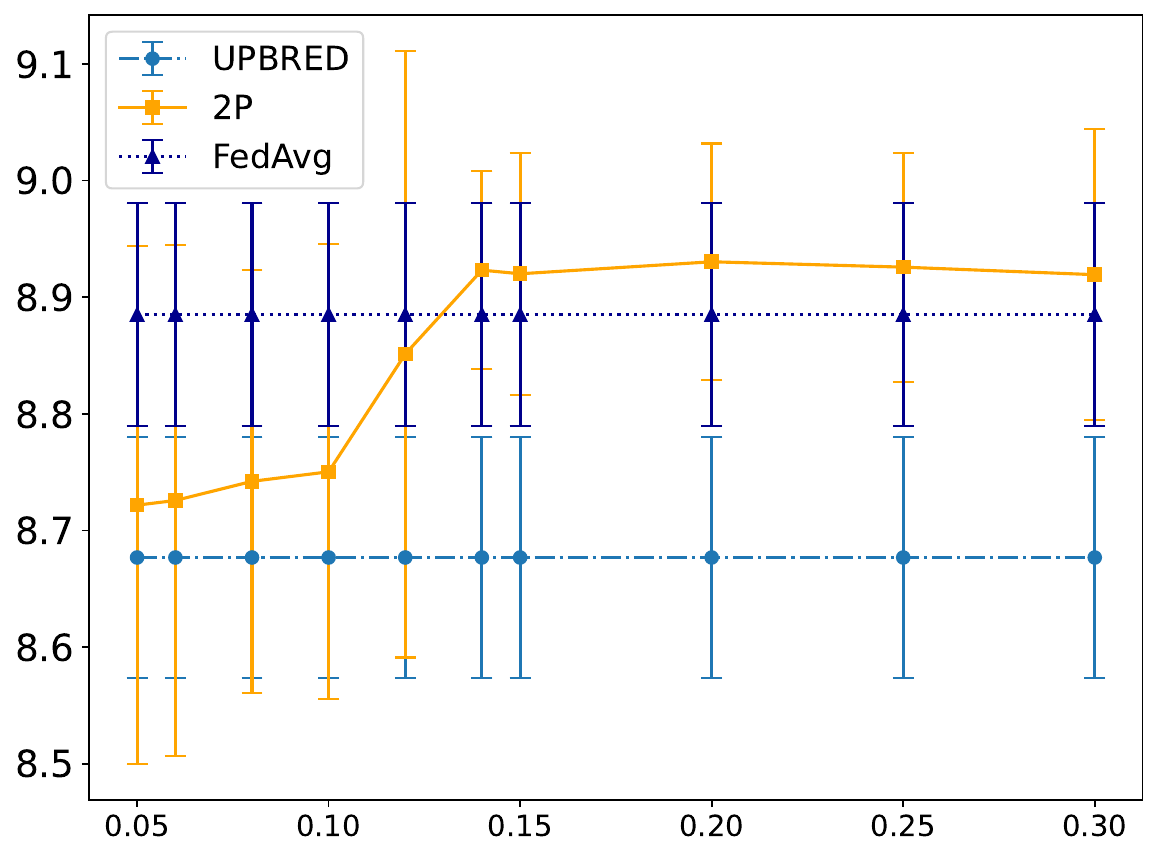}
    \includegraphics[width=0.32\linewidth]{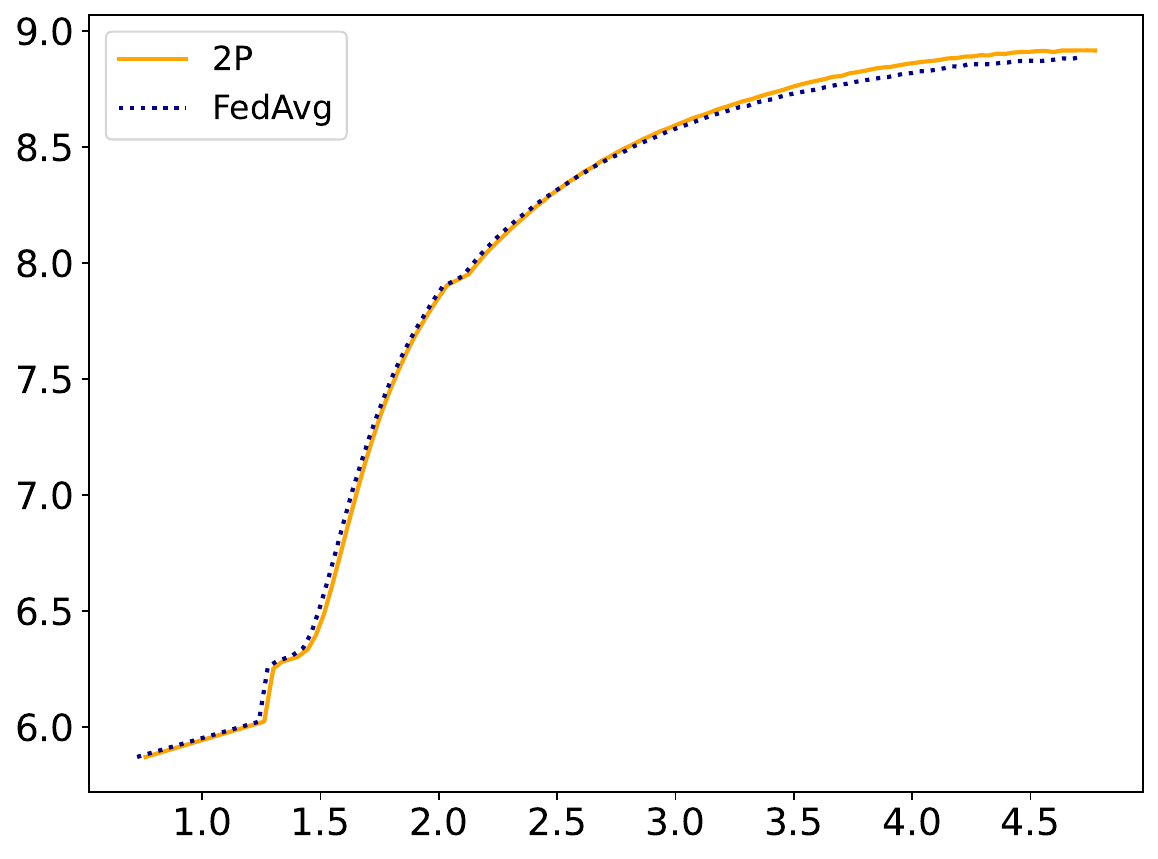}
    \caption{FEMNIST}
  \end{subfigure}


  \begin{subfigure}{\linewidth}
  \hspace*{0pt}
    \centering
    \includegraphics[width=0.32\linewidth]{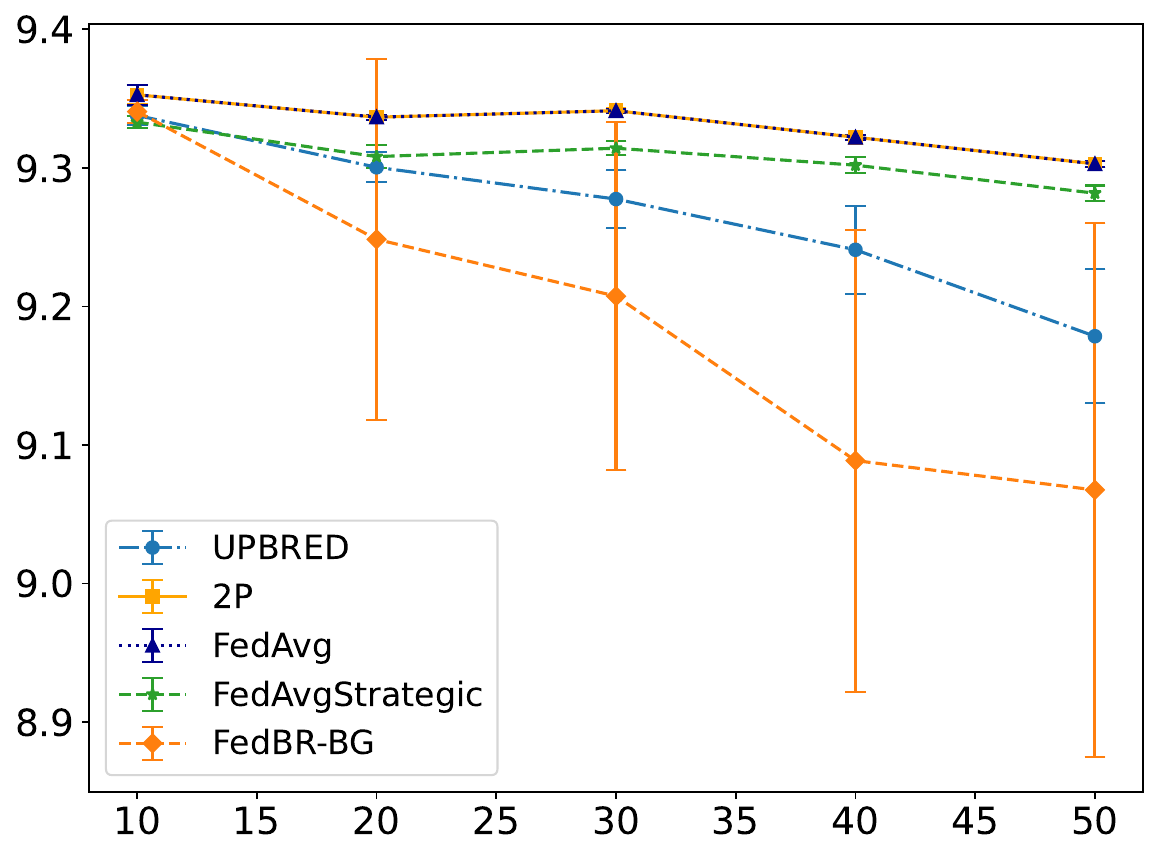}
    \includegraphics[width=0.32\linewidth]{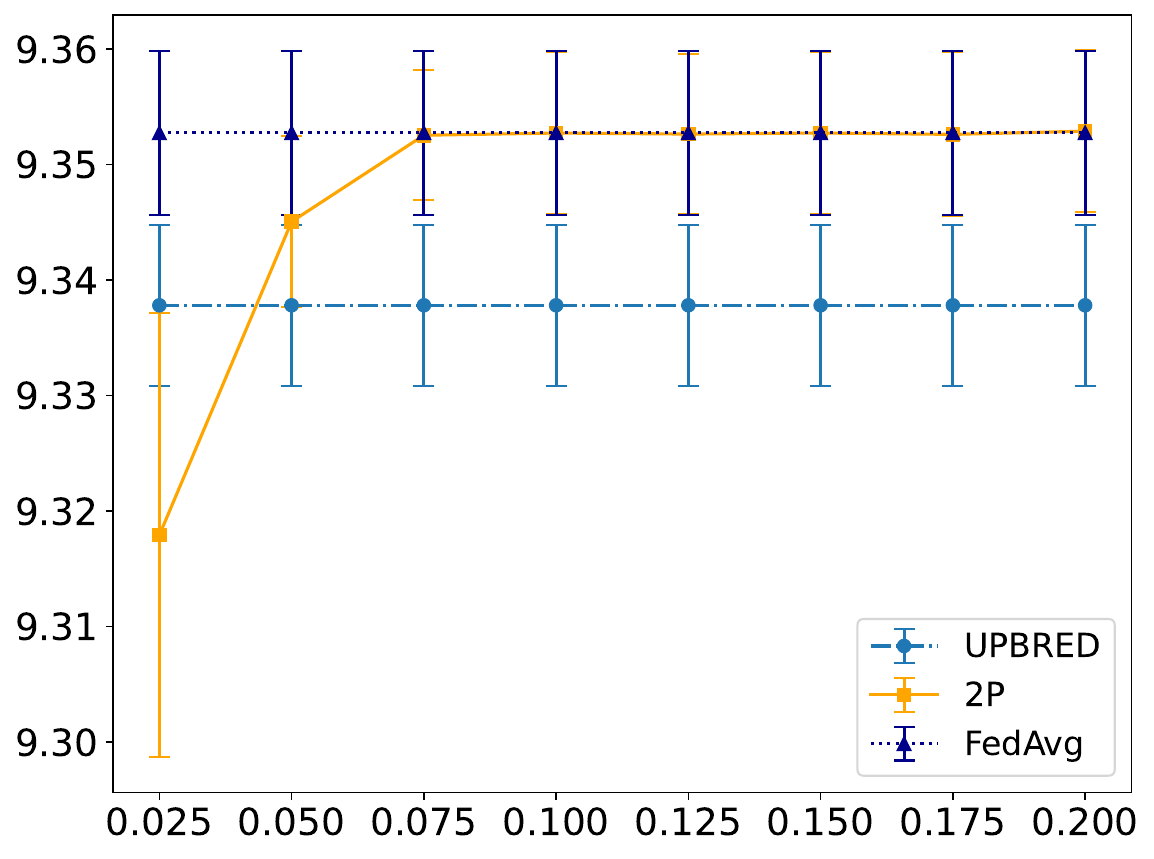}
    \includegraphics[width=0.32\linewidth]{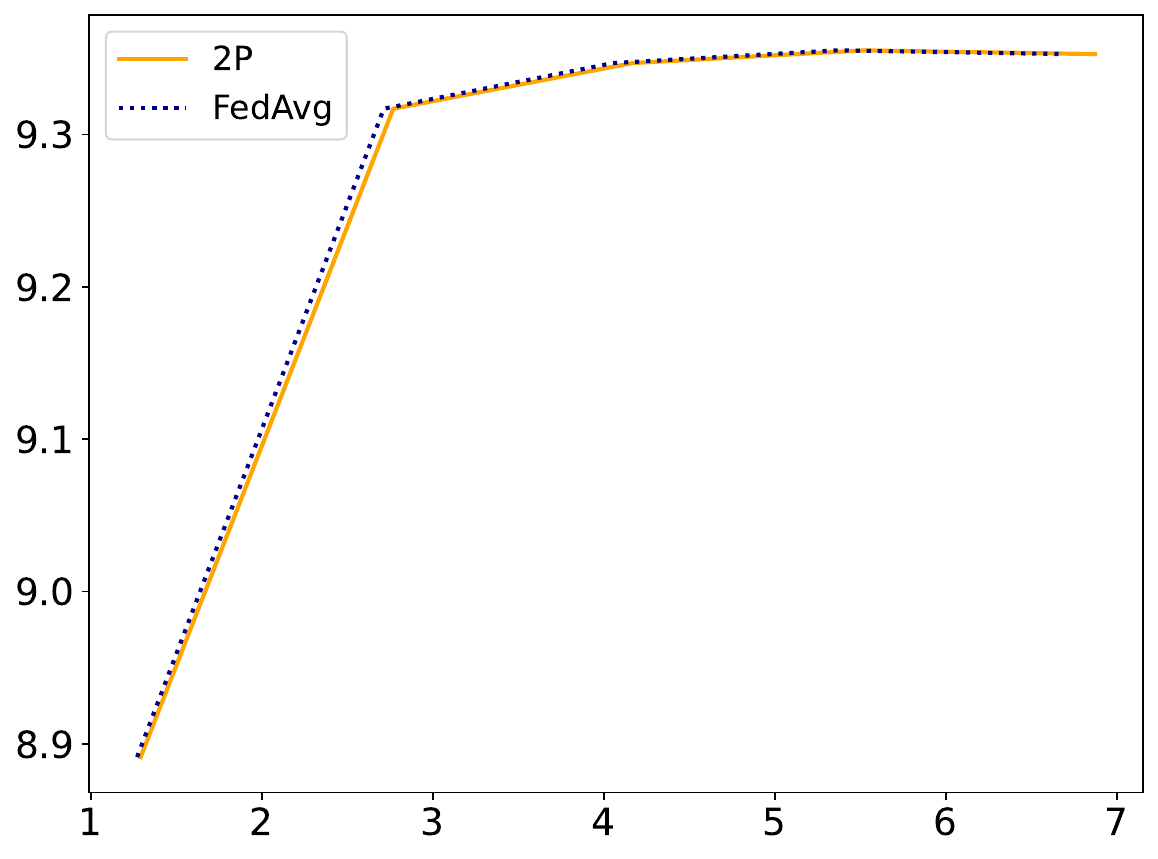}
    \caption{Twitter}
  \end{subfigure}

  \caption{The three rows in this figure correspond to the experiments with the datasets CIFAR-10, FEMNIST, and Twitter, respectively. The {\tt y-axis} in every plot is the social welfare (computed using $10-L_i(w)$, where $L_i(w)$ is the test loss for agent $i$). The {\tt x-axis} corresponds to the number of agents in the first plot, the choice of $\beta$ in the second, and the time taken (in minutes) in the third plot, respectively, in every row.}
  \label{fig:combined_plots}
  \vspace{-10pt}
\end{figure}
Our results in \Cref{sec:without-payment,sec:with-payment} guarantee convergence of \Cref{algo:main_algo,algo:2-phase} to a pure Nash equilibrium under the given assumptions. However, it is important to ask how these FL algorithms behave in practice with real datasets under more general setups. We capture results in these realistic settings in our experiments. We train models on the CIFAR-10~\citep{CIFARKrizhevsky2009LearningML}, FEMNIST~\citep{leaf2019}, and Twitter~\citep{twitterdataset} datasets and compare the social welfare of \mech{}, \mechtwo{}, and \FedAvg{} and the \murhekar{} algorithm of \citet{murhekar2023incentives}. 

We also consider another algorithm, \strategic{}, as a baseline. In this algorithm, the agents choose to contribute their data strategically as in \mech{}, and stick to this contribution in the entire training phase of \FedAvg{} (details in \Cref{sec:strategic}). 

The cost function used is linear $c_i(s_i)=c_i\cdot s_i$, where $c_i=c$ for all agents is sampled from $U[0.01,0.05]$ (sampled once and shared for all the agents for each run). We choose our valuation function to be $v_i(w^k,s^k)=r_i-{L}_i(w^k,s^k)$ where the ${L}_i(w^k,s^k)$ is the loss evaluated on agent $i$'s test dataset at the $k$-th round of training and $r_i$ is a sufficiently large constant such that $u_i$ remains positive. The social welfare is given by $\sum_{i \in N}  v_i(w^k,s^k)$. We perform strategy updates using the numerical derivative as follows: 
$s_i^{k+1}=s_i^k +\gamma \left(-\frac{L_i(w_i^k,s^k)-L_i(w^{k-1},s^{k-1})}{s_i^{k}-s_i^{k-1}}-c_i+\beta\right)$ when $s_i^{k} \neq s_i^{k-1}$ and $s_i^{k+1}=s_i^k +\gamma \left(-c_i+\beta\right)$ otherwise. $\beta=0$ for all simulations of \mech{} and \strategic{}. 

The model parameters obtained by agent $i$ after performing one training step locally on the model initialized with $w^{k-1}$ is denoted by $w_i^k$. We use the cross-entropy loss function for ${L_i}$ on agent $i$'s dataset. All the experiments reported are performed on \emph{NVIDIA RTX A5000}, \emph{NVIDIA RTX A6000}, and \emph{NVIDIA RTX 5000} GPUs with 24GB, 48GB, and 32GB of memory, respectively. For comparing social welfare against $\beta$ and number of agents, we randomize over $10$ runs. For each run we initialize a seed, which is used to sample costs, initial strategies, model initialization and dataset splits. We report the average running time across $10$ runs. We use the Flower framework to create IID splits of the datasets mentioned below. 

\smallskip
\noindent\textbf{Dataset specifics.}
For CIFAR-10 we train a ResNet-18 model with $11{,}173{,}962$ parameters for $100$ rounds. Hyperparameters: $\gamma=0.01,\ \beta=2, \ \eta=0.003$, optimizer=Adam.  

For FEMNIST we train a CNN with $6{,}603{,}710$ parameters for $100$ rounds. Hyperparameters: $\gamma=0.01,\ \beta=2, \ \eta=0.001$, optimizer=Adam. 

For Twitter we fine-tune all layers of the pretrained \texttt{distilbert-base-uncased} model \citep{sanh2020distilbertdistilledversionbert} with a sequence-classification head for the three sentiment classes of TweetEval~\citep{twitterdataset} ($66{,}955{,}779$ parameters), truncating inputs to $128$ tokens, and train for $5$ rounds. Hyperparameters: $\gamma=0.1,\ \beta=2, \ \eta=0.00002$, optimizer=Adam. Unlike the CNNs used in the other datasets, which are trained from scratch over $50$--$100$ rounds, the pre-trained DistilBERT model allows us to achieve full convergence within just $5$ communication rounds.

All results are summarized in \Cref{fig:combined_plots}, where the agents' valuations, and hence the social welfare, are computed with respect to the test loss. We provide the social welfare computed using test accuracy in \Cref{fig:combined_plots_accuracy} (\Cref{app:accuracy-welfare}). We provide more details of the model architecture for all models in \Cref{appendix:experiments_architecture}.

We observe that \mechtwo{} and \FedAvg{} have the same social welfare performance w.r.t.\ the number of agents (first column), and running times are nearly the same for all the datasets (third column). \mechtwo{} outperforms \mech{} in nearly all the runs with respect to social welfare. Additionally, we present a study on how the convergence of \mechtwo{} is affected across different values of $\beta$ and observe that low values of $\beta$ are sufficient to ensure performance that matches \FedAvg{}. \murhekar{} performs suboptimally in terms of social welfare compared to \FedAvg{} and \mechtwo{}. We further comment on results regarding the utilities and payments for agents in \Cref{app:experiment_utility}.
%



\section{Conclusions and future work}

In this paper, we proposed \mech{}, which ensures convergence to a Nash equilibrium with model parameters $w$, and its modification \mechthree{}, which enforces truthful reporting. Since $w$ may deviate from the optimal $\wopt$, we introduced a two-phase mechanism, \mechtwo{}, that guarantees full agent contribution and convergence to $\wopt$ via internal agent transfers that are BB, along with its truthful version \mechfour{}, which elicits truthful contribution reports, makes gradient reports approximately truthful, and still ensures full contribution, with transfers that are budget balanced on the truthful path. Its guarantee for contribution reports relies on agents observing their realized valuations after training; \Cref{prop:no-evidence} shows that some evidence of this kind cannot be avoided. Looking ahead, our framework opens several promising avenues for future research. One natural extension is incorporating agent budget constraints to preserve approximate welfare guarantees, as well as adapting the mechanism to clustered FL architectures. Furthermore, relaxing current trust assumptions, for instance, by designing verification protocols for gradient computation or incorporating center-side learning techniques for unknown agent valuation functions $v_i$, presents a compelling frontier for building fully decentralized and robust FL mechanisms.


\bibliographystyle{plainnat}
\bibliography{references}
\appendix
\section{\texorpdfstring{{Proofs for \Cref{sec:without-payment}}}{Proofs for \ref{sec:without-payment}}}

\subsection{\texorpdfstring{{Proof of \Cref{thm:convergence-genl}}}{Proof of \ref{thm:convergence-genl}}}
\label{sec:convergence-genl-app}
\begin{proof}
Throughout this proof we write $\Delta s^t := s^{t+1}-s^t$ and $\Delta w^t := w^{t+1}-w^t$ for the two increments produced by one round of \Cref{algo:main_algo}, and we abbreviate $g^t := g(w^t,s^t,\mu^t)$ and $\tilde g^t := \tilde g(w^t,s^t)$. We also write $S := \prod_{i\in N}S_i$, $z^t := (w^t,s^t)$, and $\nabla_s u(z)\in\mathbb{R}^n$ for the pseudo-gradient with entries $[\nabla_s u(z)]_i := \pd{s_i}u_i(z)$. We make no assumption that the iterates avoid the boundary of $S$.
 
Under \Cref{eq:mu-def}, the strategy update of \Cref{algo:main_algo} is the projected gradient step
\begin{equation}
    \label{eq:proj-step}
    s^{t+1} = \Pi_S\big(s^t + \gamma\,\nabla_s u(z^t)\big),
\end{equation}
where $\Pi_S$ acts coordinatewise as $[\Pi_S(x)]_i = \min\{s_i^{\max},\max\{s_i^{\min},x_i\}\}$. Two consequences of \Cref{eq:proj-step} are used throughout. First, every iterate is feasible, $s^t\in S$ for all $t$, since each $s^{t+1}$ is the image of a point under $\Pi_S$ and the parameter $w$ is unconstrained. Second, the update rules \Cref{eq:update_u_w} give
\begin{equation}
    \label{eq:proj-increments}
    \Delta s^t = \gamma\,g^t, \qquad \Delta w^t = \eta\,\tilde g^t,
\end{equation}
so that $g^t = (s^{t+1}-s^t)/\gamma$ is the displacement actually produced by the step, after clipping. We stress that $g$ is continuous but is \emph{not} differentiable at the faces of $S$, where the projection has kinks, so we never differentiate $g$ itself. We differentiate only the pseudo-gradient $\nabla_s u$, which is continuously differentiable on $\mathbb{R}^m\times S$ because $u_i$ is twice continuously differentiable by \Cref{assump:conv-NE}, and whose Jacobians with respect to $s$ and $w$ are, by definition, the matrices $G$ and $H$.
 
We use three elementary facts about the coordinatewise projection. First, since $S$ is a product of intervals, $\|x-y\|_2^2 = \sum_{i\in N}(x_i-y_i)^2$ with each $y_i$ constrained only to $S_i$, so minimizing $\|x-y\|_2$ over $y\in S$ decouples across coordinates, and $\Pi_S$ coincides with the Euclidean projection onto $S$. Second, each $\Pi_{S_i}$ is $1$-Lipschitz on $\mathbb{R}$, since clipping to an interval never increases the distance between two points; summing the squared coordinatewise bounds gives the non-expansiveness
\begin{equation}
    \label{eq:proj-nonexp-fact}
    \|\Pi_S(x)-\Pi_S(y)\|_2 \leqslant \|x-y\|_2 \qquad\text{for all } x,y\in\mathbb{R}^n .
\end{equation}
Third, for an interval $C = [\ell,h]$, the point $y = \Pi_C(x)$ is characterized by
\begin{equation}
    \label{eq:proj-vi}
    y\in C \quad\text{and}\quad (x-y)(c-y)\leqslant 0 \ \text{ for all } c\in C,
\end{equation}
as one checks in three cases: if $x\in C$ then $y=x$ and the product vanishes; if $x>h$ then $y=h$, so $x-y>0$ and $c-y\leqslant0$; and if $x<\ell$ then $y=\ell$, so $x-y<0$ and $c-y\geqslant0$.
 
\textbf{Observation 1:} $\lambda \leqslant L$ and $\tilde\lambda \leqslant \tilde L$. To see this, apply condition~\ref{cond:concavity} of \Cref{assump:conv-NE} with the test vector $v = e_i$, the $i$-th standard basis vector of $\mathbb{R}^n$. This gives $e_i^\top\big(G(w,s)+\lambda \mathbb{I}\big)e_i \leqslant 0$, that is, $G(w,s)_{ii} + \lambda \leqslant 0$, and therefore $\lambda \leqslant -G(w,s)_{ii} \leqslant |G(w,s)_{ii}| \leqslant L$, the last step being condition~\ref{cond:bound-deriv}(i). The argument for $\tilde\lambda \leqslant \tilde L$ is identical, using condition~\ref{cond:bound-deriv}(ii) in place of (i). Combining $\lambda\leqslant L$ with the standing hypothesis $\lambda > \tilde P$ of the theorem gives the chain $\tilde P < \lambda \leqslant L \leqslant nL$, and hence
\begin{equation}
    \label{eq:denominator-positive}
    n^2L^2 - \tilde P^2 > 0,
\end{equation}
so that the step-size bound stated in the theorem is a positive quantity. The symmetric chain $P < \tilde\lambda \leqslant \tilde L \leqslant m\tilde L$ gives $m^2\tilde L^2 - P^2 > 0$.

\textbf{Observation 2:} each $u_i$ is $\lambda$-strongly concave in its own contribution, and the average valuation is $\tilde\lambda$-strongly concave in the parameter. The computation of Observation~1 gives $G(w,s)_{ii} = \ppd{s_i}{s_i}u_i(w,s)\leqslant-\lambda<0$ for every $(w,s)$, which is $\lambda$-strong concavity of $u_i$ in $s_i$ with $w$ and $s_{-i}$ held fixed. Since $\tilde G$ is the Hessian of $\frac1n\sum_{i\in N}v_i(\cdot,s)$, the second half of condition~\ref{cond:concavity} states that this function is $\tilde\lambda$-strongly concave in $w$. Observation 2 is not used in the contraction argument of Steps~1--8; it is what gives a vanishing gradient mapping its equilibrium content in Step~9.

\textit{Step 1: absorbing the projection.}
We wish to relate the gradient mappings at round $t+1$ to their values at round $t$. Fix $t\geqslant0$. By \Cref{eq:proj-step} applied at rounds $t+1$ and $t$, both $s^{t+2}$ and $s^{t+1}$ are images under $\Pi_S$:
\begin{equation*}
    s^{t+2} = \Pi_S\big(s^{t+1} + \gamma\nabla_s u(z^{t+1})\big), \qquad s^{t+1} = \Pi_S\big(s^{t} + \gamma\nabla_s u(z^{t})\big).
\end{equation*}
Subtracting the second identity from the first and applying the non-expansiveness \Cref{eq:proj-nonexp-fact} to the two arguments, we obtain
\begin{equation}
    \label{eq:proj-nonexpansive}
    \|\Delta s^{t+1}\|_2 \;\leqslant\; \Big\|\,\Delta s^{t} + \gamma\big(\nabla_s u(z^{t+1}) - \nabla_s u(z^{t})\big)\Big\|_2 .
\end{equation}
The right-hand side is the increment of the smooth map $\nabla_s u$ rather than of the clipped map $g$. It is this step that allows the argument to proceed without any assumption that the iterates stay in the interior of $S$.
 
\textit{Step 2: an exact expansion of the smooth maps.}
Since both arguments of $\nabla_s u$ change between rounds $t$ and $t+1$, we first decompose the joint increment into two increments in which only one argument varies at a time, by adding and subtracting the intermediate value $\nabla_s u(w^t,s^{t+1})$:
\begin{equation}
    \label{eq:telescope-g}
    \begin{split}
    \nabla_s u(z^{t+1}) - \nabla_s u(z^t)
    &= \underbrace{\Big[\nabla_s u(w^{t+1},s^{t+1}) - \nabla_s u(w^t,s^{t+1})\Big]}_{\text{only } w \text{ varies}}
    + \underbrace{\Big[\nabla_s u(w^t,s^{t+1}) - \nabla_s u(w^t,s^t)\Big]}_{\text{only } s \text{ varies}}.
    \end{split}
\end{equation}
Each bracket is the increment of a continuously differentiable map along a straight segment, and both segments lie in the domain on which \Cref{assump:conv-NE} is stated: in the first, $s$ is held at the feasible point $s^{t+1}$; in the second, $s$ ranges over the segment between the feasible points $s^t$ and $s^{t+1}$, which lies in $S$ because $S$ is convex. This is where the feasibility of the iterates guaranteed by \Cref{eq:proj-step} is used; were the iterates allowed to leave $S$, part of a segment could fall outside the domain of \Cref{assump:conv-NE}.
 
To each bracket we apply the integral form of the mean value theorem, which for a continuously differentiable map $F$ with Jacobian $DF$ reads
\begin{equation*}
    F(b) - F(a) = \left(\int_0^1 DF\big(a + \tau(b-a)\big)\,d\tau\right)(b-a),
\end{equation*}
and follows from the fundamental theorem of calculus applied to $\tau\mapsto F(a+\tau(b-a))$. We emphasize that we use the integral form rather than the more familiar pointwise form $F(b)-F(a) = DF(\xi)(b-a)$. The pointwise form is valid only for scalar-valued $F$: for a vector-valued map the mean value theorem must be applied coordinatewise, producing a different intermediate point for each coordinate, and the resulting matrix, whose rows are evaluated at different points, is not $G(w,s)$ at any single point $(w,s)$. Since \Cref{assump:conv-NE} is a statement about $G$ at a point, it would not apply to such a matrix. The integral form avoids this, as \Cref{lem:averaged-matrix} below makes precise.
 
Accordingly, define the averaged matrices
\begin{equation}
    \label{eq:averaged-matrices}
    \begin{split}
    \overline G^{\,t} &:= \int_0^1 G\big(w^t,\, s^t + \tau \Delta s^t\big)\, d\tau \in \mathbb{R}^{n\times n},\\
    \overline H^{\,t} &:= \int_0^1 H\big(w^t + \tau \Delta w^t,\, s^{t+1}\big)\, d\tau \in \mathbb{R}^{n\times m},\\
    \overline{\tilde G}^{\,t} &:= \int_0^1 \tilde G\big(w^t + \tau \Delta w^t,\, s^t\big)\, d\tau \in \mathbb{R}^{m\times m},\\
    \overline{\tilde H}^{\,t} &:= \int_0^1 \tilde H\big(w^{t+1},\, s^t + \tau \Delta s^t\big)\, d\tau \in \mathbb{R}^{m\times n}.
    \end{split}
\end{equation}
Applying the integral form to each bracket of \Cref{eq:telescope-g}, and to the analogous decomposition of $\tilde g$ obtained by adding and subtracting $\tilde g(w^{t+1},s^t)$, we obtain the exact identities
\begin{equation}
    \label{eq:exact-expansion}
    \begin{split}
        \nabla_s u(z^{t+1}) - \nabla_s u(z^t) &= \overline G^{\,t}\,\Delta s^t + \overline H^{\,t}\,\Delta w^t,\\
        \tilde g^{t+1} - \tilde g^t &= \overline{\tilde G}^{\,t}\,\Delta w^t + \overline{\tilde H}^{\,t}\,\Delta s^t.
    \end{split}
\end{equation}
The segments for $\tilde g$ also have their $s$-coordinate in $S$, by the same convexity argument. A dimension check confirms the shapes: $H(w,s)_{ik} = \ppd{w_k}{s_i}u_i(w,s)$ is the derivative of the $i$-th entry of $\nabla_s u$ with respect to $w_k$, so $H$ is $n\times m$ and $\overline H^{\,t}\Delta w^t \in \mathbb{R}^n$ as required; symmetrically $\tilde H$ is $m \times n$ and $\overline{\tilde H}^{\,t}\Delta s^t \in \mathbb{R}^m$.
 
The following lemma records that the averaged matrices inherit the properties that \Cref{assump:conv-NE} grants to the pointwise ones.
 
\begin{lemma}
    \label{lem:averaged-matrix}
    Let $A(\tau)$, $\tau\in[0,1]$, be a continuous family of matrices and let $\overline A = \int_0^1 A(\tau)\,d\tau$. If $\tfrac12\big(A(\tau)+A(\tau)^\top\big)+\lambda\mathbb{I}$ is negative semi-definite for every $\tau$, then so is $\tfrac12\big(\overline A+\overline A^\top\big)+\lambda\mathbb{I}$. If $|A(\tau)_{ij}|\leqslant L$ for every $\tau$ and all $i,j$, then $|\overline A_{ij}|\leqslant L$ for all $i,j$. If $\|A(\tau)\|_{op}\leqslant P$ for every $\tau$, then $\|\overline A\|_{op}\leqslant P$.
\end{lemma}
 
\begin{proof}
    For the first claim, fix any $v$ and observe that $v^\top \overline A v = \int_0^1 v^\top A(\tau) v \, d\tau \leqslant \int_0^1 (-\lambda\|v\|_2^2)\,d\tau = -\lambda\|v\|_2^2$, where the inequality holds pointwise in $\tau$ by hypothesis and integration preserves the inequality. The second and third claims follow from the triangle inequality for integrals: $|\overline A_{ij}| = \big|\int_0^1 A(\tau)_{ij}d\tau\big| \leqslant \int_0^1 |A(\tau)_{ij}|d\tau \leqslant L$, and for any unit vector $v$, $\|\overline Av\|_2 \leqslant \int_0^1\|A(\tau)v\|_2 d\tau \leqslant P$.
\end{proof}
 
\noindent The families in \Cref{eq:averaged-matrices} are continuous in $\tau$ because $u_i$ and $v_i$ are twice continuously differentiable. We read condition~\ref{cond:concavity} of \Cref{assump:conv-NE} as a statement about the quadratic form $v^\top(G+\lambda\mathbb{I})v$, equivalently about the symmetric part $\tfrac12(G+G^\top)+\lambda\mathbb{I}$. This is the relevant reading because $G(w,s)_{ij} = \ppd{s_j}{s_i}u_i(w,s)$ assembles second derivatives of $n$ \emph{different} functions and is not symmetric in general, and the proof uses the condition only through quadratic forms, which see $G$ only through its symmetric part. The matrix $\tilde G$ is the Hessian of $\tfrac1n\sum_{i\in N}v_i$ in $w$ and is symmetric, so no such care is needed for it.
 
\textit{Step 3: substituting the algorithm's update rules.}
Substituting the first line of \Cref{eq:exact-expansion} into \Cref{eq:proj-nonexpansive}, and then using \Cref{eq:proj-increments} to write $\Delta s^t = \gamma g^t$ and $\Delta w^t = \eta\tilde g^t$, the vector inside the norm on the right-hand side of \Cref{eq:proj-nonexpansive} becomes
\begin{equation*}
    \gamma g^t + \gamma\big(\gamma\overline G^{\,t}g^t + \eta\overline H^{\,t}\tilde g^t\big)
    = \gamma\Big[\big(\mathbb{I}_{n\times n}+\gamma\overline G^{\,t}\big)g^t + \eta\,\overline H^{\,t}\tilde g^t\Big],
\end{equation*}
where we have factored out the common $\gamma$ and grouped the terms in $g^t$. On the left-hand side, \Cref{eq:proj-increments} gives $\|\Delta s^{t+1}\|_2 = \gamma\|g^{t+1}\|_2$. Dividing both sides of \Cref{eq:proj-nonexpansive} by $\gamma>0$ therefore gives
\begin{equation*}
    \|g^{t+1}\|_2 \leqslant \big\|\big(\mathbb{I}_{n\times n}+\gamma\overline G^{\,t}\big)g^t + \eta\,\overline H^{\,t}\tilde g^t\big\|_2 .
\end{equation*}
For the parameter block there is no projection, so substituting $\Delta w^t = \eta\tilde g^t$ and $\Delta s^t = \gamma g^t$ into the second line of \Cref{eq:exact-expansion} and grouping the terms in $\tilde g^t$ gives the exact identity
\begin{equation*}
    \tilde g^{t+1} = \big(\mathbb{I}_{m\times m} + \eta\overline{\tilde G}^{\,t}\big)\tilde g^t + \gamma\,\overline{\tilde H}^{\,t} g^t .
\end{equation*}
Taking norms in the latter and applying the triangle inequality to both, we arrive at
\begin{equation}
    \label{eq:pre_youngs}
    \begin{split}
        \|g^{t+1}\|_2 &\leqslant \big\|\big(\mathbb{I}_{n\times n}+\gamma\overline G^{\,t}\big)g^t\big\|_2 + \eta\big\|\overline H^{\,t}\tilde g^t\big\|_2,\\
        \|\tilde g^{t+1}\|_2 &\leqslant \big\|\big(\mathbb{I}_{m\times m}+\eta\overline{\tilde G}^{\,t}\big)\tilde g^t\big\|_2 + \gamma\big\|\overline{\tilde H}^{\,t} g^t\big\|_2.
    \end{split}
\end{equation}
The remainder of the argument consists in bounding these four terms.
 
\textit{Step 4: bounding the diagonal terms.}
Consider the square of the first term on the right-hand side of the first line of \Cref{eq:pre_youngs}. Expanding the squared norm of a sum in the usual way,
\begin{equation}
    \label{eq:norm-square}
    \big\|\big(\mathbb{I}+\gamma\overline G^{\,t}\big)g^t\big\|_2^2
    = \|g^t\|_2^2 + \gamma^2\big\|\overline G^{\,t}g^t\big\|_2^2 + 2\gamma\, (g^t)^\top \overline G^{\,t} g^t.
\end{equation}
We bound the second and third terms in turn. For the second term, we use the fact that the operator norm of a matrix is dominated by its Frobenius norm, so that $\|\overline G^{\,t}g^t\|_2^2 \leqslant \|\overline G^{\,t}\|_F^2\|g^t\|_2^2$, where $\|A\|_F := \sqrt{\sum_i\sum_j|A_{ij}|^2}$. By \Cref{lem:averaged-matrix} applied to condition~\ref{cond:bound-deriv}(i) of \Cref{assump:conv-NE}, every entry of $\overline G^{\,t}$ is bounded in absolute value by $L$, and since $\overline G^{\,t}$ has $n^2$ entries we conclude that $\|\overline G^{\,t}\|_F^2 \leqslant n^2L^2$. For the third term, we use \Cref{lem:averaged-matrix} applied to condition~\ref{cond:concavity} of \Cref{assump:conv-NE}: the averaged matrix satisfies $\tfrac12(\overline G^{\,t}+\overline G^{\,t\top})+\lambda\mathbb{I}\preceq 0$, and testing this quadratic form against the particular vector $v = g^t$ gives
\begin{equation}
    \label{eq:concavity}
    (g^t)^\top \overline G^{\,t} g^t \leqslant -\lambda\|g^t\|_2^2 .
\end{equation}
Substituting both bounds into \Cref{eq:norm-square} and collecting the common factor $\|g^t\|_2^2$,
\begin{equation}
    \label{eq:diag-bound-g}
    \big\|\big(\mathbb{I}+\gamma\overline G^{\,t}\big)g^t\big\|_2^2
    \leqslant \big(1+\gamma^2n^2L^2-2\gamma\lambda\big)\|g^t\|_2^2 .
\end{equation}
Write $a := 1+\gamma^2n^2L^2-2\gamma\lambda$ for the bracketed constant. Before taking square roots we must check that $a \geqslant 0$; this requires no additional hypothesis. Viewing $a$ as a quadratic in $\gamma$ with positive leading coefficient, its discriminant is $4\lambda^2-4n^2L^2$, which is non-positive because Observation~1 gives $\lambda\leqslant L\leqslant nL$. Hence $a\geqslant 0$ for every choice of $\gamma$, and we may write
\begin{equation}
    \label{eq:diag-root-g}
    \big\|\big(\mathbb{I}+\gamma\overline G^{\,t}\big)g^t\big\|_2 \leqslant \sqrt a\,\|g^t\|_2 .
\end{equation}
The identical computation applied to the first term of the second line of \Cref{eq:pre_youngs}, using conditions~\ref{cond:concavity} and~\ref{cond:bound-deriv}(ii) of \Cref{assump:conv-NE} in place of \ref{cond:bound-deriv}(i), yields
\begin{equation}
    \label{eq:diag-root-gtilde}
    \big\|\big(\mathbb{I}+\eta\overline{\tilde G}^{\,t}\big)\tilde g^t\big\|_2 \leqslant \sqrt{\tilde a}\,\|\tilde g^t\|_2,
    \qquad \tilde a := 1+\eta^2m^2\tilde L^2-2\eta\tilde\lambda,
\end{equation}
with $\tilde a \geqslant 0$ by the same discriminant argument applied to $\tilde\lambda \leqslant \tilde L \leqslant m\tilde L$.
 
\textit{Step 5: bounding the cross terms.}
The two remaining terms of \Cref{eq:pre_youngs} are controlled directly by the operator norm bounds. By the definition of the operator norm and \Cref{lem:averaged-matrix} applied to condition~\ref{cond:bound-deriv}(iii) of \Cref{assump:conv-NE},
\begin{equation*}
    \eta\big\|\overline H^{\,t}\tilde g^t\big\|_2 \leqslant \eta\big\|\overline H^{\,t}\big\|_{op}\|\tilde g^t\|_2 \leqslant \eta P\|\tilde g^t\|_2 = \sqrt b\,\|\tilde g^t\|_2, \qquad b := P^2\eta^2,
\end{equation*}
and symmetrically, using condition~\ref{cond:bound-deriv}(iv),
\begin{equation*}
    \gamma\big\|\overline{\tilde H}^{\,t} g^t\big\|_2 \leqslant \gamma \tilde P\|g^t\|_2 = \sqrt{\tilde b}\,\|g^t\|_2, \qquad \tilde b := \tilde P^2\gamma^2 .
\end{equation*}
 
\textit{Step 6: putting them together and establishing the contraction factor.}
Substituting the four bounds of Steps~4 and~5 into \Cref{eq:pre_youngs}, we obtain the coupled system
\begin{equation}
    \label{eq:tighter-bound-norm}
    \begin{split}
        \|g^{t+1}\|_2 &\leqslant \sqrt a\,\|g^t\|_2 + \sqrt b\,\|\tilde g^t\|_2,\\
        \|\tilde g^{t+1}\|_2 &\leqslant \sqrt{\tilde a}\,\|\tilde g^t\|_2 + \sqrt{\tilde b}\,\|g^t\|_2 .
    \end{split}
\end{equation}
Adding the two inequalities and grouping the right-hand side by which of $\|g^t\|_2$ and $\|\tilde g^t\|_2$ each term multiplies, we get
\begin{equation*}
    \|g^{t+1}\|_2 + \|\tilde g^{t+1}\|_2 \leqslant \big(\sqrt a + \sqrt{\tilde b}\big)\|g^t\|_2 + \big(\sqrt{\tilde a}+\sqrt b\big)\|\tilde g^t\|_2,
\end{equation*}
and bounding each of the two coefficients by their maximum,
\begin{equation}
    \label{eq:contraction}
    \|g^{t+1}\|_2+\|\tilde g^{t+1}\|_2 \leqslant \max\big\{\sqrt a+\sqrt{\tilde b},\ \sqrt{\tilde a}+\sqrt b\big\}\Big(\|g^t\|_2+\|\tilde g^t\|_2\Big).
\end{equation}
Recalling the definitions preceding \Cref{thm:convergence-genl}, the two quantities inside the maximum are precisely $W_1 = \sqrt a + \sqrt{\tilde b}$ and $W_2 = \sqrt{\tilde a}+\sqrt b$, and $W = \max\{W_1,W_2\}$. It is worth noting that $W_1$ depends only on $\gamma$ and $W_2$ only on $\eta$, so the two step sizes may be chosen independently of one another.
 
\textit{Step 7: ensuring the contraction factor $W<1$.}
It remains to determine when $W_1 < 1$ and $W_2<1$. We treat $W_1$; the argument for $W_2$ is obtained by the substitutions $\gamma\mapsto\eta$, $n\mapsto m$, $L\mapsto\tilde L$, $\lambda\mapsto\tilde\lambda$, $\tilde P\mapsto P$.
 
Since $\gamma>0$ and $\tilde P\geqslant 0$ we have $\sqrt{\tilde b}=\gamma\tilde P$, so the requirement $W_1<1$ reads
\begin{equation}
    \label{eq:W1-condition}
    \sqrt{1+\gamma^2n^2L^2-2\gamma\lambda} < 1 - \gamma\tilde P .
\end{equation}
A necessary condition for \Cref{eq:W1-condition} is that the right-hand side be positive, which is the requirement $\gamma < 1/\tilde P$. Given that, both sides of \Cref{eq:W1-condition} are non-negative, so squaring is an equivalence rather than merely an implication, and \Cref{eq:W1-condition} holds if and only if
\begin{equation*}
    1+\gamma^2n^2L^2-2\gamma\lambda < 1 - 2\gamma\tilde P + \gamma^2\tilde P^2 .
\end{equation*}
Cancelling the constant $1$ from both sides and then dividing through by $\gamma > 0$,
\begin{equation*}
    \gamma n^2L^2 - 2\lambda < -2\tilde P + \gamma\tilde P^2 .
\end{equation*}
Collecting the terms carrying $\gamma$ on the left and the constants on the right,
\begin{equation*}
    \gamma\big(n^2L^2-\tilde P^2\big) < 2(\lambda-\tilde P),
\end{equation*}
and dividing by $n^2L^2-\tilde P^2$, which is strictly positive by \Cref{eq:denominator-positive}, we arrive at the exact characterization
\begin{equation}
    \label{eq:gamma-exact}
    W_1 < 1 \quad\Longleftrightarrow\quad \gamma < \frac{2(\lambda-\tilde P)}{n^2L^2-\tilde P^2} .
\end{equation}
Note that the hypothesis $\lambda>\tilde P$ of the theorem is exactly what makes the right-hand side of \Cref{eq:gamma-exact} positive, so that a feasible $\gamma$ exists at all.
 
The auxiliary condition $\gamma<1/\tilde P$ identified above is implied by \Cref{eq:gamma-exact} and therefore need not be imposed separately. Indeed, it suffices to check that $2\tilde P(\lambda-\tilde P) \leqslant n^2L^2-\tilde P^2$, that is, $2\tilde P\lambda - \tilde P^2 \leqslant n^2L^2$. Completing the square on the left, $2\tilde P\lambda-\tilde P^2 = \lambda^2 - (\lambda-\tilde P)^2 \leqslant \lambda^2$, and $\lambda^2 \leqslant L^2 \leqslant n^2L^2$ by Observation~1, which gives the claim. By the symmetric computation, $\eta < 2(\tilde\lambda-P)/(m^2\tilde L^2-P^2)$ likewise implies $\eta<1/P$.
 
Consequently, choosing
\begin{equation}
    \label{eq:step-sizes}
    \gamma < \frac{2(\lambda-\tilde P)}{n^2L^2-\tilde P^2} \quad \text{and} \quad \eta < \frac{2(\tilde\lambda-P)}{m^2\tilde L^2-P^2}
\end{equation}
guarantees $W_1<1$ and $W_2<1$, hence $W = \max\{W_1,W_2\}<1$, and \Cref{eq:contraction} becomes a genuine contraction:
\begin{equation}
    \label{eq:one-step-contraction}
    \|g^{t+1}\|_2+\|\tilde g^{t+1}\|_2 \leqslant W\Big(\|g^t\|_2+\|\tilde g^t\|_2\Big), \qquad W<1 .
\end{equation}
 
\textit{Step 8: iterating the contraction over $t$.}
Since \Cref{eq:one-step-contraction} holds for every $t\geqslant 0$, applying it repeatedly, starting from $t=0$ and iterating $T$ times, gives
\begin{equation}
    \label{eq:geometric-decay}
    \|g^T\|_2+\|\tilde g^T\|_2 \leqslant W^T\Big(\|g^0\|_2+\|\tilde g^0\|_2\Big) = W^T E,
\end{equation}
where $E := \|g(w^0,s^0,\mu^0)\|_2+\|\tilde g(w^0,s^0)\|_2$ as defined preceding \Cref{thm:convergence-genl}. Here $g^0 = (s^1-s^0)/\gamma$ is the clipped step at the initial point, so that $E$ depends on $\gamma$; this is harmless, since $E$ is finite for every initialization. To make the right-hand side of \Cref{eq:geometric-decay} smaller than $\epsilon$ we require $W^TE<\epsilon$. Taking logarithms of both sides and using that $\ln W<0$, so that dividing by it reverses the inequality, this is equivalent to
\begin{equation*}
    T \ln\tfrac1W > \ln\tfrac E\epsilon, \qquad \text{that is,} \qquad T > T_0(w^0,s^0) := \frac{\ln\left(\nicefrac E\epsilon\right)}{\ln\left(\nicefrac 1W\right)} .
\end{equation*}
Hence for every $T>T_0(w^0,s^0)$ we have $\|g^T\|_2+\|\tilde g^T\|_2<\epsilon$, and since both summands are non-negative, each is individually smaller than $\epsilon$:
\begin{equation}
    \label{eq:eps-stationary}
    \|g(w^T,s^T,\mu^T)\|_2 < \epsilon \quad\text{and}\quad \|\tilde g(w^T,s^T)\|_2<\epsilon .
\end{equation}

\textit{Step 9: from a small gradient mapping to a Nash equilibrium.}
\Cref{eq:eps-stationary} states that the step actually taken by each agent, after clipping, is small. We now convert this into the equilibrium statements of the theorem. Because $[g]_i$ is the clipped step rather than the derivative $\pd{s_i}u_i$, a small $[g]_i$ does not imply a small derivative: at a best response on the boundary of $S_i$ the derivative need not vanish at all. The right certificate of equilibrium is therefore the gradient mapping itself, and the following two lemmas make this precise, first qualitatively and then quantitatively.
 
\begin{lemma}[Fixed points of the projected step are Nash equilibria]
    \label{lem:fixed-point-ne}
    Under \Cref{assump:conv-NE}, for every $\gamma>0$, $w\in\mathbb{R}^m$ and $s\in S$, we have $g(w,s,\mu) = 0$ if and only if $s$ is a Nash equilibrium at the parameter $w$ in the sense of \Cref{def:nash}.
\end{lemma}
 
\begin{proof}
    Since $\Pi_S$ acts coordinatewise, $g(w,s,\mu)=0$ holds if and only if
    $[g(w,s,\mu)]_i=0$ for every $i\in N$. Fix an agent $i$ and write
    $\varphi(x) := u_i(w,x,s_{-i})$ for its utility as a function of its own
    contribution alone, so that $\pd{s_i}u_i(w,s) = \varphi'(s_i)$. By
    \Cref{eq:mu-def}, $[g(w,s,\mu)]_i = \big(\Pi_{S_i}(s_i + \gamma\varphi'(s_i))
    - s_i\big)/\gamma$ is the clipped step divided by $\gamma$, so
    \begin{equation*}
        [g(w,s,\mu)]_i = 0 \quad\Longleftrightarrow\quad
        s_i = \Pi_{S_i}\big(s_i + \gamma\varphi'(s_i)\big),
    \end{equation*}
    that is, the fixed-point condition states precisely that $s_i$ is the
    projection of $s_i + \gamma\varphi'(s_i)$ onto $S_i$. We now apply the
    characterization \Cref{eq:proj-vi}, which is an equivalence, with $C = S_i$,
    $x = s_i + \gamma\varphi'(s_i)$ and $y = s_i$. Its membership requirement
    $y\in C$ holds because $s\in S$ by hypothesis, and $x - y = \gamma\varphi'(s_i)$,
    so the fixed-point condition is equivalent to $\gamma\varphi'(s_i)(c-s_i)
    \leqslant 0$ for all $c\in S_i$. Since $\gamma>0$ the factor $\gamma$ may be
    dropped:
    \begin{equation}
        \label{eq:first-order-vi}
        \varphi'(s_i)\,(c - s_i) \leqslant 0 \quad\text{for all } c\in S_i .
    \end{equation}
    It remains to show that \Cref{eq:first-order-vi} characterizes the maximizers of $\varphi$ over $S_i$, and this is where Observation~2 enters: $\varphi$ is concave on $S_i$. For a concave function the first-order condition \Cref{eq:first-order-vi} is sufficient for optimality over a convex set, by concavity of $\varphi$:
    \begin{align*}
    \varphi(c) &\leqslant \varphi(s_i) + \varphi'(s_i)(c-s_i) \\ 
    &\leqslant \varphi(s_i) \qquad \qquad  \text{ using \Cref{eq:first-order-vi}} 
    \end{align*}
    for every $c\in S_i$. It is also necessary: if $\varphi'(s_i)(c-s_i)>0$ for some $c\in S_i$, then a small move from $s_i$ toward $c$, which stays in $S_i$ by convexity, strictly increases $\varphi$. Hence \Cref{eq:first-order-vi} holds if and only if $s_i\in\argmax_{x\in S_i}u_i(w,x,s_{-i})$, and as this holds for every $i$, the condition $g=0$ is equivalent to every agent playing a best response, which is \Cref{def:nash}.
\end{proof}
 
\begin{lemma}[Error bound for the gradient mapping]
    \label{lem:error-bound}
    Under \Cref{assump:conv-NE}, for every $\gamma>0$, $w\in\mathbb{R}^m$, $s\in S$ and $i\in N$, let $s_i^\star(w,s) := \argmax_{x\in S_i}u_i(w,x,s_{-i})$ be agent $i$'s best response and $s^\star(w,s)$ the profile of best responses. Then
    \begin{equation}
        \label{eq:error-bound}
        \big|s_i - s_i^\star(w,s)\big| \;\leqslant\; \frac{1+\gamma L}{\lambda}\,\big|[g(w,s,\mu)]_i\big|,
    \end{equation}
    and consequently $\|s - s^\star(w,s)\|_2 \leqslant \frac{1+\gamma L}{\lambda}\,\|g(w,s,\mu)\|_2$.
\end{lemma}
 
\begin{proof}
As in \Cref{lem:fixed-point-ne} write $\varphi(x) := u_i(w,x,s_{-i})$, set $F := -\varphi'$, and abbreviate $s_i^\star := s_i^\star(w,s)$. Two properties of $F$ follow from \Cref{assump:conv-NE}. It is strongly monotone with modulus $\lambda$: by the mean value theorem, which is valid here since $F$ is scalar-valued, 

\begin{align*}
    (F(x)-F(y))(x-y) = -\varphi''(\xi)(x-y)^2\geqslant\lambda(x-y)^2
\end{align*} using $\varphi''\leqslant-\lambda$ from Observation~2. And it is Lipschitz with constant $L$, since $|F'| = |\varphi''| = |G_{ii}|\leqslant L$ by condition~\ref{cond:bound-deriv}(i).
 
    Write $x := s_i$, let $x^+ := \Pi_{S_i}(x-\gamma F(x))$ be the projected step, and let $r := x - x^+$, so that $|r| = \gamma|[g]_i|$ by \Cref{eq:proj-increments}. We combine two inequalities. First, the characterization \Cref{eq:proj-vi} of $x^+$, tested at the point $c = s_i^\star$, gives $\big(x-\gamma F(x)-x^+\big)(s_i^\star-x^+)\leqslant0$, which rearranges to $\gamma F(x)(x^+-s_i^\star)\leqslant r(x^+-s_i^\star)$. Second, the optimality of $s_i^\star$, in the form \Cref{eq:first-order-vi} tested at the point $c = x^+$, gives $\gamma F(s_i^\star)(x^+-s_i^\star)\geqslant0$. Subtracting the second from the first,
    \begin{equation*}
        \gamma\big(F(x)-F(s_i^\star)\big)(x^+-s_i^\star) \;\leqslant\; r\,(x^+-s_i^\star).
    \end{equation*}
    We now substitute $x^+-s_i^\star = (x-s_i^\star)-r$ on both sides and expand. The left-hand side becomes $\gamma(F(x)-F(s_i^\star))(x-s_i^\star) - \gamma(F(x)-F(s_i^\star))\,r$, where the first term satisfies 
    $\gamma(F(x)-F(s_i^\star))(x-s_i^\star)\geq \gamma\lambda|x-s_i^\star|^2 $  by strong monotonicity and whose second term satisfies $- \gamma(F(x)-F(s_i^\star)) r\geq -\gamma L|x-s_i^\star|\,|r|$ by the Lipschitz bound. The right-hand side becomes $r(x-s_i^\star)-r^2\leqslant|r|\,|x-s_i^\star|$. Combining,
    \begin{equation*}
        \gamma\lambda|x-s_i^\star|^2 - \gamma L|x-s_i^\star|\,|r| \;\leqslant\; |r|\,|x-s_i^\star| .
    \end{equation*}
    If $x=s_i^\star$ there is nothing to prove. Otherwise, dividing by $\gamma\lambda|x-s_i^\star|>0$ and collecting terms gives $|x-s_i^\star|\leqslant\frac{1+\gamma L}{\gamma\lambda}|r|$, and substituting $|r|=\gamma|[g]_i|$ yields \Cref{eq:error-bound}. The vector form follows by squaring \Cref{eq:error-bound}, summing over $i\in N$, and taking square roots.
\end{proof}
 
With these two lemmas in hand we complete the proof. For convergence of the iterates, \Cref{eq:proj-increments} and \Cref{eq:geometric-decay} give $\|\Delta s^t\|_2+\|\Delta w^t\|_2\leqslant\max\{\gamma,\eta\}W^tE$. Since $W<1$ the geometric series $\sum_{t\geqslant0}W^t$ converges, so the increments of both sequences are absolutely summable, and $(s^t)_t$ and $(w^t)_t$ are Cauchy sequences in finite-dimensional spaces; {since $\mathbb{R}^n$ and $\mathbb{R}^m$ are complete,} they therefore converge, to limits $s^*$ and $w^*$, and $s^*\in S$ because $S$ is closed. The map $z\mapsto\big(\Pi_S(s+\gamma\nabla_s u(z))-s\big)/\gamma$ is continuous, as a composition of the continuous maps $\Pi_S$ and $\nabla_s u$, and so is $\tilde g$. Passing to the limit in \Cref{eq:geometric-decay} therefore gives $g(w^*,s^*,\mu^*)=0$ and $\tilde g(w^*,s^*)=0$.
 
For the equilibrium property, \Cref{lem:fixed-point-ne} shows that $s^*$ is a Nash equilibrium at the parameter $w^*$. For the second, Observation~2 states that $\frac1n\sum_{i\in N}v_i(\cdot,s^*)$ is $\tilde\lambda$-strongly concave, and a stationary point of a strongly concave function is its unique global maximizer; hence $w^*$ is the unique maximizer of $\frac1n\sum_{i\in N}v_i(\cdot,s^*)$.
 
Finally, \Cref{lem:error-bound} quantifies the equilibrium property at any finite $T$. For every $T>T_0(w^0,s^0)$, combining \Cref{eq:error-bound} with \Cref{eq:eps-stationary} gives
\begin{equation}
    \label{eq:approx-NE}
    \big\|s^T - s^\star(w^T,s^T)\big\|_2 \;\leqslant\; \frac{1+\gamma L}{\lambda}\,\|g^T\|_2 \;<\; \frac{1+\gamma L}{\lambda}\,\epsilon,
\end{equation}
that is, every agent's contribution lies within $\frac{1+\gamma L}{\lambda}\epsilon$ of its best response to the others at the parameter $w^T$. This completes the proof.
\end{proof}

\subsection{\texorpdfstring{{Proof of \Cref{thm:EPIC}}}{Proof of \ref{thm:EPIC}}}
\label{app:EPIC-proof}
 
\begin{proof}
Throughout, \emph{truthful reporting} in round $t$ means
$\hat s_i^{t+1} = s_i^{t+1}$, $\hat d_i^{t+1} = \nabla_w v_i(w^t,s^t)$ and
$\hat v_i^{t+1} = v_i(w^{t+1},s^{t+1})$. The proof uses three consequences of the timing
of a round, which play the role of (T1)--(T3) in the proof of \Cref{thm:truthful_T2P}.
\begin{enumerate}[leftmargin=22pt, itemsep=0pt, parsep=0pt, topsep=0pt]
    \item[(R1)] Agent $i$'s new contribution $s_i^{t+1}$ is computed in step~2 from
    $(w^t,s_i^t,\hat s_{-i}^t)$, before any report of round $t$ is sent, so no report of
    round $t$ affects it.
    \item[(R2)] The model $w^{t+1}$ and the charge $p^{3,t}$ depend only on the gradient
    reports $\hat d^{t+1}$.
    \item[(R3)] The contribution and valuation reports of round $t$ affect the round-$t$
    utility only through $p^{2,t}$. They also enter the other agents' updates in round
    $t+1$, which a myopic agent does not take into account; see the remark after the
    proof.
\end{enumerate}
 
\paragraph{Part 1 (truthful contribution and valuation reports).}
Fix a round $t$, an agent $i$, an arbitrary history and an arbitrary profile of gradient
reports, so that $w^{t+1}$ and the actual contributions $s^{t+1}$ are fixed. Suppose the
other agents report $\hat s_{-i}^{t+1}=s_{-i}^{t+1}$ and
$\hat v_j^{t+1}=v_j(w^{t+1},s^{t+1})$, which they can do by
\Cref{assump:realized-round}. Write $\hat s := (\hat s_i^{t+1},s_{-i}^{t+1})$.
 
\emph{Step 1: only $p^{2,t}$ changes.} By (R1), agent $i$'s reports do not affect
$s^{t+1}$; by (R2), they do not affect $w^{t+1}$ or $p^{3,t}$. Hence its valuation
$v_i(w^{t+1},s^{t+1})$, its cost $c_i(s_i^{t+1})$ and its gradient charge are the same
under every report $(\hat s_i^{t+1},\hat v_i^{t+1})$, and by (R3) the difference between
its round utility under such a report and under truthful reporting equals the
difference in $p_i^{2,t}$.
 
\emph{Step 2: $p^{2,t}$ vanishes under truthful reporting.} Under truthful reporting the
reported profile is $s^{t+1}$. Agent $i$'s own report $v_i(w^{t+1},s^{t+1})$ coincides
with the center's evaluation, so the quadratic term vanishes, and every other agent's
report coincides with the center's evaluation as well, so $\chi_i^t = 0$. Combining
with Step~1,
\begin{equation}
\label{eq:tupbred-diff}
    u_i^t - u_i^{t,\mathrm{truth}}
    = -\big(\hat v_i^{t+1} - v_i(w^{t+1},\hat s)\big)^2 - K\chi_i^t .
\end{equation}
 
\emph{Case 1: $\hat s_i^{t+1} = s_i^{t+1}$.} Then $\hat s = s^{t+1}$, each other agent's
report equals the center's evaluation at $\hat s$, and so $\chi_i^t = 0$. The difference
\Cref{eq:tupbred-diff} reduces to $-\big(\hat v_i^{t+1}-v_i(w^{t+1},s^{t+1})\big)^2
\leqslant 0$, with equality if and only if agent $i$ reports its true valuation.
 
\emph{Case 2: $\hat s_i^{t+1}\neq s_i^{t+1}$.} By \Cref{assump:strict-interdep}, applied
at the parameter $w^{t+1}$ with the other agents' contributions $s_{-i}^{t+1}$, there is
an agent $j\neq i$ with $v_j(w^{t+1},\hat s)\neq v_j(w^{t+1},s^{t+1})$. Agent $j$ reports
the latter, so its report fails the center's check and $\chi_i^t = 1$, whatever agent $i$
reports, since $\chi_i^t$ does not depend on $\hat v_i^{t+1}$. Substituting into
\Cref{eq:tupbred-diff} and dropping the non-positive quadratic term,
$u_i^t - u_i^{t,\mathrm{truth}}\leqslant -K<0$.
 
In both cases truthful reporting is optimal, and every misreport of the contribution
loses at least $K$. Since the history and the gradient reports were arbitrary, this
holds in every round.
 
\paragraph{Part 2 (approximately truthful gradients).}
Fix a round $t$ and an agent $i$, and suppose the other agents report
$d_j^{t+1} := \nabla_w v_j(w^t,s^t)$. Write $w := w^t$, $b := \sum_{j\neq i}d_j^{t+1}$,
$\hat v_i(\cdot) := v_i(\cdot,s^{t+1})$, and let $x\in\mathbb{R}^m$ be agent $i$'s report.
 
\emph{Step 1: the round utility as a function of the report.} By (R1) and (R2), the
report $x$ affects only $w^{t+1} = w+\frac{\eta}{n}(x+b)$ and the charge $p_i^{3,t}$;
the contributions $s^{t+1}$, and therefore the cost, are already fixed. In
\Cref{eq:tupbred-grad} the rebate depends only on the other agents' reports. When the
contribution and valuation reports are truthful, as Part~1 ensures, $p_i^{2,t}=0$
whatever the model $w^{t+1}$ is, by Step~2 of Part~1. Dropping the terms that do not
depend on $x$, agent $i$'s round utility is, up to a constant,
\begin{equation*}
    U(x) = \hat v_i\Big(w+\frac{\eta}{n}(x+b)\Big) - \frac{\eta}{2n}\|x\|_2^2 .
\end{equation*}
 
\emph{Step 2: strong concavity.} By the chain rule,
$\nabla U(x) = \frac{\eta}{n}\nabla\hat v_i\big(w+\frac{\eta}{n}(x+b)\big)-\frac{\eta}{n}x$.
The first condition of \Cref{assump:agent-smooth-joint}, applied at the fixed profile
$s^{t+1}$, states that $\nabla\hat v_i$ is $M_i$-Lipschitz. Exactly as in Part~3, Step~2,
of the proof of \Cref{thm:truthful_T2P}, the Cauchy--Schwarz inequality then gives
$\langle\nabla U(x)-\nabla U(y),x-y\rangle\leqslant-\mu\|x-y\|_2^2$ with
$\mu := \frac{\eta}{n}\big(1-\frac{\eta M_i}{n}\big)$, which is positive because
$\eta M_i<n$. Hence $U$ is $\mu$-strongly concave, and, by the argument leading to
\Cref{eq:app-PL}, $U(x)-U(y)\leqslant\|\nabla U(y)\|_2^2/(2\mu)$ for all $x$ and $y$.
 
\emph{Step 3: the gradient at the truthful report.} At the truthful report
$x = d_i^{t+1} = \nabla_w v_i(w^t,s^t)$, the other agents' truthfulness gives
$x+b = n\,\tilde g(w^t,s^t) = n\,\tilde g^t$, so the new model is $w^{t+1} = w^t+\eta\tilde
g^t$. Substituting into the formula of Step~2,
\begin{equation*}
    \nabla U(d_i^{t+1})
    = \frac{\eta}{n}\Big(\nabla_w v_i(w^{t+1},s^{t+1}) - \nabla_w v_i(w^t,s^t)\Big).
\end{equation*}
This is where the joint dynamics differ from \mechfour{}: between the two gradients both
the parameter and the contributions change. Adding and subtracting
$\nabla_w v_i(w^t,s^{t+1})$ and applying the triangle inequality, then the two conditions
of \Cref{assump:agent-smooth-joint},
\begin{align*}
    \|\nabla U(d_i^{t+1})\|_2
    &\leqslant \frac{\eta}{n}\Big(\big\|\nabla_w v_i(w^{t+1},s^{t+1})-\nabla_w v_i(w^t,s^{t+1})\big\|_2
      + \big\|\nabla_w v_i(w^t,s^{t+1})-\nabla_w v_i(w^t,s^t)\big\|_2\Big)\\
    &\leqslant \frac{\eta}{n}\Big(M_i\,\eta\|\tilde g^t\|_2 + Q_i\|s^{t+1}-s^t\|_2\Big),
\end{align*}
where the first term uses $w^{t+1}-w^t = \eta\tilde g^t$.
 
\emph{Step 4: the per-round bound.} Squaring the bound of Step~3 and dividing by
$2\mu = \frac{2\eta(n-\eta M_i)}{n^2}$, the factors $n^2$ cancel and one factor of $\eta$
cancels, so for every report $x$
\begin{equation*}
    U(x)-U(d_i^{t+1})\;\leqslant\;\frac{\eta\big(\eta M_i\|\tilde g^t\|_2+Q_i\|s^{t+1}-s^t\|_2\big)^2}{2(n-\eta M_i)}
    = \varepsilon_i^t .
\end{equation*}
 
\emph{Step 5: the bound along the truthful path.} On the truthful path the iterates are
those of \mech{} (Part~3), so \Cref{eq:proj-increments} gives
$s^{t+1}-s^t = \gamma g^t$. Bounding each coefficient by the larger of the two,
\begin{equation*}
    \eta M_i\|\tilde g^t\|_2 + \gamma Q_i\|g^t\|_2
    \leqslant \max\{\eta M_i,\gamma Q_i\}\big(\|\tilde g^t\|_2+\|g^t\|_2\big)
    \leqslant \max\{\eta M_i,\gamma Q_i\}\,W^tE,
\end{equation*}
where the second inequality is the geometric decay \Cref{eq:geometric-decay} established
in the proof of \Cref{thm:convergence-genl}. Substituting into $\varepsilon_i^t$
gives the stated bound proportional to $W^{2t}$. Since $W<1$ under the step-size
conditions, the geometric series $\sum_{t\geqslant0}W^{2t}$ equals $\frac{1}{1-W^2}$,
which gives the bound on the sum. As in \Cref{thm:truthful_T2P}, this sum bounds one-shot
deviations along the truthful path, not deviations in several rounds.
 
\paragraph{Part 3 (learning).}
\emph{Truthful reports.} Under truthful reporting, each agent's contribution update in
step~2 uses $\hat s_{-i}^t = s_{-i}^t$, so it is the update of \Cref{algo:main_algo}, and
the model update uses the gradients $\nabla_w v_i(w^t,s^t)$, so it is $w^{t+1} =
w^t+\eta\tilde g^t$. The iterates of \mechthree{} therefore coincide with those of
\mech{}, and \Cref{thm:convergence-genl} applies.
 
\emph{Equilibrium gradient reports: existence and uniqueness.} Fix a round $t$, so that
$w^t$ and $s^{t+1}$ are fixed, and let $F^t(u) := -\frac1n\sum_{j\in N}v_j(u,s^{t+1})$.
By Step~2 of Part~2, which did not use the truthfulness of the other agents, agent $i$'s
round utility is strictly concave in its report for any reports of the others, so its
best response is the unique zero of its gradient, namely
$x = \nabla_w v_i(w^{t+1},s^{t+1})$ with $w^{t+1} = w^t+\frac{\eta}{n}(x+\sum_{j\neq
i}\hat d_j^{t+1})$. Exactly as in Part~4 of the proof of \Cref{thm:truthful_T2P}, a report
profile is an equilibrium if and only if $w^{t+1}$ solves $u = w^t-\eta\nabla F^t(u)$,
which is the optimality condition of $\min_u\{F^t(u)+\frac{1}{2\eta}\|u-w^t\|_2^2\}$.
Condition~\ref{cond:concavity} of \Cref{assump:conv-NE} states that
$\frac1n\sum_jv_j(\cdot,s)$ is $\tilde\lambda$-strongly concave for every $s$, so $F^t$ is
$\tilde\lambda$-strongly convex, the proximal objective has a unique minimizer, and the
round has a unique equilibrium.
 
\emph{Equilibrium gradient reports: convergence under separable valuations.} Suppose now
that the valuations are separable, $v_j(w,s) = \varphi_j(w)+\sum_k\psi_{jk}(s_k)$. Two
simplifications follow. First, $F^t(u) = -\frac1n\sum_j\varphi_j(u)$ up to a constant,
so the proximal step minimizes the same function $F := -\frac1n\sum_j\varphi_j$ in every
round, whose unique minimizer $w^*$ is the maximizer of $\frac1n\sum_jv_j(\cdot,s)$ for
every $s$. The rate argument of Part~4 of the proof of \Cref{thm:truthful_T2P}, with the
strong-convexity constant $\tilde\lambda$ of $F$ in place of $\nu$, gives
$\|w^{t+1}-w^*\|_2\leqslant(1+\eta\tilde\lambda)^{-1}\|w^t-w^*\|_2$, and iterating gives
the stated rate. Second, since $\psi_{ik}$ depends only on $s_k$ and $\varphi_i$ does not depend on $s$,
the marginal utility $\pd{s_i}u_i(w,s) = \psi_{ii}'(s_i)-c_i'(s_i)$ depends neither on
the model nor on the other agents' contributions. With truthful contribution reports
(Part~1), the contribution dynamics are therefore exactly those of \mech{}, whatever the
gradient reports. Under separability every mixed derivative $\ppd{w_k}{s_i}u_i$ vanishes,
so $H\equiv 0$, $P=0$ and $b=0$, and the first line of \Cref{eq:tighter-bound-norm} reads
$\|g^{t+1}\|_2\leqslant\sqrt a\,\|g^t\|_2$, whatever the model updates are. {Iterating this inequality gives $\|g^t\|_2\leqslant(\sqrt a)^t\|g^0\|_2$, where
$\sqrt a\leqslant W_1<1$ under the step-size conditions. Since
$s^{t+1}-s^t=\gamma g^t$ by \Cref{eq:proj-increments}, the increments are absolutely
summable, so $(s^t)_t$ is a Cauchy sequence in $\mathbb{R}^n$. Because $\mathbb{R}^n$
is complete, it converges to some $s^\infty$, and $s^\infty\in S$ because $S$ is
closed. Under separability the gradient mapping does not depend on $w$, and it is
continuous in $s$ as a composition of $\Pi_S$ and $\nabla_s u$; passing to the limit
therefore gives $g(s^\infty)=0$. Finally, \Cref{assump:conv-NE} makes each $u_i$
concave in $s_i$ (Observation~2), so \Cref{lem:fixed-point-ne} applies and shows
that $s^\infty$ is a PSNE. It is} the same PSNE as
under truthful reporting, because the contribution dynamics are the same.
 
\paragraph{Part 4 (budget).}
In every round, summing \Cref{eq:tupbred-grad} over $i$ and exchanging the order of
summation, exactly as in Part~5 of the proof of \Cref{thm:truthful_T2P}, gives
$\sum_ip_i^{3,t}=0$ for every report profile. Both terms of \Cref{eq:tupbred-cc} are
non-positive because $K>0$ and $\chi_i^t\geqslant 0$, so $\sum_ip_i^{2,t}\leqslant0$ for
every report profile, and both vanish on the truthful path by Step~2 of Part~1. The total
transfer in every round is therefore zero on the truthful path and non-positive in
general, so the center never runs a deficit.
\end{proof}

\section{\texorpdfstring{{Proofs for \Cref{sec:with-payment}}}{Proofs for \ref{sec:with-payment}}}

\subsection{\texorpdfstring{{Proof of \Cref{lem:incr_utility}}}{Proof of \ref{lem:incr_utility}}}
\label{app:proof-incr}
\begin{proof}
    Using \Cref{eq:payment}, we get the derivative of the utility function for $s_i \in (s_i^{\min},s_i^{\max})$ (note that this holds only for the interior of $S_i$); at the interior, the updates are given by
    $[g(w,s,\mu)]_i = \pd{s_i}u_i(w,s) = \pd{s_i}v_i(w,s) - c_i'(s_i) + \beta$, where $c_i'(s_i) = \frac{d}{ds_i}c_i(s_i)$. Using \Cref{assump:derivatives} and choosing in \Cref{eq:payment} the parameter $\beta > \zeta+\tau$, we have
    \begin{align*}
        [g(w,s,\mu)]_i &= \pd{s_i}v_i(w,s)-c_i'(s_i)+\beta \geqslant -\tau - \zeta +\beta > 0.
    \end{align*}
The last inequality holds since $\beta>\zeta + \tau$, $\forall i \in N$.
\end{proof}

\subsection{\texorpdfstring{{Proof of \Cref{thm:two_phase}}}{Proof of \ref{thm:two_phase}}}
\label{app:proof-two-phase}
\begin{proof}
\textbf{Phase 1.}
Fix $i\in N$ and write $\Delta:=\beta-\zeta-\tau>0$ and
$l_i:=s_i^{\max}-s_i^0\ge 0$. Throughout Phase 1 the model is fixed at $w^0$.

\emph{Step 1: each update increases $s_i$ by at least $\gamma\Delta$, up to the cap.}
Let $s_i':=s_i^k+\gamma\,\pd{s_i}u_i(w^0,s^k)$ be the unconstrained step. By
\Cref{lem:incr_utility}, $\pd{s_i}u_i(w^0,s^k)\ge\Delta$, so
\begin{equation}
  \label{eq:phase1-unconstrained}
  s_i'\;\ge\; s_i^k+\gamma\Delta\;>\;s_i^k\;\ge\; s_i^{\min}.
\end{equation}
Hence $s_i'$ never falls below $s_i^{\min}$, and the projection can only clip from above:
\begin{equation}
  \label{eq:phase1-step}
  s_i^{k+1}=\Pi_{S_i}(s_i')=\min\{s_i^{\max},\,s_i'\}
  \;\ge\;\min\{s_i^{\max},\,s_i^k+\gamma\Delta\}.
\end{equation}
Equivalently, $[g(w^0,s^k,\mu^k)]_i=\big(s_i^{k+1}-s_i^k\big)/\gamma\ge 0$, and this
is at least $\Delta$ whenever the step is not clipped.

\emph{Step 2: a lower bound on $s_i^k$.}
We claim that for all $k\ge 0$,
\begin{equation}
  \label{eq:phase1-induction}
  s_i^k\;\ge\;\min\{s_i^{\max},\,s_i^0+\gamma\Delta k\}.
\end{equation}
For $k=0$ this holds because $s_i^0\le s_i^{\max}$. Suppose it holds for $k$.
Since $x\mapsto\min\{s_i^{\max},x+\gamma\Delta\}$ is nondecreasing,
\eqref{eq:phase1-step} gives
\begin{align*}
  s_i^{k+1}
  &\ge \min\big\{s_i^{\max},\,\min\{s_i^{\max},\,s_i^0+\gamma\Delta k\}+\gamma\Delta\big\}\\
  &=\min\big\{s_i^{\max},\,s_i^{\max}+\gamma\Delta,\,s_i^0+\gamma\Delta(k+1)\big\}
  =\min\{s_i^{\max},\,s_i^0+\gamma\Delta(k+1)\},
\end{align*}
which proves \eqref{eq:phase1-induction}.

\emph{Step 3: reaching $s^{\max}$.}
Let
\[
  \kappa:=\max_{i\in N}\Big\lceil \frac{l_i}{\gamma\Delta}\Big\rceil .
\]
For $k\ge\kappa$ we have $s_i^0+\gamma\Delta k\ge s_i^0+l_i=s_i^{\max}$, so
\eqref{eq:phase1-induction} gives $s_i^k\ge s_i^{\max}$. Since $s_i^k\in S_i$
implies $s_i^k\le s_i^{\max}$, we conclude $s_i^k=s_i^{\max}$ for all $i\in N$
and all $k\ge\kappa$. In particular, $s^\kappa=s^{\max}$.

\emph{Step 4: $s^{\max}$ is a fixed point.}
At $s=s^{\max}$, \Cref{lem:incr_utility} gives $\pd{s_i}u_i>0$, so
$s_i'=s_i^{\max}+\gamma\,\pd{s_i}u_i>s_i^{\max}$. Then $\mu_i=(s_i^{\max}-s_i')/\gamma=-\pd{s_i}u_i$,
so $[g(w^0,s^{\max},\mu)]_i=\pd{s_i}u_i+\mu_i=0$ and $\Pi_{S_i}(s_i')=s_i^{\max}$. Thus
$s^{\max}$ is a fixed point of $s^{k+1}=s^k+\gamma\,g(w^0,s^k,\mu^k)$, and the
iterates remain at $s^{\max}$ from round $\kappa$ onward.

{\emph{Step 5: $s^{\max}$ is a Nash equilibrium at every parameter.}
Step~4 shows that $s^{\max}$ is a fixed point of the projected update. A fixed point
is a Nash equilibrium only if every agent is best responding, and here we cannot
invoke \Cref{lem:fixed-point-ne}, since \Cref{thm:two_phase} does not assume the
concavity of \Cref{assump:conv-NE}. We use instead the monotonicity given by
\Cref{lem:incr_utility}. Fix $w\in\mathbb{R}^m$, $i\in N$ and $s_{-i}$. By
\Cref{lem:incr_utility}, the function $x\mapsto u_i(w,x,s_{-i})$ has a strictly
positive derivative on $S_i$. For any $x\in S_i$ with $x<s_i^{\max}$, the mean value
theorem gives a point $\xi\in(x,s_i^{\max})$ with
\[
  u_i(w,s_i^{\max},s_{-i})-u_i(w,x,s_{-i})
  = \pd{s_i}u_i(w,\xi,s_{-i})\,\big(s_i^{\max}-x\big) > 0 .
\]
Hence $s_i^{\max}$ is agent $i$'s unique best response to every $s_{-i}$ at every
parameter $w$, and $s^{\max}$ satisfies \Cref{def:nash} at every $w$, in particular
at $\wopt$.}

\textbf{Phase 2}: For $t > \kappa$, we can focus on the second phase of the algorithm. From the fixed point property of $s^{\max}$, we can now focus entirely on the function $f(w,s^{\max}) = -\frac{1}{n}\sum_{i \in N} v_i(w,s^{\max}) $. From the algorithmic description, the second phase is just a simple gradient descent, run by the center, on $f(w,s^{\max})$. This is easy to see since at every iteration the center gets $d_i^{t+1}$ from all the agents $i \in N$, aggregates them and constructs $\tilde{g}(w^t,s^{\max})$, which is the gradient of $f(w,s^{\max})$ computed at $w^t$. We denote $f(w,s^{\max})$ with $f(w)$ in this part for notational cleanliness.

Note that the gradient descent is run with initialization $w^{\kappa}$ (which is the same as $w^0$ as given by the first phase of the algorithm). We exploit the strong convexity and smoothness of $f(w,s^{\max})$. Running the second phase for $T_0$ iterations and using the convergence result of \citet{Wright_Recht_2022}, we obtain
\begin{align*}
    f(w^{T_0}) - f(\wopt) \leqslant (1-\frac{\nu}{M})^{T_0}[f(w^{0}) - f(\wopt)].
\end{align*}
Hence, for $f(w^{T_0}) - f(\wopt) <\epsilon$, we require
\begin{align*}
    T_0 > \left(\ln{\ddfrac{f(w^{0})-f(\wopt)}{\epsilon}}\right)\Big/\left(\ln \ddfrac{ 1}{\Big({1-\ddfrac{\nu}{M}\Big)}}\right),
\end{align*}
{which gives the bound on the welfare gap stated in the theorem.

It remains to show that the iterates themselves converge. Since $f$ is
$\nu$-strongly convex (\Cref{assump:conv-opt-NE}), it has a unique minimizer $\wopt$,
at which $\nabla f(\wopt)=0$. Applying the definition of strong convexity
(\Cref{def:m-strict-convex}) at $\wopt$ gives
$f(w)-f(\wopt)\geqslant\frac{\nu}{2}\|w-\wopt\|_2^2$ for every $w$, and therefore
\[
  \|w^{T_0}-\wopt\|_2 \leqslant \sqrt{2\big(f(w^{T_0})-f(\wopt)\big)/\nu},
\]
which tends to zero as $T_0$ grows. Combined with Step~5, and with the fact that
$\wopt$ maximizes $\sum_{i\in N}v_i(\cdot,s^{\max})$ by \Cref{eq:social-accuracy},
the limit pair $(\wopt,s^{\max})$ is the fixed point described after
\Cref{def:nash}: $s^{\max}$ is a PSNE at $\wopt$, and $\wopt$ maximizes the welfare
at $s^{\max}$. This proves the theorem.}
\end{proof}

\subsection{\texorpdfstring{{Proof of \Cref{cor:iterate}}}{Proof of \ref{cor:iterate}}}
\label{app:proof-iterate}
\begin{proof}
    The proof follows along the same lines as that of \Cref{thm:two_phase}. Using the choice of $\beta$, we first ensure that with $\kappa$ steps, we obtain $s_i^\kappa = s^{\max}_i$ for all $i \in N$. Now, the framework is the same as minimization of a strongly convex and smooth function $f(w)$ with initialization $w^\kappa = w^0$. Using the result of \citet{Wright_Recht_2022} with $\eta = \frac{2}{M + \nu} $, we obtain the iterate convergence, namely
    $\|w^{\tilde{T}_0} - \wopt \|_2 \leqslant \left ( \frac{1-\frac{\nu}{M}}{1+\frac{\nu}{M}}\right)^{\tilde{T}_0} \|w^0 - \wopt\|_2$.
Taking logarithms of both sides implies the result.
\end{proof}

\subsection{\texorpdfstring{{Proofs of the individual rationality results}}{Proofs of the individual rationality results}}
\label{app:proofs-IR}
\begin{proof}[{Proof of \Cref{prop:IR}}]
\emph{Part 1.} We first compute the transfer that agent $i$ receives at the
full-contribution profile. Substituting $s=s^{\max}$ into \Cref{eq:payment}, and using
that the other agents' capacities sum to $\sum_{j\neq i}s_j^{\max} = n\bar s^{\max} - s_i^{\max}$
by the definition \Cref{eq:avg-capacity}, we obtain
\begin{equation*}
    p_i^1(s^{\max})
    = \beta\Big(s_i^{\max} - \frac{n\bar s^{\max} - s_i^{\max}}{n-1}\Big)
    = \frac{\beta\big((n-1)s_i^{\max} - n\bar s^{\max} + s_i^{\max}\big)}{n-1}
    = \frac{n\beta}{n-1}\big(s_i^{\max} - \bar s^{\max}\big),
\end{equation*}
where the second equality brings both terms over the common denominator $n-1$ and the
third collects the two terms in $s_i^{\max}$. Subtracting the individual learning outcome $\bar u_i$
from the utility \Cref{eq:limit-utility} and grouping the first two terms and $-\bar u_i$
into $\pi_i$, as defined in \Cref{eq:collab-gain}, gives \Cref{eq:IR-gain}.

\emph{Part 2.} By \Cref{def:IR}, the limit outcome is individually rational if and only
if the right-hand side of \Cref{eq:IR-gain} is non-negative for every agent. We examine
this condition separately for the three groups of agents, since the sign of
$s_i^{\max}-\bar s^{\max}$ determines how the condition depends on $\beta$.
For $i\in N_0$, the transfer vanishes, so the condition reads $\pi_i\geqslant 0$ and does
not involve $\beta$.
For $i\in N_-$, the factor $s_i^{\max}-\bar s^{\max}$ is negative, so dividing the
condition by $\frac{n}{n-1}(\bar s^{\max}-s_i^{\max})>0$, which preserves the inequality,
shows that it is equivalent to
$\beta\leqslant (n-1)\pi_i\big/\big(n(\bar s^{\max}-s_i^{\max})\big)$.
For $i\in N_+$, the factor is positive, and dividing by
$\frac{n}{n-1}(s_i^{\max}-\bar s^{\max})>0$ shows that the condition is equivalent to
$\beta\geqslant -(n-1)\pi_i\big/\big(n(s_i^{\max}-\bar s^{\max})\big)$.
Taking the most restrictive of the upper bounds and of the lower bounds, and recalling
that $\beta>0$ in \Cref{eq:payment}, all conditions hold simultaneously exactly when
$\pi_i\geqslant 0$ on $N_0$ and $\beta_{\mathrm{low}}\leqslant\beta\leqslant\beta_{\mathrm{high}}$,
with the thresholds defined in \Cref{eq:beta-window}.

\emph{Part 3.} \Cref{thm:two_phase} requires $\beta>\zeta+\tau$, and by Part~2 individual
rationality requires $\beta\in[\beta_{\mathrm{low}},\beta_{\mathrm{high}}]$, together with
$\pi_i\geqslant0$ on $N_0$. A $\beta$ meeting both requirements exists if and only if the
interval $[\beta_{\mathrm{low}},\beta_{\mathrm{high}}]$ is non-empty and contains a point
strictly larger than $\zeta+\tau$, which happens exactly when
$\beta_{\mathrm{low}}\leqslant\beta_{\mathrm{high}}$ and $\zeta+\tau<\beta_{\mathrm{high}}$.
If $\pi_i\geqslant 0$ for every agent, then every lower bound in \Cref{eq:beta-window} is
non-positive, so $\beta_{\mathrm{low}}=0$, the conditions on $N_0$ hold automatically, and
$\beta_{\mathrm{high}}\geqslant 0=\beta_{\mathrm{low}}$; the only remaining condition is
$\zeta+\tau<\beta_{\mathrm{high}}$.
\end{proof}

\begin{proof}[{Proof of \Cref{cor:IR-finite}}]
We show that $U_i^T$ converges to $U_i$ for each agent, and then use the strict
inequality $U_i>\bar u_i$.

First, contributions and transfers stop changing after Phase~1. By
\Cref{thm:two_phase}, $s^T = s^{\max}$ for all $T\geqslant\kappa$, so the cost term
$c_i(s_i^T)$ and the transfer $p_i^1(s^T)$ in $U_i^T$ coincide with the corresponding
terms of $U_i$ from round $\kappa$ onwards.

Second, the parameter converges to $\wopt$. Write $f(w):=f(w,s^{\max})$. By
\Cref{thm:two_phase}, $f(w^T)-f(\wopt)$ tends to $0$ as $T$ grows. Since $f$ is
$\nu$-strongly convex by \Cref{assump:conv-opt-NE} and $\wopt$ minimizes $f$, so that
$\nabla f(\wopt)=0$, the definition of strong convexity
(\Cref{def:m-strict-convex}) applied at the point $\wopt$ gives
\begin{equation*}
    f(w^T) - f(\wopt) \;\geqslant\; \langle\nabla f(\wopt), w^T-\wopt\rangle
        + \frac{\nu}{2}\|w^T-\wopt\|_2^2
    \;=\; \frac{\nu}{2}\|w^T-\wopt\|_2^2 .
\end{equation*}
Rearranging, $\|w^T-\wopt\|_2\leqslant\sqrt{2(f(w^T)-f(\wopt))/\nu}$, and the right-hand
side tends to $0$; hence $w^T\to\wopt$.

Third, by the continuity of $v_i(\cdot,s^{\max})$ at $\wopt$, it follows that
$v_i(w^T,s^{\max})\to v_i(\wopt,s^{\max})$. Combining the three observations,
$U_i^T\to U_i$ for every $i\in N$.

Finally, set $\delta := \min_{i\in N}(U_i - \bar u_i)$, which is strictly positive by
hypothesis because $N$ is finite. By the convergence just established, there is a
$\bar T\geqslant\kappa$ such that $|U_i^T - U_i|<\delta$ for every $i\in N$ and every
$T\geqslant\bar T$. For such $T$ we obtain
$U_i^T > U_i - \delta \geqslant \bar u_i$, which is individual rationality after $T$
rounds.
\end{proof}

\begin{proof}[{Proof of \Cref{cor:IR-rebate}}]
\emph{Part 1.} Summing \Cref{eq:payment-rebate} over agents, the term $\sum_i p_i^1(s)$
vanishes because $p^1$ is budget balanced, and $\sum_i\rho_i=0$ by construction; hence
$\sum_i p_i^{\rho}(s)=0$ for every $s$. Since each $\rho_i$ is a constant, it does not
depend on $s_i$, so $\pd{s_i}p_i^{\rho} = \pd{s_i}p_i^1 = \beta$. The proof of
\Cref{lem:incr_utility} uses the payment only through this derivative, so the lemma
still holds. The proof of \Cref{thm:two_phase} in turn uses the payment only through
\Cref{lem:incr_utility} in Phase~1, and Phase~2 involves no payments, so all of its
conclusions remain valid.

\emph{Part 2.} With rebates, agent $i$'s gain at the limit outcome becomes, by the same
computation as in Part~1 of \Cref{prop:IR},
\begin{equation*}
    \pi_i + p_i^1(s^{\max}) - \rho_i .
\end{equation*}
Suppose first that some choice of rebates makes the limit outcome individually
rational. Summing these gains over all agents, using $\sum_i p_i^1(s^{\max})=0$ by budget
balance and $\sum_i\rho_i=0$ by construction, gives $\sum_i\pi_i$, and this sum is
non-negative because every summand is. Conversely, suppose $\sum_i\pi_i\geqslant0$ and
choose
\begin{equation*}
    \rho_i := \pi_i + p_i^1(s^{\max}) - \frac1n\sum_{j\in N}\pi_j .
\end{equation*}
These rebates sum to zero, again because $\sum_i p_i^1(s^{\max})=0$, so they satisfy
\Cref{eq:payment-rebate}; and substituting them shows that every agent's gain equals
$\frac1n\sum_j\pi_j\geqslant 0$, so the limit outcome is individually rational. This
choice splits the total gain from collaboration equally among the agents.

\emph{Part 3.} If the capacities are known before Phase~1, the constants
$\rho_i := p_i^1(s^{\max})$ can be fixed in advance, and they sum to zero by the budget
balance of $p^1$, so they satisfy \Cref{eq:payment-rebate}. At the limit outcome each
agent's transfer is $p_i^1(s^{\max})-\rho_i = 0$, so by Part~1 of \Cref{prop:IR} its
gain is exactly $\pi_i$, and individual rationality holds if and only if
$\pi_i\geqslant 0$ for every agent.
\end{proof}

\subsection{\texorpdfstring{{Proofs of the no-evidence results}}{Proofs of the no-evidence results}}
\label{app:proofs-no-evidence}
\begin{proof}[{Proof of \Cref{prop:no-evidence}}]
Fix a transfer rule $p$, a contribution $s_i\in S_i$ and a reporting strategy
$m_i\in M_i$. Condition (E1) guarantees that every such pair is available to agent $i$,
so agent $i$ chooses from the product set $S_i\times M_i$.
 
\emph{Step 1: decomposing the utility.} By the separability of the valuations
(\Cref{def:separable}), applied at $\bar w = W(R)$ and at the contribution profile $s$
actually used, the valuation in \Cref{eq:app-ne-utility} splits as
\begin{equation*}
    v_i\big(W(R),s\big)
    = \varphi_i\big(W(R)\big) + \psi_{ii}(s_i) + \sum_{j\neq i}\psi_{ij}(s_j),
\end{equation*}
where we have separated the term $j=i$ from the sum. Substituting into
\Cref{eq:app-ne-utility} and grouping the terms that involve agent $i$'s own
contribution, the realized utility is
\begin{equation}
\label{eq:app-ne-split}
    u_i = \underbrace{\psi_{ii}(s_i) - c_i(s_i)}_{=A_i(s_i)}
      \;+\; \underbrace{\varphi_i\big(W(R)\big) + \sum_{j\neq i}\psi_{ij}(s_j)
      + p_i(R)}_{=:\,Z_i}.
\end{equation}
The first group is deterministic once $s_i$ is chosen. The second group, $Z_i$, is a
function of the pair $(R,s_{-i})$ alone: the parameter $W(R)$ and the transfer $p_i(R)$
are computed from the record, and the remaining terms involve only the other agents'
contributions.
 
\emph{Step 2: the second group does not depend on the contribution (Part 1).} Taking
expectations in \Cref{eq:app-ne-split} under the pair $(s_i,m_i)$, linearity of
expectation gives
$\mathbb{E}[u_i] = A_i(s_i) + \mathbb{E}[Z_i]$. Since $Z_i$ is a function of $(R,s_{-i})$,
its expectation is determined by the joint distribution of $(R,s_{-i})$. This is where
condition (E2) is used: for the fixed reporting strategy $m_i$, that joint distribution
is the same for every $s_i$, so $\mathbb{E}[Z_i]$ depends on $m_i$ but not on $s_i$,
and it equals $\Pi_i(m_i)$ as defined in \Cref{eq:app-ne-Pi}. The same argument,
applied to the function $p_i(R)$ alone, shows that the expected transfer
$\mathbb{E}[p_i(R)]$ does not depend on $s_i$. This proves Part~1.
 
\emph{Step 3: separate maximization (Part 2).} By Steps~1 and~2, agent $i$'s expected
utility is $A_i(s_i)+\Pi_i(m_i)$, the sum of a function of $s_i$ alone and a function
of $m_i$ alone, maximized over the product set $S_i\times M_i$. For such a sum,
$\sup_{(s_i,m_i)}\big[A_i(s_i)+\Pi_i(m_i)\big] = \sup_{s_i}A_i(s_i)+\sup_{m_i}\Pi_i(m_i)$,
because the two variables can be chosen independently. Consequently, a pair attains the
supremum if and only if each component attains its own supremum: if $s_i$ maximizes
$A_i$ and $m_i$ maximizes $\Pi_i$, the sum equals the sum of the two suprema; and if
either component falls short of its supremum by some $\delta>0$, the pair falls short of
the joint supremum by at least $\delta$. This proves the characterization of best
responses. Finally, $A_i(s_i)=\psi_{ii}(s_i)-c_i(s_i)$ involves neither the transfer
rule nor the reports, so the set of maximizers of $A_i$, which is the set of agent
$i$'s best contributions, is the same for every transfer rule; taking $p\equiv0$
identifies it with the set of best contributions without transfers. This proves Part~2.
 
\emph{Step 4: full contribution (Part 3).} By Part~2, $s_i^{\max}$ is part of a best
response under a given transfer rule if and only if $s_i^{\max}$ maximizes $A_i$.
Since this condition does not involve the transfer rule, it holds for one rule if and
only if it holds for every rule. This proves Part~3.
 
\emph{Step 5: approximate truthfulness (Part 4).} Write
$\bar\Pi := \sup_{m\in M_i}\Pi_i(m)$, which is finite by assumption. The hypothesis
states that $\bar\Pi-\varepsilon\leqslant\Pi_i\big(m_i^T(s_i)\big)$ for every $s_i$, and
by the definition of the supremum $\Pi_i\big(m_i^T(s_i)\big)\leqslant\bar\Pi$. Hence
every value $\Pi_i\big(m_i^T(x)\big)$, $x\in S_i$, lies in the interval
$[\bar\Pi-\varepsilon,\bar\Pi]$ of length $\varepsilon$, and any two numbers in an
interval of length $\varepsilon$ differ by at most $\varepsilon$. Note that Part~1 is
what makes this hypothesis meaningful: since $\Pi_i$ does not depend on the
contribution, the same function $\Pi_i$ is compared across all contributions. This
proves Part~4.
\end{proof}

\begin{proof}[{Proof of \Cref{cor:no-evidence-2P}}]
\emph{Part 1.} With the valuations of \Cref{sec:examples_utility} and linear costs,
$A_i(x) = -\frac1n x^{-\alpha_i} - c_i x$. Its derivative
$A_i'(x) = \frac{\alpha_i}{n}x^{-\alpha_i-1} - c_i$ is strictly decreasing in $x$, so
$A_i$ is strictly concave and its unconstrained maximizer is the unique root of
$A_i'(x)=0$, namely $x^* = (\alpha_i/(nc_i))^{1/(\alpha_i+1)}$. For a strictly concave
function of one variable, the maximizer over the interval $S_i$ is the projection of the
unconstrained maximizer onto $S_i$, which is $s_i^{\mathrm{nt}}$. By Part~2 of
\Cref{prop:no-evidence}, this is agent $i$'s contribution under every transfer rule,
and in particular under $p^1$ for every $\beta$.
 
\emph{Part 2.} We compare two reporting strategies $m_i^T(x)$ and $m_i^T(y)$ that report
gradients truthfully and differ only in the contribution reported, $x$ versus $y$. Three
observations show that only the transfer changes. First, the final parameter is computed
from the gradient reports with uniform weights, and by separability the gradient reports
$\nabla\varphi_i(w^t)$ do not depend on the reported contribution, so $W(R)$ is the same
under both strategies. Second, the other agents' contributions are the same: by
separability, agent $j$'s marginal utility in Phase~1,
$\psi_{jj}'(s_j)-c_j'(s_j)+\beta$, does not involve agent $i$'s reported contribution,
so agent $j$'s updates are unaffected by it. Third, under $p^1$ the transfer differs only
through agent $i$'s own report, and $p_i^1$ is linear in that report with slope $\beta$,
by \Cref{eq:payment}. Therefore
\begin{equation*}
    \Pi_i\big(m_i^T(x)\big)-\Pi_i\big(m_i^T(y)\big) = \beta\,(x-y)
    \qquad\text{for all } x,y\in S_i .
\end{equation*}
Now let agent $i$'s true contribution be $s_i^{\min}$. By Part~2 of
\Cref{prop:no-evidence}, its expected utility is $A_i(s_i^{\min})+\Pi_i(m_i)$, so
reporting $s_i^{\max}$ instead of the truthful $s_i^{\min}$ raises its expected utility by
$\Pi_i\big(m_i^T(s_i^{\max})\big)-\Pi_i\big(m_i^T(s_i^{\min})\big)
= \beta(s_i^{\max}-s_i^{\min})$. Truthful reporting is therefore $\varepsilon$-optimal only
if $\varepsilon\geqslant\beta(s_i^{\max}-s_i^{\min})$, and $\beta>\zeta+\tau$ gives the
final inequality.
\end{proof}

\subsection{\texorpdfstring{{Proof of \Cref{thm:truthful_T2P}}}{Proof of \ref{thm:truthful_T2P}}}
\label{app:T2P-proof}

\paragraph{Notation and timing.}
Throughout, $\bar w := w^{T_0}$ denotes the final model, $p^1,p^2,p^3$ are the transfers
of \Cref{eq:2PT_part1,eq:payment_truthful_2P,eq:grad-charge}, and $s=(s_i)_{i\in N}$
denotes the contributions \emph{actually used} in training, as opposed to the reported
contributions $\hat s$. In the elicitation stage, \emph{truthful reporting} by agent $i$
means $\hat s_i = s_i$ and $\hat v_i = v_i(\bar w,s)$, its realized valuation. We write
$v_i(\cdot) := v_i(\cdot,s^{\max})$ whenever the contributions are fixed at $s^{\max}$,
and $f(w) := -\frac1n\sum_{i\in N}v_i(w)$, as in the statement.

The proof repeatedly uses the order in which \Cref{algo:truthful_welfare} collects
reports, so we record three consequences of it.
\begin{enumerate}[leftmargin=22pt, itemsep=0pt, parsep=0pt, topsep=0pt]
    \item[(T1)] Contribution reports sent during Phase~1 are used only in the other agents'
    Phase~1 updates; they enter no transfer and do not affect the model.
    \item[(T2)] Gradient reports sent in round $t$ of Phase~2 affect only the next model
    $w^{t+1}$ and the round-$t$ charges $p^3(\hat d^t)$.
    \item[(T3)] The elicitation reports $(\hat s,\hat v)$ are sent after $\bar w$ has been
    computed and after all Phase~2 charges have been paid; they affect only $p^1$ and
    $p^2$.
\end{enumerate}
We now prove \Cref{thm:truthful_T2P}, taking its five parts in turn.

\paragraph{Part 1 (truthful contribution and valuation reports).}
Fix an agent $i$ and an arbitrary history of Phase~1 and Phase~2 reports, so that the
final model $\bar w$ and the actual contributions $s$ are fixed. Suppose the other agents
report truthfully in the elicitation stage, that is, $\hat s_{-i} = s_{-i}$ and
$\hat v_j = v_j(\bar w,s)$ for every $j\neq i$; that the latter is possible is the first
part of \Cref{assump:realized}, which states that each agent observes its realized
valuation at the contributions actually used. We compare agent $i$'s utility under an
arbitrary report $(\hat s_i,\hat v_i)$, with $\hat s_i\in S_i$, with its utility under
truthful reporting.

\emph{Step 1: only $p^1$ and $p^2$ change.} By (T3), agent $i$'s elicitation report does
not affect $\bar w$, and therefore does not affect its valuation $v_i(\bar w,s)$, which
depends only on $\bar w$ and the actual contributions. It does not affect its cost
$c_i(s_i)$, which depends only on its actual contribution, nor any Phase~2 charge, which
has already been paid. Consequently the difference between agent $i$'s utility under the
report $(\hat s_i,\hat v_i)$ and under truthful reporting equals the difference between
the corresponding values of $p_i^1+p_i^2$.

\emph{Step 2: the difference in $p^1$.} The other agents' reports $\hat s_{-i}=s_{-i}$
are the same in both cases, so in \Cref{eq:2PT_part1} only the first term changes, and
\begin{equation*}
    p_i^1(\hat s_i,s_{-i}) - p_i^1(s_i,s_{-i}) = \beta\,(\hat s_i - s_i).
\end{equation*}

\emph{Step 3: the value of $p^2$ under truthful reporting.} Under truthful reporting the
reported profile is $s$, so the center evaluates $v_j(\bar w,s)$ for every $j$. Agent
$i$'s own report $v_i(\bar w,s)$ coincides with this evaluation, so the quadratic term of
\Cref{eq:payment_truthful_2P} vanishes; and every other agent's report $v_j(\bar w,s)$
coincides with it as well, so $\chi_i = 0$. Hence $p_i^2 = 0$ under truthful reporting.

Combining Steps~1--3, and writing $\hat s := (\hat s_i,s_{-i})$ for the reported profile,
the utility difference is
\begin{equation}
\label{eq:app-cc-difference}
    u_i - u_i^{\mathrm{truth}}
    = \beta\,(\hat s_i - s_i)
      - \big(\hat v_i - v_i(\bar w,\hat s)\big)^2
      - \chi_i\,\big(\beta\hat s_i + K\big).
\end{equation}
We now distinguish two cases according to the contribution report.

\emph{Case 1: $\hat s_i = s_i$.} Then $\hat s = s$, and each other agent's report
$\hat v_j = v_j(\bar w,s)$ equals the center's evaluation $v_j(\bar w,\hat s)$, so
$\chi_i = 0$. The first term of \Cref{eq:app-cc-difference} also vanishes, and the
difference reduces to $-\big(\hat v_i - v_i(\bar w,s)\big)^2\leqslant 0$, with equality
if and only if $\hat v_i = v_i(\bar w,s)$, that is, if and only if agent $i$ reports its
true valuation.

\emph{Case 2: $\hat s_i \neq s_i$.} Here we use \Cref{assump:strict-interdep}, applied at
the parameter $\bar w$ with the other agents' contributions $s_{-i}$ and the pair
$\hat s_i\neq s_i$: there is an agent $j\neq i$ with
$v_j(\bar w,\hat s)\neq v_j(\bar w,s)$. Since agent $j$ reports $\hat v_j = v_j(\bar w,s)$,
its report differs from the center's evaluation $v_j(\bar w,\hat s)$, and by the
definition of $\chi_i$ in \Cref{eq:payment_truthful_2P}, $\chi_i = 1$. This conclusion
does not involve agent $i$'s valuation report, because $\chi_i$ depends only on the other
agents' reports and on the center's evaluations at $\hat s$. Substituting $\chi_i=1$ into
\Cref{eq:app-cc-difference}, the terms $\beta\hat s_i$ cancel, and
\begin{equation*}
    u_i - u_i^{\mathrm{truth}}
    = -\beta s_i - K - \big(\hat v_i - v_i(\bar w,\hat s)\big)^2
    \leqslant -\beta s_i - K < 0,
\end{equation*}
where the first inequality drops a non-positive term and the second uses $\beta>0$,
$s_i\geqslant 0$ (since $S_i\subseteq[0,\infty)$) and $K>0$.

In both cases truthful reporting is optimal, and every report with $\hat s_i\neq s_i$ is
worse by at least $\beta s_i + K$. Since $\bar w$ and the history were arbitrary, the
conclusion holds for every history of gradient reports. Finally, the profile $s$ in this
argument is the profile actually used in training. An agent that trains on $s_i$ samples
but reports $\hat s_i\neq s_i$, in particular one that trains on fewer samples than it
reports, is therefore in Case~2 and is penalized.

\paragraph{Part 2 (full contribution).}
\emph{Step 1: the payoff that drives contributions.} By Part~1, at the elicitation stage
agent $i$ reports its actual contribution, receives $p_i^1$ evaluated at the actual
contributions, and pays $p_i^2 = 0$. The contribution-phase utility to which the Phase~1
dynamics of \Cref{algo:truthful_welfare} apply is therefore
\begin{equation*}
    u_i(w,s) = v_i(w,s) - c_i(s_i) + p_i^1(s).
\end{equation*}
This utility contains no term
$\sum_{j\neq i}v_j$, so it coincides with the contribution-phase utility of
\mechtwo{}.

\emph{Step 2: a uniform lower bound on the marginal utility.} Differentiating with respect
to $s_i$, the three terms contribute $\pd{s_i}v_i(w,s)$, $-c_i'(s_i)$, and, by
\Cref{eq:2PT_part1}, exactly $\beta$. By \Cref{assump:derivatives}, the first two are
bounded below by $-\tau$ and $-\zeta$ respectively; this is the computation of
\Cref{lem:incr_utility}. Hence
\begin{equation*}
    \pd{s_i}u_i(w,s)\;\geqslant\;\beta-\tau-\zeta \;=\; \Delta \;>\; 0
    \qquad\text{for every } w\in\mathbb{R}^m \text{ and every } s\in\textstyle\prod_{j}S_j ,
\end{equation*}
where $\Delta>0$ because $\beta>\zeta+\tau$ by hypothesis.

\emph{Step 3: dominance.} Fix $w$ and $s_{-i}$. By Step~2, the function
$x\mapsto u_i(w,x,s_{-i})$ has a strictly positive derivative on $S_i$, so by the mean
value theorem it is strictly increasing on $S_i$ and attains its maximum only at
$s_i^{\max}$. Since $w$ and $s_{-i}$ were arbitrary, contributing $s_i^{\max}$ maximizes
$u_i$ whatever the other agents do, which is the first claim of Part~2.

\emph{Step 4: Phase~1 reaches $s^{\max}$.} Each Phase~1 update is a projected step
$s_i^{k+1} = \Pi_{S_i}\big(s_i^k + \gamma\,G_i^k\big)$, where $G_i^k$ is agent $i$'s
marginal utility evaluated at the broadcast information $(w^0,\hat s^k)$. By Step~2,
$G_i^k\geqslant\Delta$, whatever the broadcast reports are, because the lower bound holds
at every $(w,s)$. Hence $s_i^k + \gamma G_i^k\geqslant s_i^k+\gamma\Delta>s_i^k\geqslant
s_i^{\min}$, and projecting a point above $s_i^{\min}$ onto $[s_i^{\min},s_i^{\max}]$
only caps it at $s_i^{\max}$. Since the projection is monotone,
\begin{equation*}
    s_i^{k+1}\;\geqslant\;\min\big\{s_i^{\max},\,s_i^k+\gamma\Delta\big\}.
\end{equation*}
By induction on $k$, $s_i^k\geqslant\min\{s_i^{\max},\,s_i^0+k\gamma\Delta\}$, so
$s_i^k = s_i^{\max}$ as soon as $k\gamma\Delta\geqslant s_i^{\max}-s_i^0$, that is, for
all $k\geqslant\lceil(s_i^{\max}-s_i^0)/(\gamma\Delta)\rceil$. Once reached, $s_i^{\max}$
is kept, because the step from $s_i^{\max}$ is positive and the projection returns
$s_i^{\max}$. Taking the maximum over $i\in N$ gives \Cref{eq:kappa-prime}.

\emph{Step 5: misreports in Phase~1 are harmless.} Step~4 used only the lower bound
$G_i^k\geqslant\Delta$, which holds at every broadcast profile. Hence, consistently with
(T1), contribution reports sent during Phase~1, truthful or not, change neither any
agent's best response nor the number of steps needed to reach $s^{\max}$.

\paragraph{Part 3 (approximately truthful gradients).}
Fix a round $t$ of Phase~2 and an agent $i$, and suppose the other agents report their
true gradients $d_j^t := \nabla v_j(w^t)$. Write $w := w^t$ and
$b := \sum_{j\neq i}d_j^t$, and let $x\in\mathbb{R}^m$ be agent $i$'s report.

\emph{Step 1: agent $i$'s round utility as a function of its report.} By (T2), agent
$i$'s report affects only the next model $w^{t+1} = w + \frac{\eta}{n}(x+b)$ and the
round-$t$ charge. Agents are myopic in Phase~2, so agent $i$'s round payoff is its
valuation of $w^{t+1}$ plus $p_i^3(\hat d^t)$. In \Cref{eq:grad-charge}, the rebate term
$\frac{\eta}{2n(n-1)}\sum_{j\neq i}\|d_j^t\|_2^2$ depends only on the other agents'
reports, and the cost $c_i(s_i^{\max})$ depends only on the contribution, which was fixed
in Phase~1. Dropping these terms, which do not depend on $x$, changes neither the
maximizer nor any utility difference, and leaves
\begin{equation*}
    U(x) = v_i\Big(w + \frac{\eta}{n}(x+b)\Big) - \frac{\eta}{2n}\|x\|_2^2 .
\end{equation*}

\emph{Step 2: $U$ is $\mu$-strongly concave.} By the chain rule, since the inner map
$x\mapsto w+\frac{\eta}{n}(x+b)$ has Jacobian $\frac{\eta}{n}\mathbb{I}$,
\begin{equation}
\label{eq:app-gradU}
    \nabla U(x) = \frac{\eta}{n}\,\nabla v_i\Big(w+\frac{\eta}{n}(x+b)\Big)
                - \frac{\eta}{n}\,x .
\end{equation}
For any $x,y$, write $z_x := w+\frac{\eta}{n}(x+b)$ and $z_y := w+\frac{\eta}{n}(y+b)$,
so that $z_x - z_y = \frac{\eta}{n}(x-y)$. Then
\begin{equation*}
    \big\langle\nabla U(x)-\nabla U(y),\,x-y\big\rangle
    = \frac{\eta}{n}\big\langle\nabla v_i(z_x)-\nabla v_i(z_y),\,x-y\big\rangle
      - \frac{\eta}{n}\|x-y\|_2^2
    \leqslant \Big(\frac{\eta^2 M_i}{n^2} - \frac{\eta}{n}\Big)\|x-y\|_2^2
    = -\mu\,\|x-y\|_2^2 ,
\end{equation*}
where the inequality applies the Cauchy--Schwarz inequality and the $M_i$-Lipschitz
continuity of $\nabla v_i$ from \Cref{assump:agent-smooth}, which give
$\langle\nabla v_i(z_x)-\nabla v_i(z_y),x-y\rangle\leqslant M_i\|z_x-z_y\|_2\|x-y\|_2
= \frac{\eta M_i}{n}\|x-y\|_2^2$, and where $\mu := \frac{\eta}{n}\big(1-\frac{\eta
M_i}{n}\big)$. The condition $\eta M_i<n$ of \Cref{assump:agent-smooth} ensures
$\mu>0$. This strong monotonicity of $-\nabla U$ is equivalent to $\mu$-strong concavity
of $U$. (When $v_i$ is twice differentiable, the same conclusion reads
$\nabla^2U = \frac{\eta^2}{n^2}\nabla^2v_i-\frac{\eta}{n}\mathbb{I}\preceq-\mu\mathbb{I}$.)

\emph{Step 3: the gain from any report is controlled by the gradient at the truthful
report.} For a $\mu$-strongly concave function, for all $x$ and $y$,
$U(x)\leqslant U(y)+\langle\nabla U(y),x-y\rangle-\frac{\mu}{2}\|x-y\|_2^2$. Maximizing
the right-hand side over $x-y$, the maximum of $\langle a,z\rangle-\frac{\mu}{2}\|z\|^2$
being $\|a\|^2/(2\mu)$, attained at $z=a/\mu$, gives
\begin{equation}
\label{eq:app-PL}
    U(x)-U(y)\;\leqslant\;\frac{\|\nabla U(y)\|_2^2}{2\mu}
    \qquad\text{for all } x\in\mathbb{R}^m .
\end{equation}
We apply \Cref{eq:app-PL} with $y = d_i^t$, agent $i$'s truthful report.

\emph{Step 4: the gradient at the truthful report.} Substitute $x = d_i^t = \nabla v_i(w)$
in \Cref{eq:app-gradU}. Because the other agents are truthful,
$d_i^t+b = \sum_{j\in N}\nabla v_j(w) = n\,\tilde g^t$, so the model argument becomes
$w+\eta\tilde g^t = w+\Delta w$, where $\Delta w := \eta\tilde g^t$ is the update under
truthful reporting. The second term of \Cref{eq:app-gradU} becomes
$-\frac{\eta}{n}\nabla v_i(w)$. Hence
\begin{equation*}
    \nabla U(d_i^t) = \frac{\eta}{n}\big(\nabla v_i(w+\Delta w)-\nabla v_i(w)\big),
\end{equation*}
and, by the $M_i$-Lipschitz continuity of $\nabla v_i$,
\begin{equation*}
    \|\nabla U(d_i^t)\|_2 \leqslant \frac{\eta}{n}M_i\|\Delta w\|_2
    = \frac{\eta^2}{n}M_i\|\tilde g^t\|_2 .
\end{equation*}

\emph{Step 5: the per-round bound.} Squaring the bound of Step~4 and dividing by
$2\mu = \frac{2\eta(n-\eta M_i)}{n^2}$, as \Cref{eq:app-PL} prescribes, the factors $n^2$
cancel and one factor of $\eta$ cancels, so for every report $x$
\begin{equation*}
    U(x)-U(d_i^t)\;\leqslant\;
    \frac{\eta^4M_i^2\|\tilde g^t\|_2^2/n^2}{2\eta(n-\eta M_i)/n^2}
    = \frac{\eta^3M_i^2}{2(n-\eta M_i)}\|\tilde g^t\|_2^2 = \varepsilon_i^t .
\end{equation*}

\emph{Step 6: the bound along the truthful path.} Along the truthful path,
$\tilde g^t = \frac1n\sum_j\nabla v_j(w^t) = -\nabla f(w^t)$ by the definition of $f$.
By the descent lemma for the $M$-smooth function $f$ (\Cref{assump:conv-opt-NE}), applied
at $w^t$ with step $1/M$,
$f(\wopt)\leqslant f\big(w^t-\frac1M\nabla f(w^t)\big)\leqslant f(w^t)-\frac{1}{2M}\|\nabla
f(w^t)\|_2^2$, where the first inequality holds because $\wopt$ minimizes $f$. Rearranging,
\begin{equation*}
    \|\tilde g^t\|_2^2 = \|\nabla f(w^t)\|_2^2 \leqslant 2M\big(f(w^t)-f(\wopt)\big).
\end{equation*}
Under truthful reports Phase~2 is gradient descent on $f$ with step $\eta=1/M$, so the
Phase~2 argument in the proof of \Cref{thm:two_phase} gives
$f(w^t)-f(\wopt)\leqslant(1-\nicefrac{\nu}{M})^t\big(f(w^0)-f(\wopt)\big)$; Phase~1 does
not move the model, so Phase~2 starts at $w^0$. Substituting $\eta=1/M$ into
$\varepsilon_i^t$ gives $\varepsilon_i^t = \frac{M_i^2\|\tilde g^t\|_2^2}{2M^2(nM-M_i)}$,
and combining the last two displays,
\begin{equation*}
    \varepsilon_i^t\;\leqslant\;\frac{M_i^2\big(f(w^0)-f(\wopt)\big)}{M(nM-M_i)}
    \Big(1-\frac{\nu}{M}\Big)^t = C_i\Big(1-\frac{\nu}{M}\Big)^t .
\end{equation*}
Since $0<\nu\leqslant M$, the ratio $1-\nicefrac{\nu}{M}$ lies in $[0,1)$, and the geometric
series satisfies $\sum_{t\geqslant 0}(1-\nicefrac{\nu}{M})^t = \nicefrac{M}{\nu}$. Summing
the per-round bounds over the rounds of Phase~2 therefore gives
$\sum_{t<T_0}\varepsilon_i^t\leqslant C_i\nicefrac{M}{\nu}
= \frac{M_i^2(f(w^0)-f(\wopt))}{\nu(nM-M_i)}$. This total is the sum of the gains from
one-shot deviations along the truthful path; a deviation in round $t$ changes the later
iterates, so the sum does not bound the gain from deviating in several rounds.

\paragraph{Part 4 (optimal parameter).}
\emph{Truthful reports.} If every agent reports $\nabla v_j(w^t)$, the update is
$w^{t+1} = w^t+\frac{\eta}{n}\sum_j\nabla v_j(w^t) = w^t-\eta\nabla f(w^t)$, which is
gradient descent on $f$ with $\eta=1/M$. The claim is then the Phase~2 part of
\Cref{thm:two_phase}.

\emph{Equilibrium reports: characterization.} Fix a round $t$ and arbitrary reports
$\hat d_{-i}^t$ of the other agents. The argument of Part~3, Step~2, did not use the
truthfulness of the other agents, since $b$ only shifts the argument of $v_i$; hence agent
$i$'s round utility is $\mu$-strongly concave in its report for any $\hat d_{-i}^t$. A
strongly concave differentiable function has a unique maximizer, characterized by a zero
gradient. By \Cref{eq:app-gradU}, agent $i$'s best response is therefore the unique $x$
with
\begin{equation}
\label{eq:app-BR}
    x = \nabla v_i(w^{t+1}),
    \qquad w^{t+1} = w^t+\frac{\eta}{n}\Big(x+\sum_{j\neq i}\hat d_j^t\Big).
\end{equation}
A report profile $\hat d^t$ is a Nash equilibrium of the round if and only if
\Cref{eq:app-BR} holds for every agent, that is, $\hat d_i^t = \nabla v_i(w^{t+1})$ for
all $i$, with $w^{t+1} = w^t+\frac{\eta}{n}\sum_j\hat d_j^t$.

\emph{Equilibrium reports: reduction to a proximal step.} We show that equilibria
correspond one-to-one to solutions $u$ of the equation
\begin{equation}
\label{eq:app-implicit}
    u = w^t-\eta\nabla f(u).
\end{equation}
If $\hat d^t$ is an equilibrium, substituting $\hat d_j^t = \nabla v_j(w^{t+1})$ into the
update gives $w^{t+1} = w^t+\frac{\eta}{n}\sum_j\nabla v_j(w^{t+1}) =
w^t-\eta\nabla f(w^{t+1})$, so $u = w^{t+1}$ solves \Cref{eq:app-implicit}. Conversely, if
$u$ solves \Cref{eq:app-implicit}, set $\hat d_j^t := \nabla v_j(u)$ for every $j$; then
$w^t+\frac{\eta}{n}\sum_j\hat d_j^t = w^t-\eta\nabla f(u) = u$, so the update returns
$w^{t+1}=u$, and every agent's report equals $\nabla v_j(w^{t+1})$, which is the
equilibrium condition. Equation \Cref{eq:app-implicit} is the first-order optimality
condition of
\begin{equation*}
    \min_{u\in\mathbb{R}^m}\ \Phi(u),\qquad \Phi(u) := f(u)+\frac{1}{2\eta}\|u-w^t\|_2^2 ,
\end{equation*}
since $\nabla\Phi(u) = \nabla f(u)+\frac1\eta(u-w^t)$. By \Cref{assump:conv-opt-NE}, $f$ is
$\nu$-strongly convex, so $\Phi$ is $(\nu+\frac1\eta)$-strongly convex. A strongly convex
differentiable function has exactly one minimizer, and for a convex differentiable
function the stationary points are exactly the minimizers. Hence \Cref{eq:app-implicit}
has exactly one solution, the round has exactly one equilibrium, and the equilibrium
update is the proximal-point step $w^{t+1}=\operatorname*{arg\,min}_u\Phi(u)$.

\emph{Equilibrium reports: rate.} Since $\wopt$ minimizes $f$, $\nabla f(\wopt)=0$, and
\Cref{eq:app-implicit} with $u=w^{t+1}$ can be written as
\begin{equation*}
    w^{t+1}-\wopt = (w^t-\wopt)-\eta\big(\nabla f(w^{t+1})-\nabla f(\wopt)\big).
\end{equation*}
Taking the inner product of both sides with $e := w^{t+1}-\wopt$ gives
$\|e\|_2^2 = \langle w^t-\wopt,e\rangle - \eta\langle\nabla f(w^{t+1})-\nabla
f(\wopt),e\rangle$. By $\nu$-strong convexity, the gradient of $f$ is strongly monotone,
$\langle\nabla f(u)-\nabla f(u'),u-u'\rangle\geqslant\nu\|u-u'\|_2^2$, so the last inner
product is at least $\nu\|e\|_2^2$. Moving it to the left-hand side and applying the
Cauchy--Schwarz inequality to the remaining inner product,
\begin{equation*}
    (1+\eta\nu)\|e\|_2^2\;\leqslant\;\langle w^t-\wopt,e\rangle
    \;\leqslant\;\|w^t-\wopt\|_2\,\|e\|_2 .
\end{equation*}
If $e=0$ there is nothing to prove; otherwise dividing by $\|e\|_2$ gives
$\|w^{t+1}-\wopt\|_2\leqslant(1+\eta\nu)^{-1}\|w^t-\wopt\|_2$, and iterating from $w^0$
gives $\|w^t-\wopt\|_2\leqslant(1+\eta\nu)^{-t}\|w^0-\wopt\|_2$.

\paragraph{Part 5 (budget).}
\emph{$p^1$ is budget balanced for every report profile.} Summing \Cref{eq:2PT_part1}
over $i$,
\begin{equation*}
    \sum_{i\in N}p_i^1(\hat s) = \beta\sum_{i\in N}\hat s_i
    - \frac{\beta}{n-1}\sum_{i\in N}\sum_{j\neq i}\hat s_j
    = \beta\sum_{i\in N}\hat s_i - \frac{\beta}{n-1}(n-1)\sum_{j\in N}\hat s_j = 0,
\end{equation*}
where the middle equality exchanges the order of summation: each $\hat s_j$ appears in the
inner sum of the $n-1$ agents $i\neq j$.

\emph{$p^3$ is budget balanced in every round.} Write $Q := \sum_{j\in N}\|\hat
d_j^t\|_2^2$. Summing \Cref{eq:grad-charge} over $i$, the first terms give
$-\frac{\eta}{2n}Q$, and by the same exchange of summation the rebates give
$\frac{\eta}{2n(n-1)}(n-1)Q = \frac{\eta}{2n}Q$. Hence $\sum_i p_i^3(\hat d^t) = 0$ for
every report profile.

\emph{$p^2$ is non-positive, and zero on the truthful path.} In
\Cref{eq:payment_truthful_2P}, the quadratic term is non-positive, and the cross-check
term is non-positive because $\chi_i\geqslant 0$, $\hat s_i\geqslant 0$ for reports in
$S_i$, $\beta>0$ and $K>0$. Hence $p_i^2\leqslant 0$ for every agent and every report
profile, and $\sum_i(p_i^1+p_i^2) = \sum_i p_i^2\leqslant 0$. On the truthful path both
terms vanish, as shown in Step~3 of Part~1, so $p^2\equiv 0$ there.

\emph{Conclusion.} The total transfer to the agents is
$\sum_i\big(p_i^1+p_i^2\big)+\sum_{t<T_0}\sum_ip_i^3$. By the three facts above, it is
zero on the truthful path and non-positive for every report profile, so the center never
runs a deficit. \hfill$\square$

\section{\texorpdfstring{{Adversarial agents}}{Adversarial agents}}
Our results for \mechtwo{} assume that agents are honest and report their gradients truthfully. However, the agents may be adversaries who misreport their gradients. Prior work~\citep{yin2018byzantine} describes some simple ways to deal with this. When the ratio of adversarial agents in the system is low, we use \mechtwo{} along with the trimmed mean algorithm of \citet{yin2018byzantine}. 

Let $\alpha^{adv}$ be the fraction of adversarial agents in the system. Phase $1$ of the algorithm remains the same with agents first converging to $s^{\max}$. In the second phase, which is a classic FedAvg algorithm, the agents only report gradients computed on their training data. Instead of aggregating all gradients, for each coordinate of the gradients, the values of the gradient corresponding to that coordinate are sorted. The lowest and highest $\alpha^{adv}$ fraction of these values are discarded and the remaining gradient coordinates are used to update the parameter. We detail the algorithm in \Cref{algo:2Ptrimmedmean}. In \Cref{app:experiments_adversarial} we show that this algorithm is robust to adversaries when the fraction of adversarial agents is small. 

\label{sec:adversarial}
\begin{algorithm}[t]
\DontPrintSemicolon
\LinesNumbered
\SetStartEndCondition{ }{}{}
\caption{Trimmed mean for \mechtwo{} (trim-\mechtwo{})}
\label{algo:2Ptrimmedmean}
\KwIn{Step sizes $\gamma,\eta$; initialization $w^0$, $s_i^0\in S_i$ for $i\in N$;
      iteration counts $\kappa,T_0$ as given by \Cref{thm:two_phase};
      trimming fraction $\beta^{trim}$}
\KwOut{$w^{T_0}$}
Set $b=\lfloor n\beta^{trim} \rfloor$\;
\tcp{Phase 1: contributions, with the model fixed at $w^0$ (identical to \mechtwo{})}
\textbf{Center} broadcasts $w^0,s^0$; set $k=0$\;
\While{$k<\kappa$}{
  \textbf{Center} broadcasts $s^k$\;
  \ForEach{\textbf{agent} $i\in N$ in parallel}{
    $s_i^{k+1}=s_i^k+\gamma\,[g(w^0,s^k,\mu^k)]_i$; send $s_i^{k+1}$ to center\;
  }
  $k\gets k+1$\;
}
\tcp{Under \Cref{eq:payment}, $s^\kappa=s^{\max}$}
\tcp{Phase 2: model learning at $s^{\max}$ with trimmed-mean aggregation}
\For{$t=0$ \KwTo $T_0-1$}{
  \textbf{Center} broadcasts $w^t$\;
  \ForEach{\textbf{agent} $i\in N$ in parallel}{
    Compute $d_i^{t+1}=\nabla_w v_i(w^t,s^{\max})$ (adversarial agents may send arbitrary $d_i^{t+1}$); send $d_i^{t+1}$ to center\;
  }
  \ForEach{coordinate $j\in[m]$}{
    Sort $\{d_{ij}^{t+1}\}_{i\in N}$ in ascending order as $c_{(1)}\le\dots\le c_{(n)}$\;
    $w_j^{t+1}=w_j^t+\dfrac{\eta}{n-2b}\sum_{r=b+1}^{n-b}c_{(r)}$\;
  }
}
\Return $w^{T_0}$\;
\end{algorithm}
 
\section{Appendix to \Cref{sec:experiments}}

\subsection{Social welfare results using test accuracy}
\label{app:accuracy-welfare}
\begin{figure}[t]
  \centering
  \begin{subfigure}{\linewidth}
  \hspace*{0pt}
    \centering
    \includegraphics[width=0.32\linewidth]{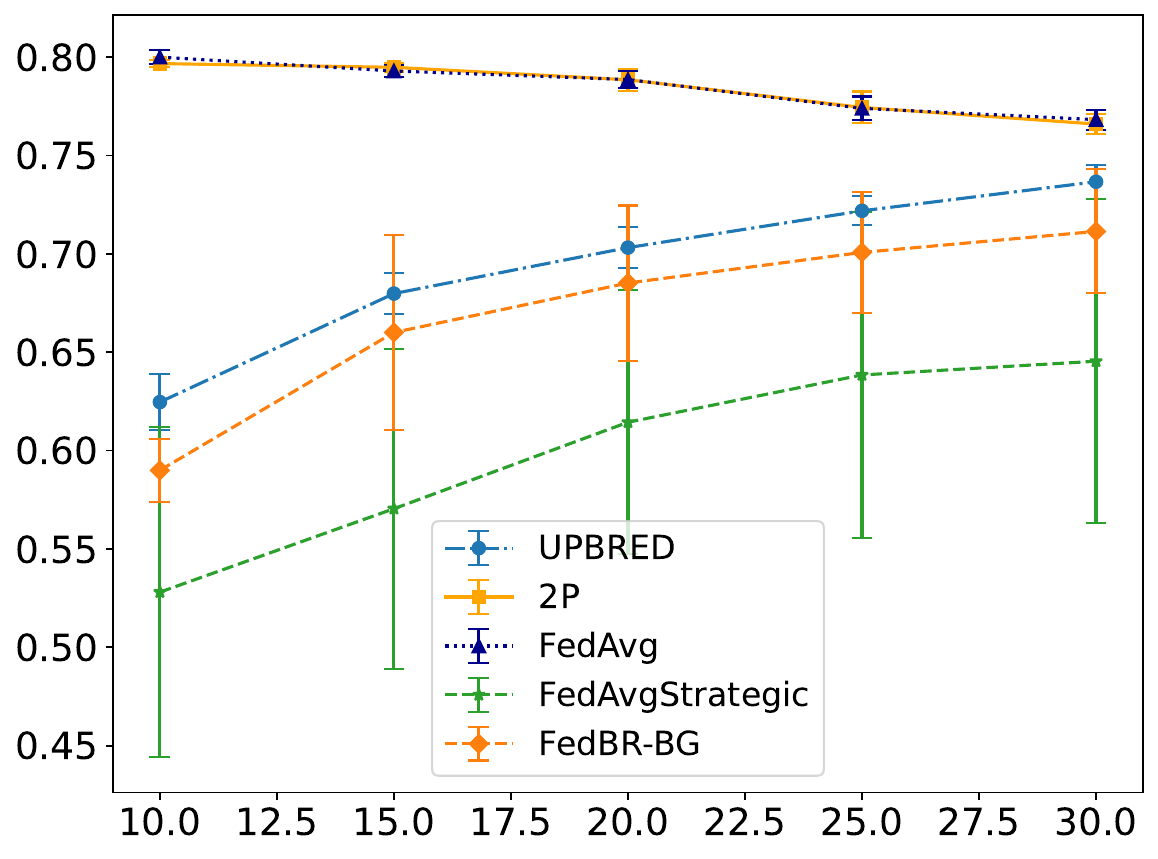}
    \includegraphics[width=0.32\linewidth]{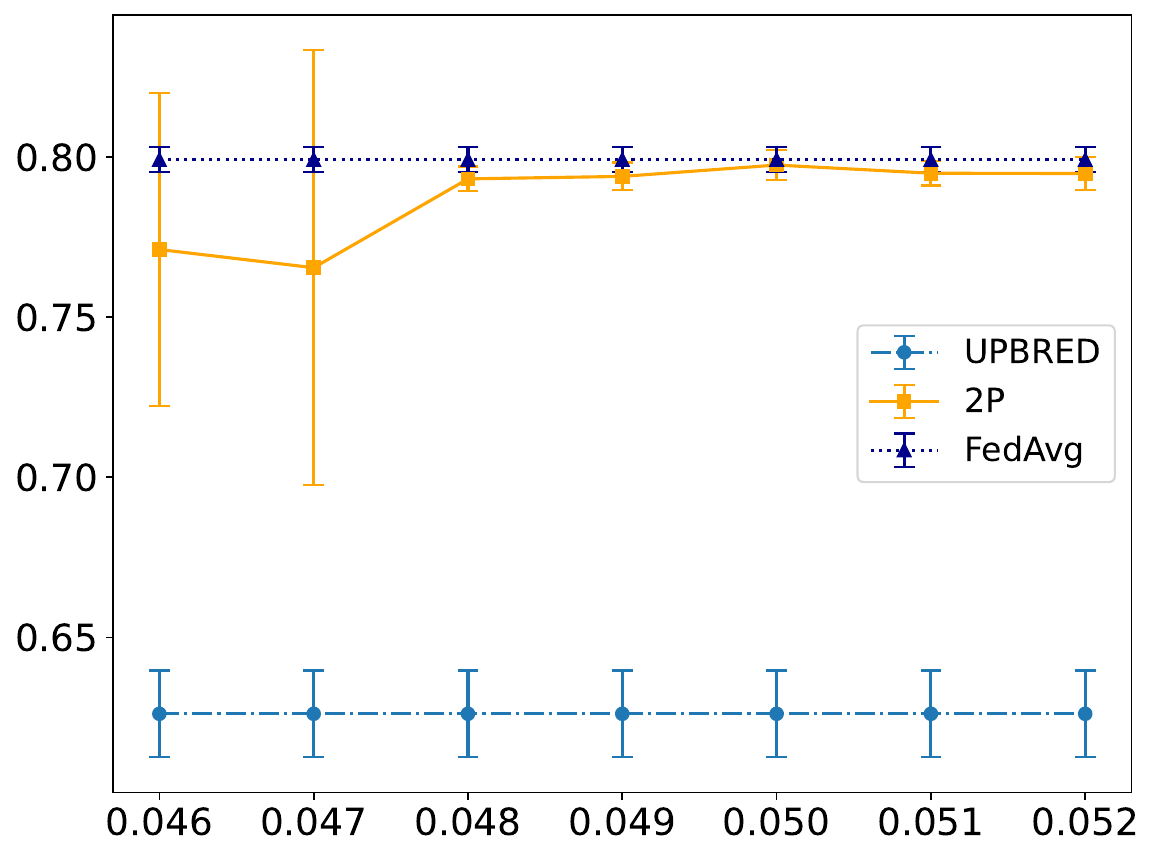}
    \includegraphics[width=0.32\linewidth]{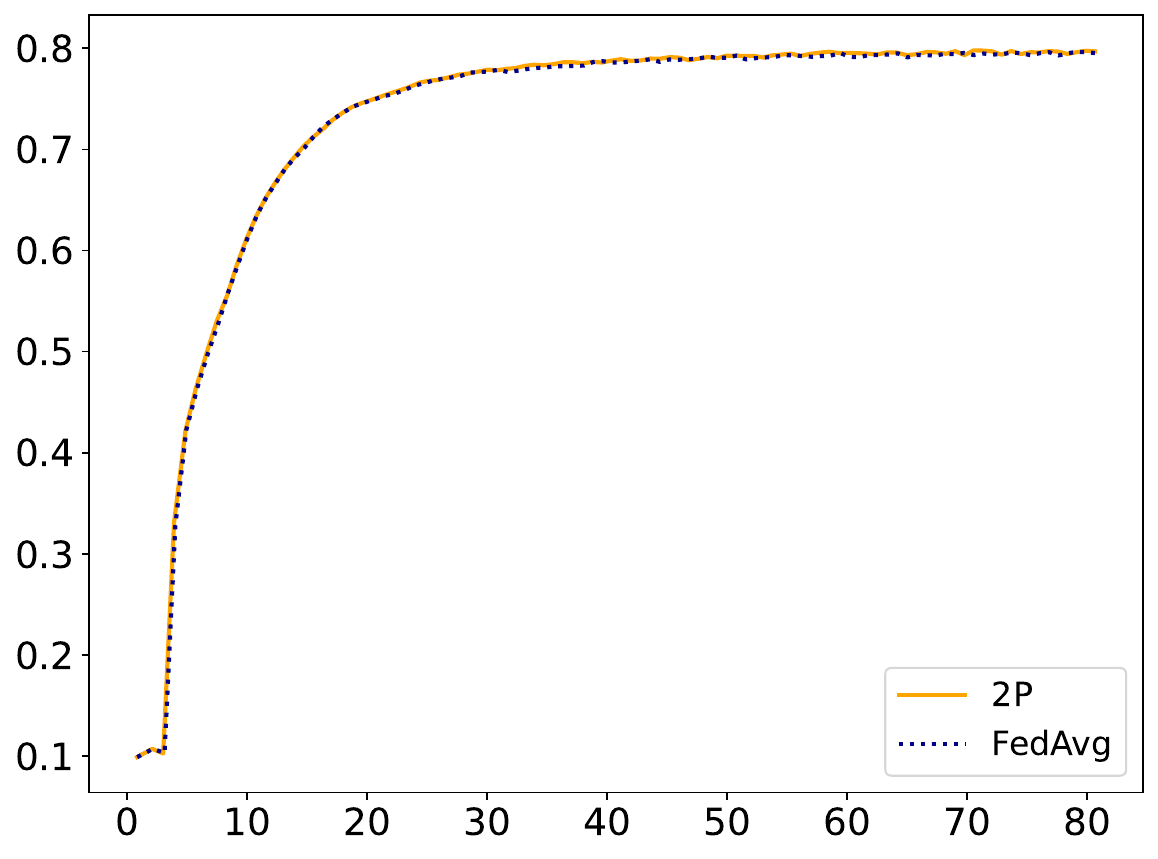}
    \caption{CIFAR-10}
  \end{subfigure}


  \begin{subfigure}{\linewidth}
  \hspace*{0pt}
    \centering
    \includegraphics[width=0.32\linewidth]{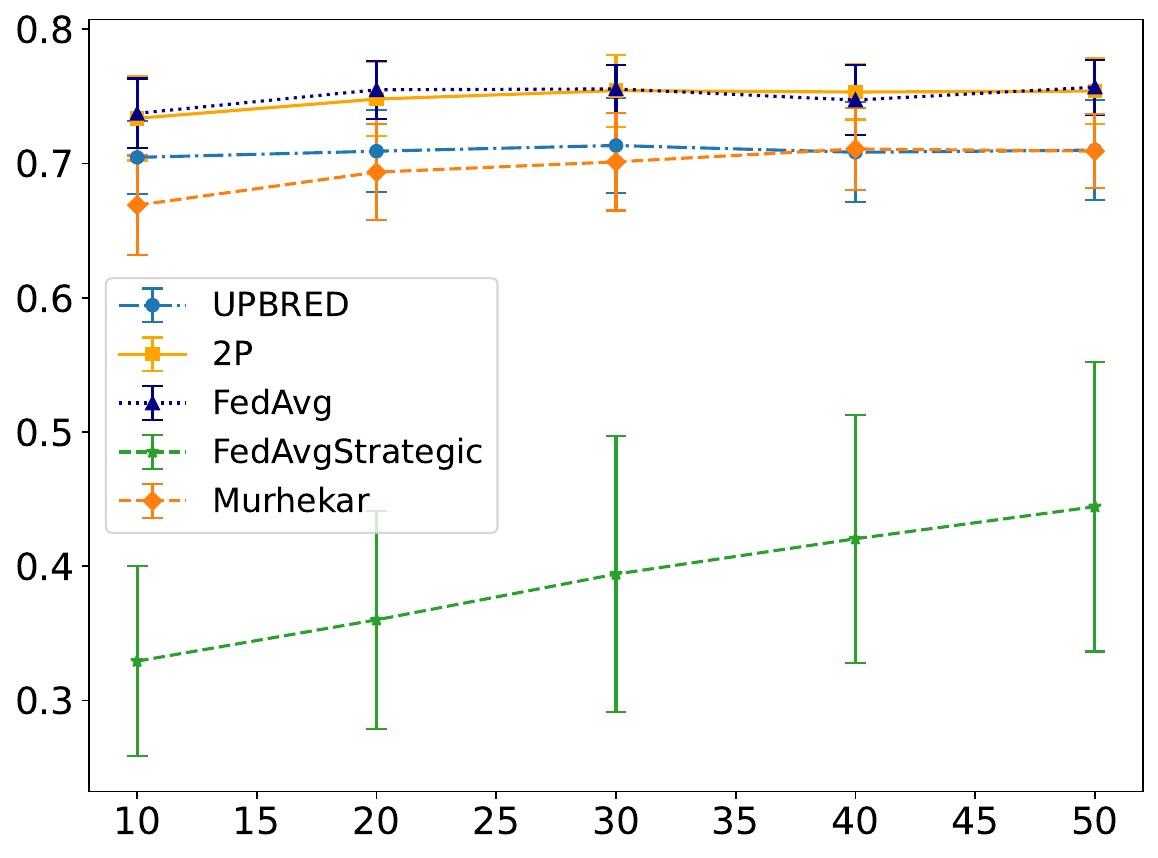}
    \includegraphics[width=0.32\linewidth]{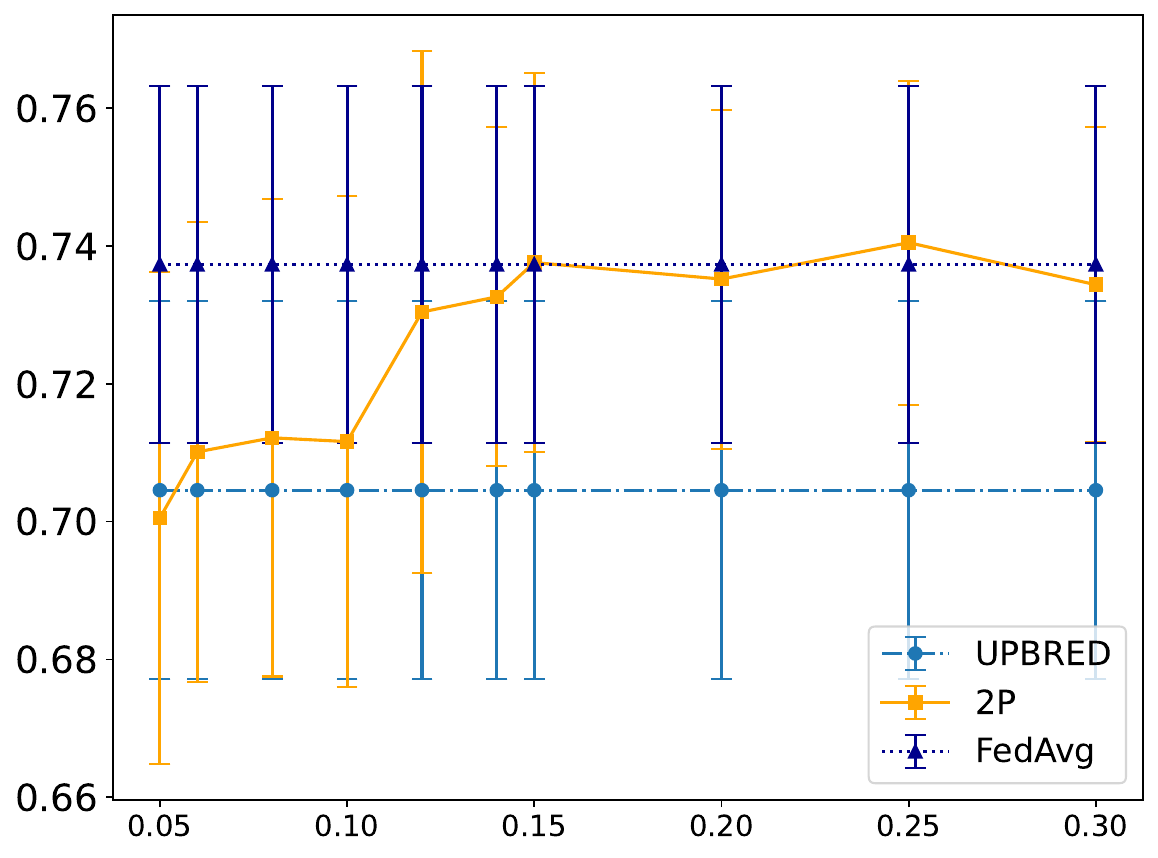}
    \includegraphics[width=0.32\linewidth]{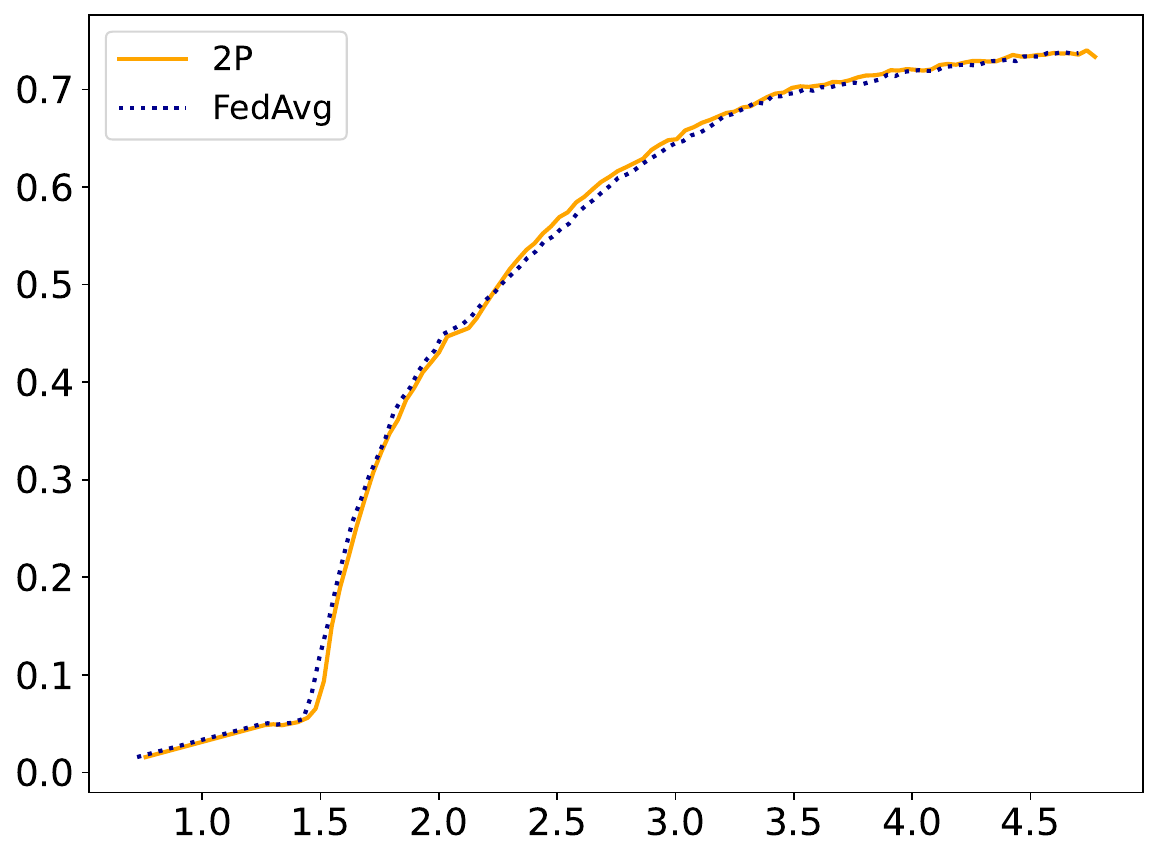}
    \caption{FEMNIST}
  \end{subfigure}


  \begin{subfigure}{\linewidth}
  \hspace*{0pt}
    \centering
    \includegraphics[width=0.32\linewidth]{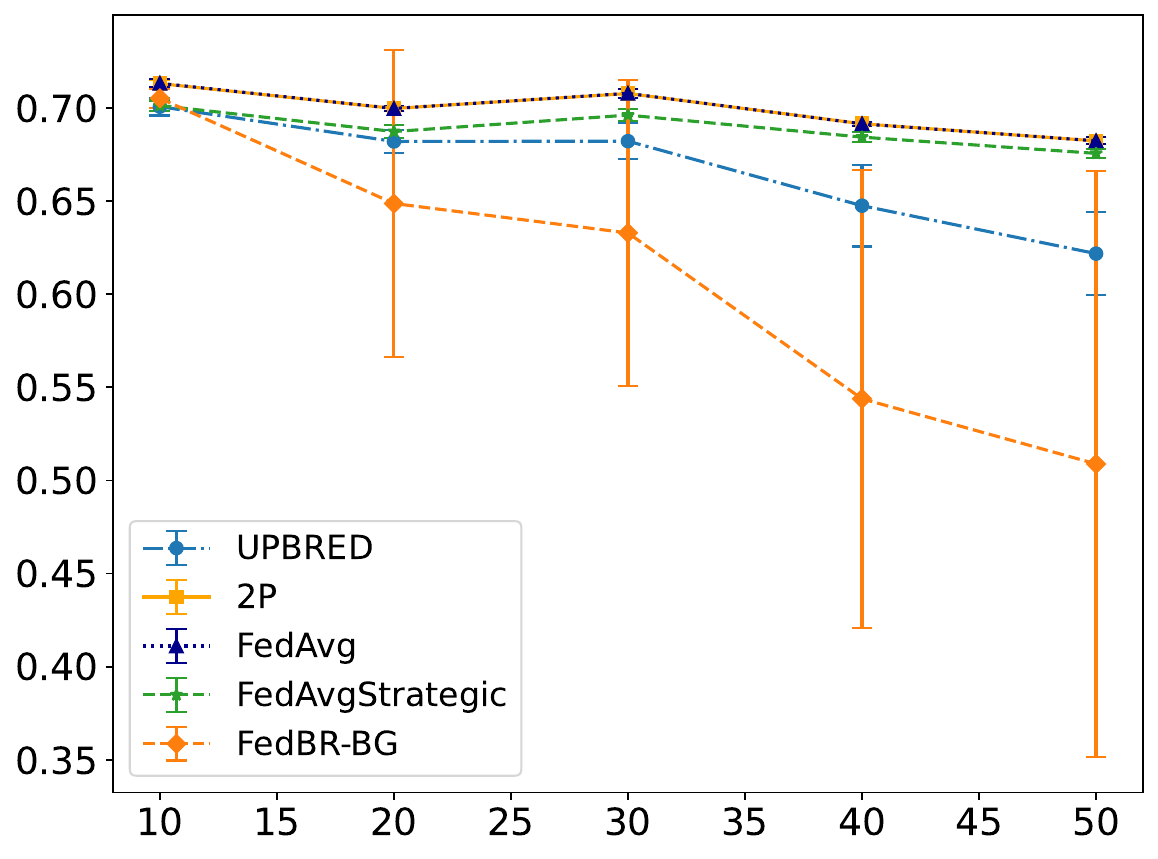}
    \includegraphics[width=0.32\linewidth]{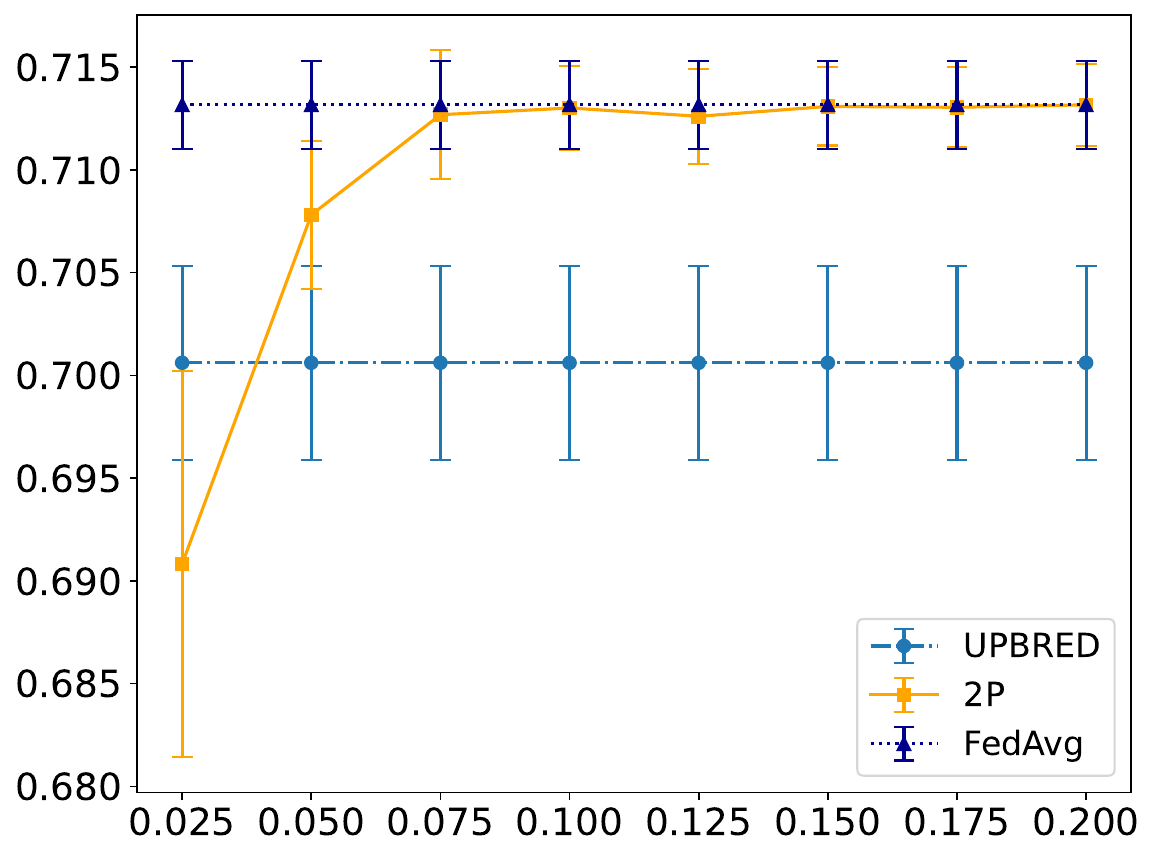}
    \includegraphics[width=0.32\linewidth]{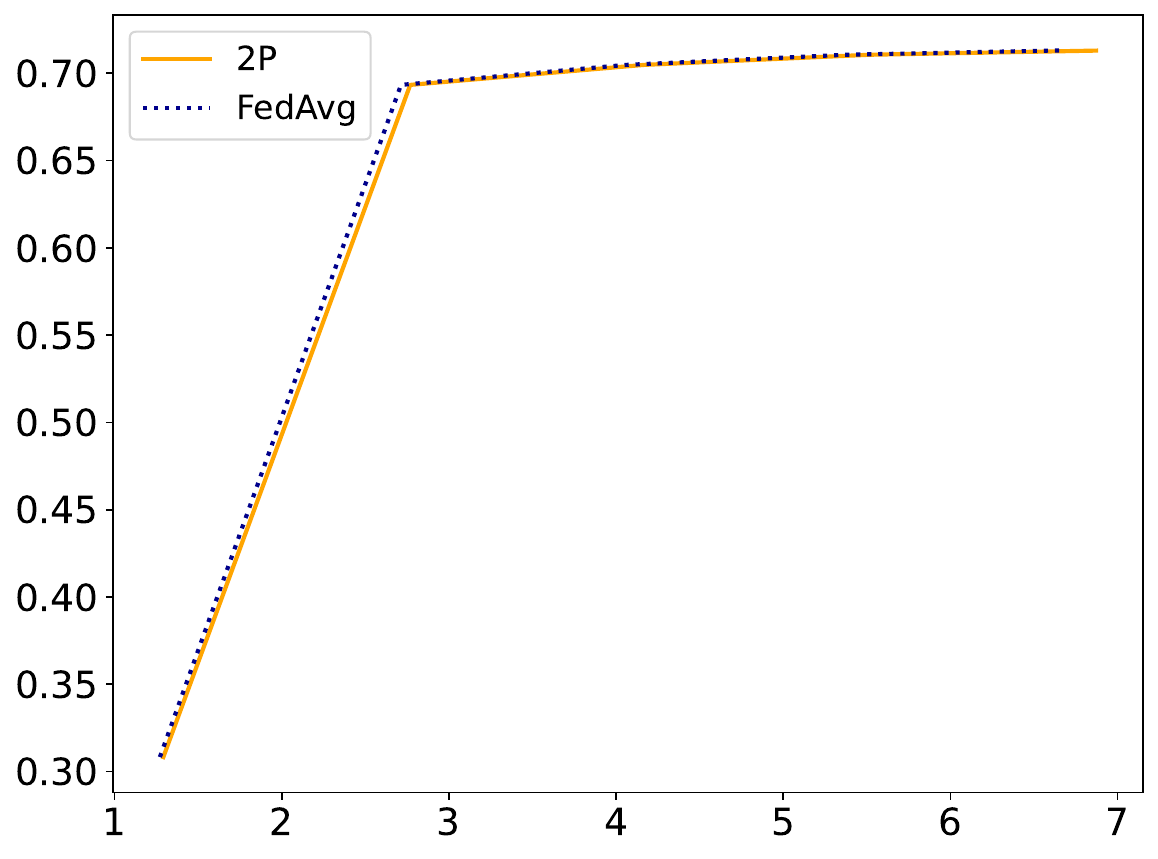}
    \caption{Twitter}
  \end{subfigure}

  \caption{The three rows in this figure correspond to the experiments with the datasets CIFAR-10, FEMNIST, and Twitter, respectively. The {\tt y-axis} in every plot is the social welfare (computed using test accuracy for each agent). The {\tt x-axis} corresponds to the number of agents in the first plot, the choice of $\beta$ in the second, and the time taken (in minutes) in the third plot, respectively, in every row.}
  \label{fig:combined_plots_accuracy}
  \vspace{-10pt}
\end{figure}
\subsection{Algorithm for \strategic{}}
\label{sec:strategic}
In the \strategic{} algorithm, the agents choose to contribute their data strategically as in \mech{}, and stick to this contribution in the entire training phase of \FedAvg{}.
\strategic{} is formally given by \Cref{algo:fedavgstrategic}.

\begin{algorithm}[t]

\DontPrintSemicolon
\LinesNumbered
\SetStartEndCondition{ }{}{}
\caption{\strategic\label{algo:fedavgstrategic}}

\KwIn{Step size $\gamma, \eta$, initialization $w^0$, $s_i^0$ for $i \in N$, number of iterations $T$, and a tolerance $\epsilon$}
\KwOut{$w^T$}
\textbf{Center} broadcasts $w^0, s^0$, set $t=0$\;
 \While{$\exists i \in N \text{ such that } \|[g(w^0,s^t,\mu^t)]_i \| \ \geq \epsilon$ \ }{
    \textbf{Center} broadcasts $s^t$\;
    \For{\textbf{agent} $i \in N$ in parallel \ }{
        $s_i^{t+1} = s_i^t + \gamma [g(w^0,s^t,\mu^t)]_i$\;
        Send $s_i^{t+1}$ to the center\;
    }
    $t = t+1$}
$s^*=s^t$\;
\For{$t=0$ \textbf{to} $T$}
{
    \textbf{Center} broadcasts $w^t$\;
    \For{\textbf{agent} $i \in N$ in parallel \ }{
        Compute local gradient: $d_i^{t+1} = \nabla_w \:v_i(w^{t},s_i^{*},s_{-i}^{*})$\;
        Send $d_i^{t+1}$ to the center\;
    }
    \textbf{Center} updates: $w^{t+1} = w^t + \frac{\eta}{n} \sum_{i \in N} d_i^{t+1} = w^t + \eta \tilde{g}(w^t, s^{*})$\;
}
\Return $w^T$\;

\end{algorithm}

\subsection{Additional experiments run using the Flower framework}
\begin{figure}[t]
     \centering
     \begin{subfigure}[b]{0.32\linewidth}
         \centering
         \includegraphics[width=\linewidth]{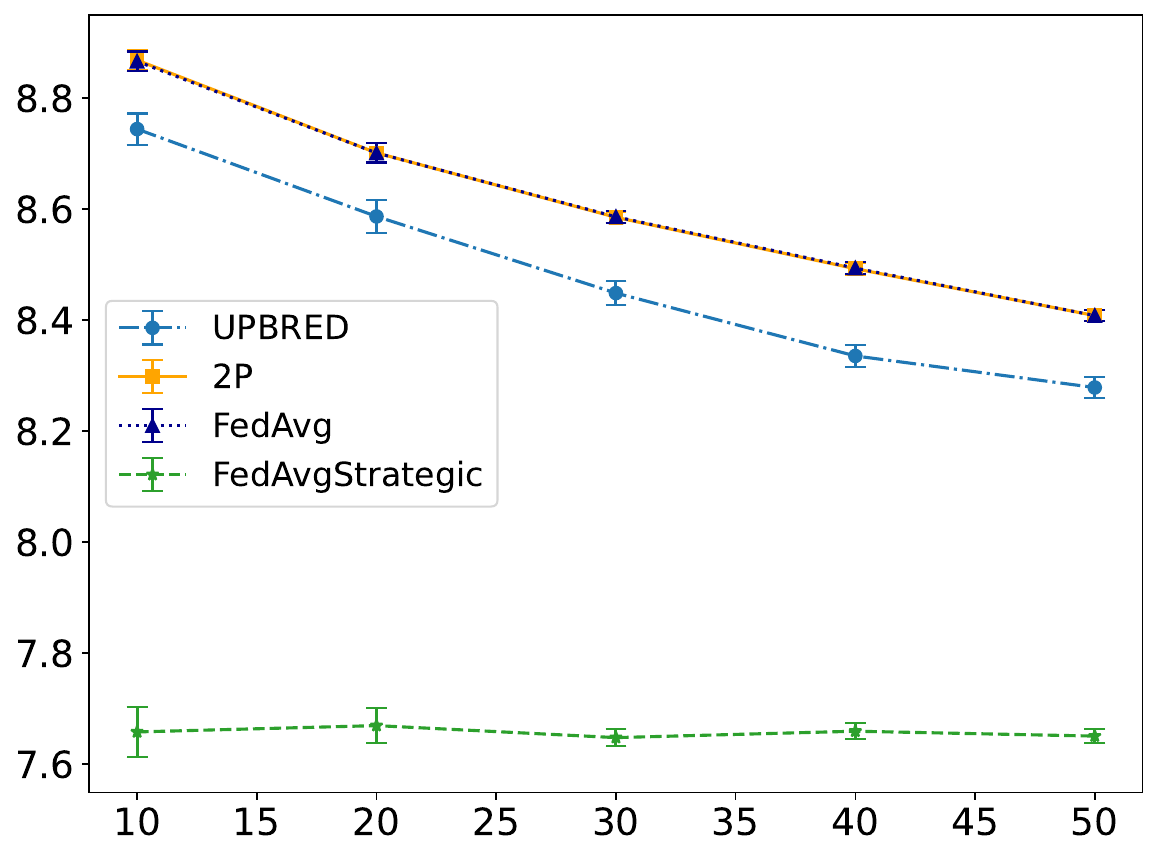}
         \caption{IID Split}
         \label{fig:iid_split}
     \end{subfigure}
     \begin{subfigure}[b]{0.32\linewidth}
         \centering
         \includegraphics[width=\linewidth]{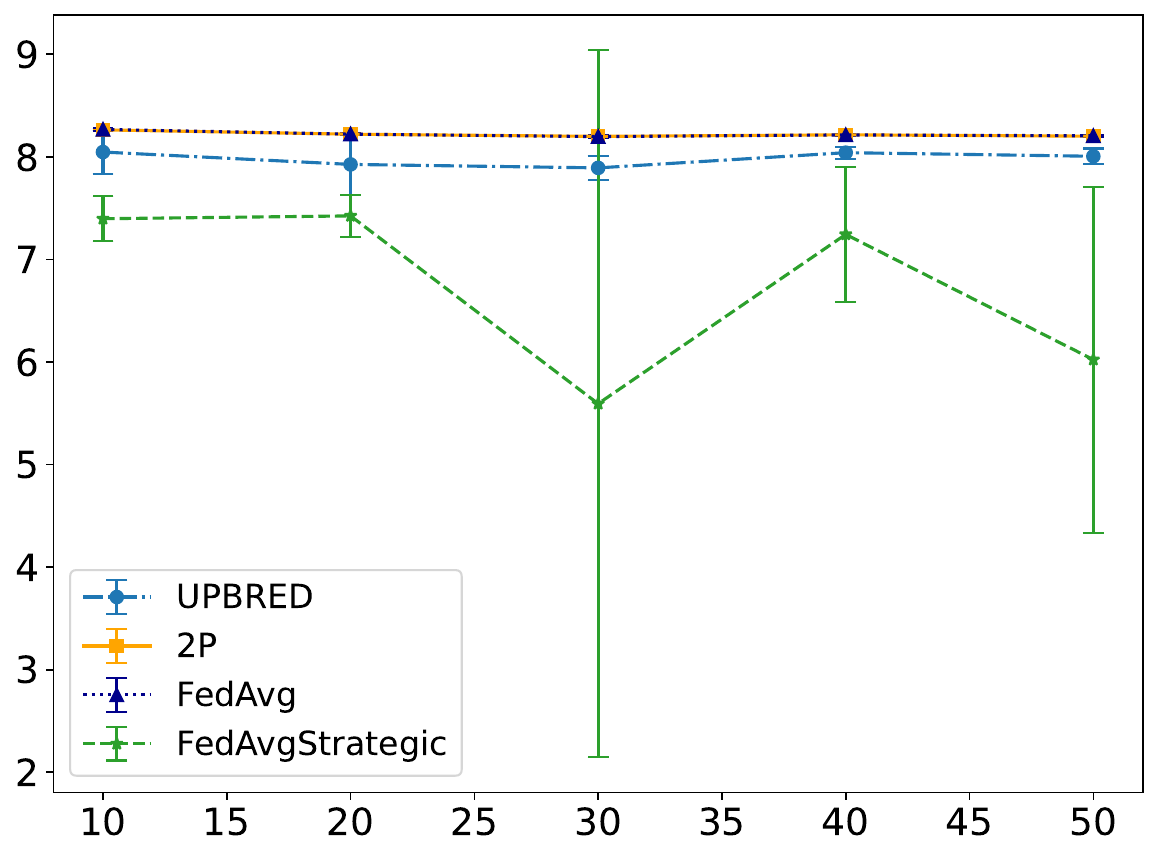}
         \caption{Dirichlet Split}
         \label{fig:dirichlet_split}
     \end{subfigure}
     \begin{subfigure}[b]{0.32\linewidth}
         \centering
         \includegraphics[width=\linewidth]{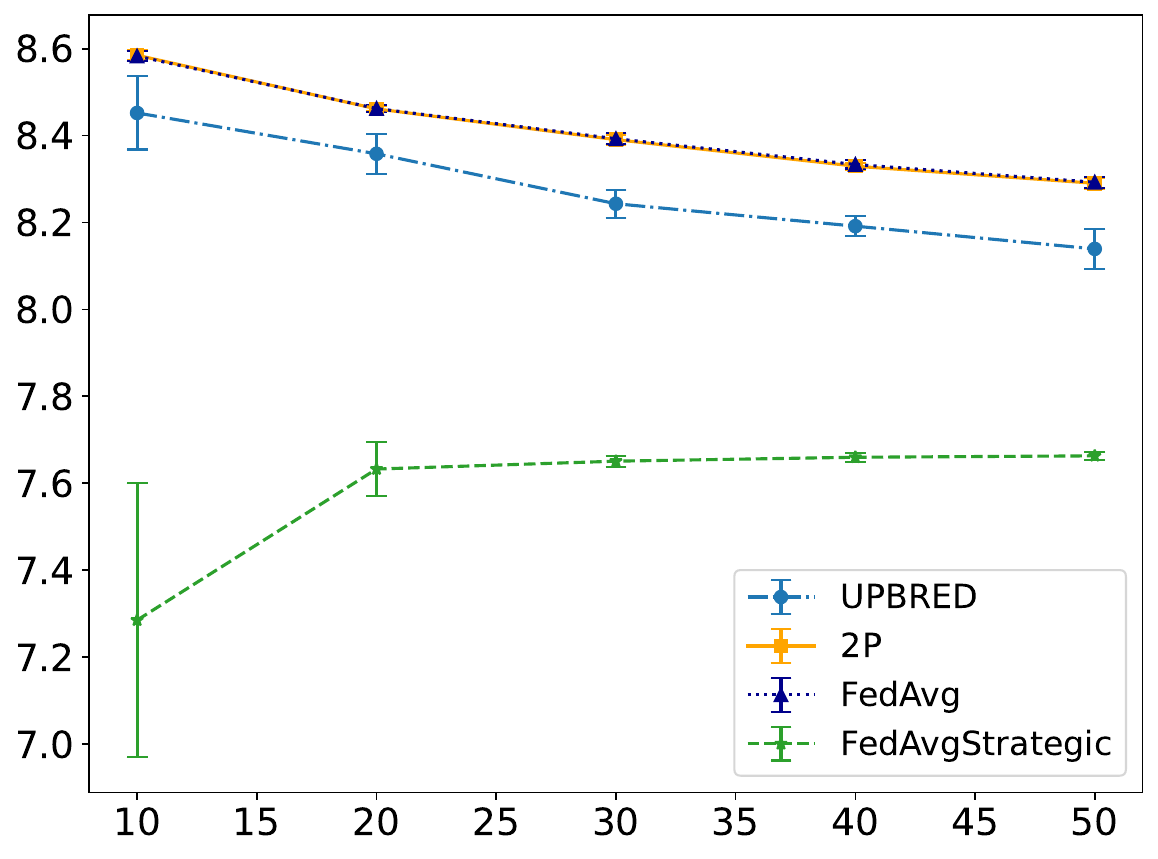}
         \caption{Pathological Split}
         \label{fig:pathological_split}
     \end{subfigure}
     
     \caption{Social welfare vs. number of clients for different CIFAR-10 dataset splits using the Flower framework.}
     \label{fig:flower}
      \vspace{-10pt}
\end{figure}

\begin{figure}[ht]
    \centering
    \begin{subfigure}[t]{0.43\linewidth}
        \centering
        \includegraphics[width=\linewidth]{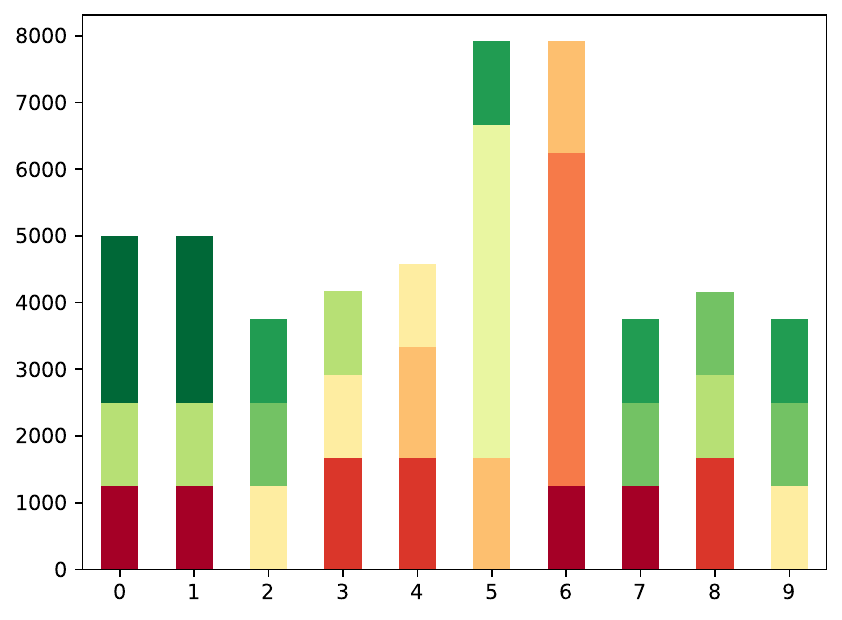}
        \caption{Pathological partitioner ($3$ classes per client).}
        \label{fig:pathological_dist}
    \end{subfigure}
    \hfill
    \begin{subfigure}[t]{0.55\linewidth}
        \centering
        \includegraphics[width=\linewidth]{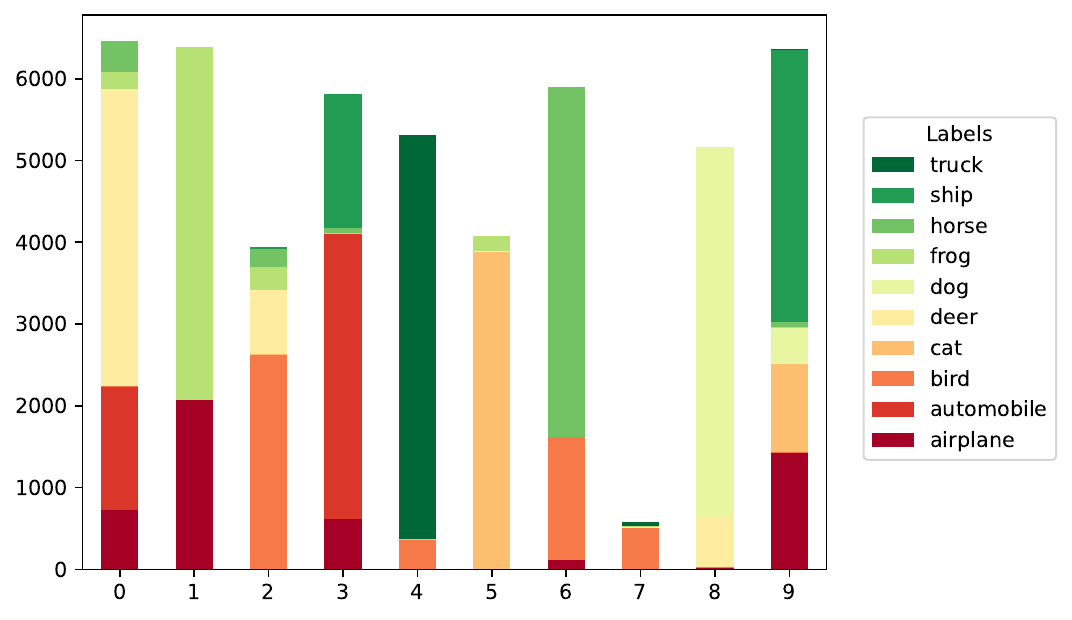}
        \caption{Dirichlet partitioner ($\alpha=0.1$).}
        \label{fig:dirichlet_dist}
    \end{subfigure}

    \caption{Label distributions for different data partitioning strategies using Flower with $10$ agents. The {\tt x-axis} represents the client ID and the {\tt y-axis} represents the number of data points of the given label.}
    \label{fig:dataset_distribution}
    \vspace{-10pt}
\end{figure}

Using the Flower framework, we partitioned the CIFAR-10 dataset into three regimes: IID, Dirichlet (concentration parameter $\alpha^{D}=0.1$, self-balanced), and Pathological (3 classes per partition, random assignment). Distribution profiles for Pathological and Dirichlet are shown in \Cref{fig:dataset_distribution}. Hyperparameter details: $\beta=2, \gamma=1, \eta=0.0001$, optimizer=Adam. We compare the algorithms \mech{}, \mechtwo{}, \strategic{} and \FedAvg{}. The results are summarized in \Cref{fig:flower}, averaged across $5$ runs.

In \Cref{app:experiments_adversarial} we show that \mechtwo{} with trimmed mean (described in \Cref{sec:adversarial}) is robust to adversarial agents. 

\subsection{Utilities and payments for agents}
\label{app:experiment_utility}
\begin{figure}
    \centering
    \includegraphics[width=\linewidth]{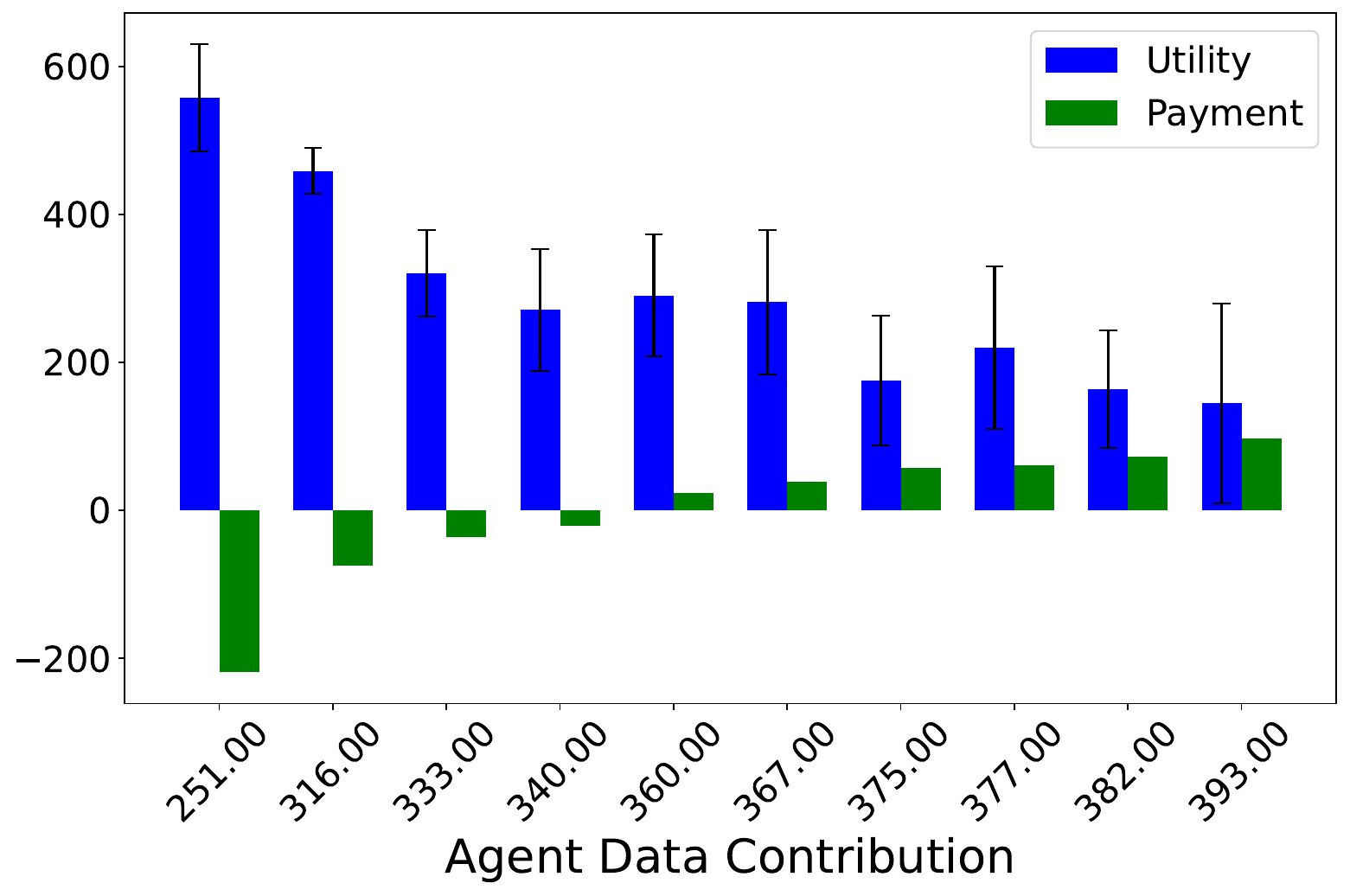}
    \caption{Utilities and payments received by agents for \mechtwo{} based on their data contributions for FEMNIST (averaged over 10 runs).}
    \label{fig:utility_payment}
    \vspace{-0.5cm}
\end{figure}

We also note that the data contributors get paid and data consumers pay in \mechtwo{}, and their payments are monotone non-decreasing with data contribution (\Cref{fig:utility_payment}). The utilities vary because the dataset on which the model is tested to obtain the accuracy varies for the agents.

\subsection{Model architecture}
We provide details of the model architecture used in our experiments in \Cref{tab:cifar10_model,tab:femnist_model}. 
\label{appendix:experiments_architecture}

\begin{table}[h]
\centering
\setlength{\tabcolsep}{2pt} 
\begin{tabular}{|l|c|c|}
\hline
\textbf{Layer (type:depth-idx)} & \textbf{Output Shape} & \textbf{Param \#} \\ \hline
Cifar10ResNet                   & {[}1, 10{]}          & --                \\ \hline
Conv2d: 1-1                     & {[}1, 64, 32, 32{]}  & 1,728             \\ \hline
BatchNorm2d: 1-2                & {[}1, 64, 32, 32{]}  & 128               \\ \hline
ReLU: 1-3                       & {[}1, 64, 32, 32{]}  & --                \\ \hline
Identity: 1-4                   & {[}1, 64, 32, 32{]}  & --                \\ \hline
Sequential (layer1): 1-5        & {[}1, 64, 32, 32{]}  & 147,968           \\ \hline
Sequential (layer2): 1-6        & {[}1, 128, 16, 16{]} & 525,568           \\ \hline
Sequential (layer3): 1-7        & {[}1, 256, 8, 8{]}   & 2,099,712         \\ \hline
Sequential (layer4): 1-8        & {[}1, 512, 4, 4{]}   & 8,393,728         \\ \hline
AdaptiveAvgPool2d: 1-9          & {[}1, 512, 1, 1{]}   & --                \\ \hline
Linear: 1-10                    & {[}1, 10{]}          & 5,130             \\ \hline
\textbf{Total Params}           &                      & \textbf{11,173,962} \\ \hline
\textbf{Trainable Params}       &                      & \textbf{11,173,962} \\ \hline
\textbf{Non-trainable Params}   &                      & \textbf{0}         \\ \hline
\textbf{Total Mult-Adds (GFLOPs)}&                     & \textbf{0.56}      \\ \hline
\textbf{Input Size (MB)}        &                      & \textbf{0.01}      \\ \hline
\textbf{Forward/Backward Pass Size (MB)} &             & \textbf{5.84}      \\ \hline
\textbf{Params Size (MB)}       &                      & \textbf{42.63}     \\ \hline
\textbf{Estimated Total Size (MB)} &                   & \textbf{48.48}     \\ \hline
\end{tabular}
\caption{Layer-wise Summary of Cifar10ResNet for CIFAR-10}
\label{tab:cifar10_model}
\end{table}
\begin{table}[h]
\centering
\setlength{\tabcolsep}{2pt}
\begin{tabular}{|l|c|c|}
\hline
\textbf{Layer (type:depth-idx)} & \textbf{Output Shape} & \textbf{Param \#} \\ \hline
FemnistCNN                      & {[}1, 62{]}          & --                \\ \hline
Conv2d: 1-1                     & {[}1, 32, 28, 28{]}  & 832               \\ \hline
MaxPool2d: 1-2                  & {[}1, 32, 14, 14{]}  & --                \\ \hline
Conv2d: 1-3                     & {[}1, 64, 14, 14{]}  & 51,264            \\ \hline
MaxPool2d: 1-4                  & {[}1, 64, 7, 7{]}    & --                \\ \hline
Linear: 1-5                     & {[}1, 2048{]}        & 6,424,576         \\ \hline
Linear: 1-6                     & {[}1, 62{]}          & 127,038           \\ \hline
\textbf{Total Params}           &                      & \textbf{6,603,710} \\ \hline
\textbf{Trainable Params}       &                      & \textbf{6,603,710} \\ \hline
\textbf{Non-trainable Params}   &                      & \textbf{0}         \\ \hline
\textbf{Total Mult-Adds (M)}    &                      & \textbf{17.21}     \\ \hline
\textbf{Input Size (MB)}        &                      & \textbf{0.00}      \\ \hline
\textbf{Forward/Backward Pass Size (MB)} &            & \textbf{0.39}      \\ \hline
\textbf{Params Size (MB)}       &                      & \textbf{26.41}     \\ \hline
\textbf{Estimated Total Size (MB)} &                  & \textbf{26.81}     \\ \hline
\end{tabular}

\caption{Layer-wise Summary of the CNN for FEMNIST}
\label{tab:femnist_model}
\end{table}

\subsection{Experiments with adversarial agents}
\label{app:experiments_adversarial}
\begin{figure*}[ht]
    \centering
    \begin{subfigure}[b]{0.48\linewidth}
        \centering
        \includegraphics[width=\linewidth]{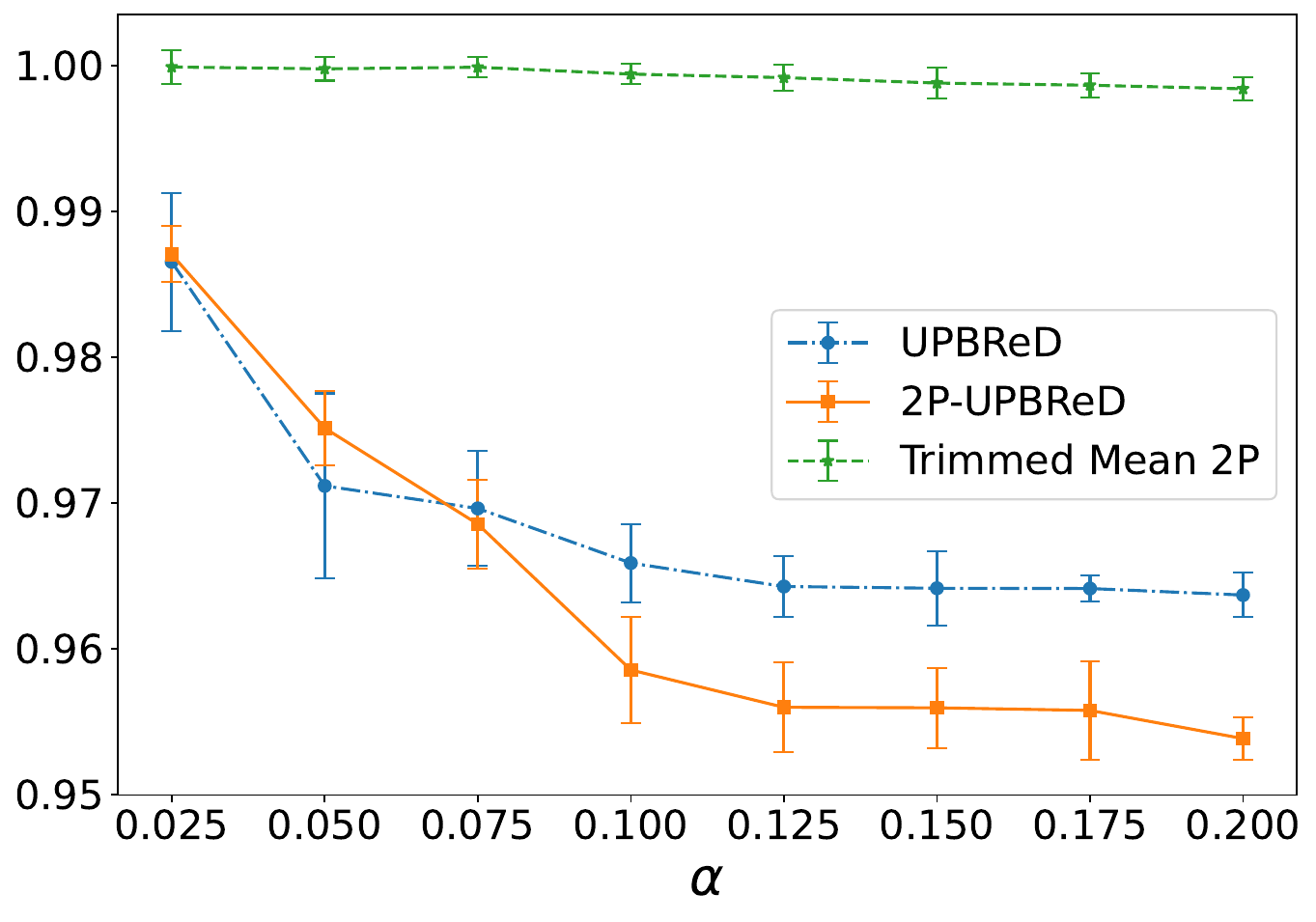}
        \caption{CIFAR-10}
        \label{fig:cifar}
    \end{subfigure}
    \hfill
    \begin{subfigure}[b]{0.48\linewidth}
        \centering
        \includegraphics[width=\linewidth]{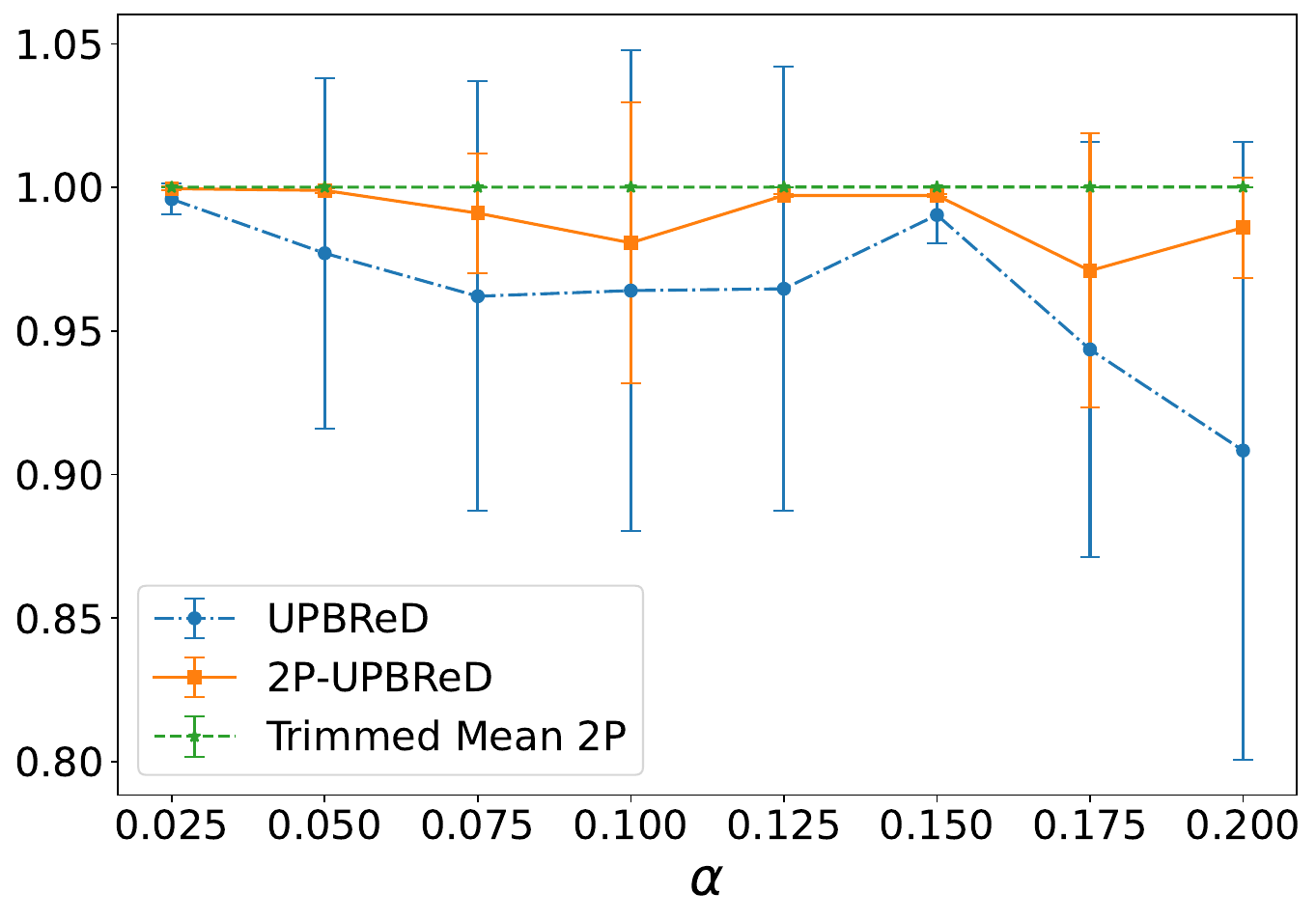}
        \caption{FEMNIST}
        \label{fig:femnist}
    \end{subfigure}
    \caption{Ratio of social welfare with adversaries and without adversaries.}
    \label{fig:adversarial_comparison}
\end{figure*}

In \Cref{sec:adversarial}, we describe the trimmed mean algorithm for \mechtwo{} designed to be robust in the presence of adversarial agents. In this section, we demonstrate this robustness on the FEMNIST and CIFAR-10 datasets. We vary the adversarial fraction $\alpha^{adv} \in \{0.025, 0.05, 0.075, 0.1, 0.125, 0.15, 0.175, 0.2\}$, selecting the trimming parameter $\beta^{trim}=\alpha^{adv}$. \Cref{fig:adversarial_comparison} shows the ratio of social welfare with and without adversaries for \mech{}, \mechtwo{} (both without trimming) and trimmed mean \mechtwo{}. While \mech{} and \mechtwo{} without trimming suffer welfare loss under adversaries, applying the trimmed mean algorithm (\Cref{algo:2Ptrimmedmean}) keeps the welfare ratio close to $1$ for trimmed mean \mechtwo{}. Adversarial agents transmit random updates generated from a standard multivariate Gaussian distribution. For a model with $m$ parameters, the adversarial gradient vector is drawn as $g \sim \mathcal{N}(0, \mathbf{I}_m)$, where each coordinate is sampled independently from $\mathcal{N}(0, 1)$.

\clearpage \pagebreak
%
%
%
%

\end{document}